\documentclass[11pt]{article}
\usepackage{cite}
\usepackage{amsmath,amsfonts,amssymb}
\usepackage[small,bf,hang]{caption}
\usepackage{slashed}
\usepackage{mathabx}
\usepackage{latexsym,epsfig}
\usepackage{stmaryrd}

\def\hybrid{
        \topmargin -20pt
        \oddsidemargin 0pt
        \headheight 0pt \headsep 0pt
        \textwidth 6.25in 
        \textheight 9.5in 
        \marginparwidth .875in
        \parskip 5pt plus 1pt \jot = 1.5ex}

\hybrid

 \csname
@addtoreset\endcsname{equation}{section}

\def\moth{\mathsurround=0pt}
\newdimen\zo \zo=0pt

\def\tick{\leaders\hrule height 0.5ex depth 0pt \hskip 0.5pt}
\def\upboxfill{$\moth \setbox\zo\hbox{\tick}%
  \hskip 3pt\hbox to 0pt{$\tick$\hss}\hrulefill \hbox to 7.5pt{$\tick$\hss}$}

\def\dtick{\leaders\hrule height .34pt depth 0.5ex \hskip 0.5pt}
\def\downboxfill{$\moth \setbox\zo\hbox{\dtick}%
  \hskip 2pt\hbox to 0pt{$\dtick$\hss}\hrulefill \hbox to 2pt{$\dtick$\hss}$}

\def\bec{\begin{center}}
\def\ec{\end{center}}

\def\d{\delta} 

\def\e{\epsilon}

\def\m{\mu}
\def\n{\nu}

 \def\det{{\rm det\,}}
\def\be{\begin{equation}}
\def\ee{\end{equation}}
\def\bea{\begin{eqnarray}}
\def\eea{\end{eqnarray}}
\def\ba{\begin{array}}
\def\ea{\end{array}}

\newcommand{\kkappa}{\gamma}
\newcommand{\KK}{{_{^{\rm KK}}}}

\begin{document}

\begin{titlepage}
\rightline{}
\rightline{\tt  MIT/CTP-4669}
\rightline{June 2015}
\begin{center}
\vskip .6cm
{\Large \bf {E$_{6(6)}$ Exceptional Field Theory:\\[1.5ex]
Review and Embedding of Type IIB}}\\
\vskip 1.2cm
{\large {Arnaud Baguet${}^1$, Olaf Hohm${}^2$ and Henning Samtleben${}^1$}}
\vskip .6cm
{\it {${}^1$Universit\'e de Lyon, Laboratoire de Physique, UMR 5672, CNRS}}\\
{\it {\'Ecole Normale Sup\'erieure de Lyon}}\\
{\it {46, all\'ee d'Italie, F-69364 Lyon cedex 07, France}}\\
arnaud.baguet@ens-lyon.fr, henning.samtleben@ens-lyon.fr
\vskip 0.2cm
{\it {${}^2$Center for Theoretical Physics}}\\
{\it {Massachusetts Institute of Technology}}\\
{\it {Cambridge, MA 02139, USA}}\\
ohohm@mit.edu

\vskip 1cm
{\bf Abstract}
\end{center}

\vskip 0.2cm

\noindent
\begin{narrower}
We review E$_{6(6)}$ exceptional field theory
with a particular emphasis on the embedding of type IIB supergravity, which  
is obtained by picking the GL$(5)\times {\rm SL}(2)$ invariant solution of the section constraint. 
We work out the precise decomposition 
of the E$_{6(6)}$ covariant fields on the one hand and the Kaluza-Klein-like 
decomposition of type IIB supergravity on the other. Matching the symmetries, this allows us to 
establish the precise dictionary between 
both sets of fields. Finally, we establish on-shell equivalence. In particular, 
we show how the self-duality constraint for the four-form potential in type IIB is reconstructed 
from the duality relations in the off-shell formulation of the E$_{6(6)}$ 
exceptional field theory.

\end{narrower}

\vskip 1.5cm

\begin{center}
Contribution to the {\it 
Proceedings of the Workshop on Quantum Fields and Strings, Corfu 2014
}
\end{center}

\end{titlepage}

\tableofcontents

\newpage

\section{Introduction}

One of the most intriguing aspects of maximal supergravity is the emergence 
of exceptional symmetry groups upon compactification on tori \cite{Cremmer:1979up}. 
For instance, compactifying 11-dimensional or type II supergravity  
to $D=5$ one obtains a rigid (continuous) E$_{6(6)}$ symmetry~\cite{Cremmer:1980gs}. 
Although these symmetries are 
understood as the supergravity 
manifestations of the (discrete) U-dualities of string-/M-theory~\cite{Hull:1994ys},   
from the point of view of conventional Riemannian geometry they are deeply mysterious. 
In fact, except for certain `geometric subgroups', the exceptional groups cannot be 
understood from the symmetries present in the conventional formulation of supergravity, although 
there is a reformulation of $D=11$ supergravity due to de~Wit and Nicolai in which the
compact subgroup ${\rm SU}(8)\subset {\rm E}_{7(7)}$ is manifest~\cite{deWit:1986mz}.
Over the decades this has led to various proposals of how to extend or embed 
the higher-dimensional theories in a way that explains the emergence of exceptional 
symmetries \cite{Koepsell:2000xg,West:2001as,Damour:2002cu,HenryLabordere:2002dk,West:2003fc,West:2004iz,Hillmann:2009ci}, 
but the complete formulation of such a theory, 
in the following called `exceptional field theory',  was only found quite recently
\cite{Hohm:2013pua,Hohm:2013vpa,Hohm:2013uia,Hohm:2014fxa},
using insights from `double field theory' 
\cite{Siegel:1993th,Hull:2009mi,Hull:2009zb,Hohm:2010jy,Hohm:2010pp,Hohm:2010xe},
subsequent generalizations to U-duality groups \cite{Berman:2010is,Berman:2011pe,Berman:2011jh,Aldazabal:2013mya}, 
and extended geometry~\cite{Hull:2007zu,Pacheco:2008ps,Coimbra:2011ky,Coimbra:2012af},   
an extension of the `generalized geometry' 
of~\cite{Hitchin:2004ut, Gualtieri:2003dx} to the case of exceptional duality groups. 
Here we will review the  E$_{6(6)}$
exceptional field theory with a particular emphasis on the explicit embedding of type IIB supergravity. 
The exceptional field theories in higher dimensions have been constructed in 
\cite{Hohm:2015xna,Abzalov:2015ega,Wang:2015hca} and the supersymmetric completions
have been given in~\cite{Godazgar:2014nqa,Musaev:2014lna}.

The formulation of exceptional field theory (EFT) is based on an extended spacetime 
that `geometrizes' the exceptional U-duality group. 
Specifically, in the E$_{6(6)}$ EFT all fields depend on $5+27$ coordinates $(x^{\mu},Y^M)$, 
where $\mu,\nu=0,\dots,4$, while lower  and upper indices 
$M,N=1,\ldots, 27$ label the (inequivalent) fundamental representations ${\bf 27}$ and  
$\bar{\bf 27}$ of E$_{6(6)}$, respectively. All functions on this extended space are subject 
to a covariant `section constraint' or `strong constraint' that implies that locally the fields
only live on a  `physical slice' of the extended space. In the present case this constraint can be written 
in terms of the invariant  symmetric $d$-symbol $d^{MNK}$ that E$_{6(6)}$ admits as 
 \be
  d^{MNK}\partial_N\partial_K A \ = \ 0\;, \qquad
  d^{MNK}\partial_NA\,\partial_K B \ = \ 0\;, 
  \label{section0}
 \ee
for arbitrary functions $A,B$ on the extended space. In particular, this constraint 
holds for all fields and gauge parameters. It was shown in \cite{Hohm:2013pua} that this constraint 
allows for (at least) two inequivalent 
solutions, in analogy to the type II double field theory \cite{Hohm:2011zr,Hohm:2011dv}. 
First, breaking E$_{6(6)}$ to GL$(6)$ the constraint is solved 
by fields depending on $6$ internal coordinates, and we recover the spacetime of 
11-dimensional supergravity. Second, breaking E$_{6(6)}$ to GL$(5)\times {\rm SL}(2)$ 
the constraint is solved by fields depending on $5$ internal coordinates, and we recover
the spacetime of type IIB supergravity. Indeed, upon picking one of these solutions 
one obtains a theory with the field content and symmetries of $D=11$ or type IIB supergravity, 
respectively, but in a non-standard formulation. These formulations are obtained from the 
standard ones by splitting the coordinates and tensor fields a la Kaluza-Klein, however, 
without truncating the coordinate dependence, as pioneered by de Wit and Nicolai \cite{deWit:1986mz}.
The full embedding of $D=11$ supergravity into EFT has been given in detail in \cite{Hohm:2013pua}.
In this article we provide all the details for the embedding of the type IIB theory.

In order to illustrate this formulation,  
an instructive analogy is the ADM formulation of, say, four-dimensional gravity, in which one singles out 
a `time direction', i.e., performs a $1+3$ split, and realizes spacetime as a one-dimensional 
foliation of three-geometries. 
One can similarly view the generalized spacetime of the E$_{6(6)}$ EFT as a five-dimensional 
foliation of a (generalized and extended) 27-dimensional geometry. However, an important 
difference is that the total 32-dimensional space cannot be viewed as a conventional manifold, because 
the gauge symmetries of EFT are governed by \textit{generalized} external and internal diffeomorphisms
satisfying an algebra that differs from the standard diffeomorphism algebra. 
Although the total space does not have a conventional geometrical interpretation, 
for the physical slices corresponding to the $D=11$ or type IIB solutions of the section constraint, 
describing \textit{inequivalent} subspaces of the extended space,  
 the generalized diffeomorphisms of EFT reduce to conventional 10 or 11-dimensional
diffeomorphisms plus tensor gauge transformations, 
thereby reconstructing the physical spacetimes in terms 
of five-dimensional foliations. 

Concretely, the E$_{6(6)}$ EFT has the following field content, 
with all fields depending on the $5+27$ coordinates $(x^{\mu},Y^M)$,  
 \be\label{EFTfields}
  g_{\mu\nu}\;, \quad {\cal M}_{MN}\;, \quad {\cal A}_{\mu}{}^{M}\;, \quad {\cal B}_{\mu\nu M}\;. 
 \ee
Here $g_{\mu\nu}$ is the external, five-dimensional metric, ${\cal M}_{MN}$ is the 
\textit{generalized} internal metric, while the tensor fields ${\cal A}_{\mu}{}^{M}$
and  ${\cal B}_{\mu\nu M}$ describe off-diagonal field components that encode,  
in particular, the interconnection between the external and internal generalized geometries. 
Upon breaking the E$_{6(6)}$ covariance by solving the section constraint,  
imposing that all fields depend only on a particular subset of the internal coordinates $Y^M$,
one can decompose the above fields in terms of their components. Modulo 
field redefinitions, these can then be interpreted as tensor fields with conventional gauge transformations. 
In this regime, and \textit{truncated} to the purely `internal' fields encoded in ${\cal M}_{MN}$, 
this formulation can be thought of as implementing what is sometimes referred to 
as extended or exceptional generalized geometry, 
which formally combines conventional tensors of different types 
into larger objects viewed as sections of extended tangent bundles~\cite{Hull:2007zu,Pacheco:2008ps}.   
For each solution of the section constraint we may thus reinterpret EFT as realizing 
a generalized geometry (enlarged, however, by including all `external' and `off-diagonal' fields in (\ref{EFTfields})
and dependence on external coordinates $x^\mu$), without additional unphysical coordinates.  
Why, then, do we insist on introducing seemingly unphysical coordinates, together with a constraint 
that eliminates most of them, as opposed to simply picking a solution from the start? 
Let us summarize several reasons why it is beneficial to 
work on such an extended space. 

\begin{itemize}

\item The theory is manifestly E$_{d(d)}$ covariant provided it is written with the extended 
derivatives $\partial_M$ properly transforming in the fundamental representation. 
For instance, the fields couple to the derivatives as 
in ${\cal A}_{\mu}{}^{M}\partial_M$. 
Thus, only this framework makes manifest the emergence 
of the E$_{d(d)}$ symmetry upon toroidal reduction by simply setting $\partial_M=0$. 

\item By defining EFT on the extended space we simultaneously cover $D=11$ supergravity 
and type IIB supergravity (and all of their Kaluza-Klein descendants). These are obtained 
by putting different solutions of the section constraint, which then determines, 
for instance, which field components in ${\cal A}_{\mu}{}^{M}\partial_M$ survive for which 
set of coordinates. In this way it is possible to describe in one single 
framework $D=11$ and type IIB supergravity, 
which are inequivalent theories and so would 
correspond to two different generalized geometries. 

\item
Although the  
coordinates beyond those of supergravity are unphysical,  at least in the currently understood formulation 
due to the strong form of the section constraint,  
in the full string theory they are actually physical and real. More precisely, at least for 
the T-duality subgroup O$(d-1,d-1)\subset {\rm E}_{d(d)}$ we known from closed string field theory
on toroidal backgrounds 
that the string field depends on momentum and winding coordinates, subject to 
the level-matching constraint that allows for a simultaneous dependence on 
all coordinates. It is thus unavoidable that eventually we come to terms with such 
extended spaces, and so it appears highly significant that much of this extended 
geometry is already visible at the level of the presently known EFT that essentially 
encodes supergravity.

\end{itemize}

Other than of conceptional interest, the manifestly covariant formulation of EFT has proven 
a rather powerful tool in order to describe consistent truncations of the standard supergravities,
in particular for sphere and hyperboloid compactifications in terms of 
generalized Scherk-Schwarz reductions~\cite{Hohm:2014qga},
see~\cite{Hohm:2011cp,Aldazabal:2011nj,Geissbuhler:2011mx,Grana:2012rr,Dibitetto:2012rk,Berman:2012uy,Musaev:2013rq,
Aldazabal:2013mya,Lee:2014mla,Blair:2014zba} for earlier related work.
This is remarkable, for although in these backgrounds there is no longer a physical E$_{d(d)}$
symmetry, the corresponding compactifications can be encoded very efficiently in terms of 
E$_{d(d)}$-valued twist matrices. The twist matrices take a universal form that is applicable to both  
solutions of the section constraint, so that, for instance, one covers in one stroke 
the sphere compactifications of $D=11$ supergravity, such as AdS$_4\times {\rm S}^7$~\cite{deWit:1986iy}
and AdS$_7\times {\rm S}^4$~\cite{Nastase:1999kf}, the AdS$_5\times {\rm S}^5$
compactification of type IIB, together with all their non-compact cousins, predicted in \cite{Hull:1988jw}.
In terms of the conventional formulation, 
this consistency requires a number of seemingly miraculous identities, 
suggesting the presence of an underlying larger structure --- 
the extended geometry of EFT. 
Combining the expressions for the E$_{6(6)}$-valued twist matrices together with the explicit dictionary
of the type IIB embedding into E$_{6(6)}$ EFT that we provide in this paper, allows to straightforwardly 
derive the non-linear reduction formulas for the full set of IIB fields on the sphere $S^5$ and hyperboloid 
$H^{p,q}$ backgrounds. We give that result in \cite{Baguet:2015xxx}.

This review article is organized as follows. In sec.~2 we briefly review the manifestly  E$_{6(6)}$ 
covariant formulation, introducing generalized diffeomorphisms and the tensor hierarchy 
governing one- and two-forms. This construction is completely rigid in that the theory is
uniquely determined by invariance under the bosonic gauge symmetries, i.e., 
internal and external generalized diffeomorphisms. In particular, nowhere is it necessary 
to refer to 11-dimensional or type IIB supergravity. The latter only emerge upon choosing a 
solution of the section constraint. In \cite{Hohm:2013vpa} it was shown in detail how $D=11$ supergravity, 
in a $5+6$ split of coordinates and tensor fields, is embedded in the E$_{6(6)}$ EFT. 
For the IIB theory, one can argue on general grounds that its embedding into EFT is guaranteed by 
the match of symmetries. In fact, it is easy to see that EFT yields the same field content 
as type IIB in the $5+5$ splitting, and we will show explicitly in sec.~3 that the 
EFT gauge algebra contains the full 10-dimensional diffeomorphism algebra. 
Together with the fact that both theories can be supersymmetrized and
reduce to the same 5-dimensional theory, it follows that EFT reduces to type IIB 
for the appropriate solution of the section constraint.
To be very explicit, in this article we work out the precise embedding formulas for IIB into EFT. 
To this end, we perform the Kaluza-Klein decomposition of type IIB without truncation in sec.~4 
and then establish the full dictionary with EFT in sec.~5. In particular, we will show how the 
duality constraints in EFT allow one to reconstruct the 3- and 4-forms that are not among the fundamental
fields of EFT but of course are present in type IIB. Finally, we review the generalized 
Scherk-Schwarz compactifications in sec.~6, which reduces the consistent embedding of five-dimensional
gauged supergravities into EFT to a set of consistency equations for the E$_{6(6)}$-valued twist matrices that 
capture the dependence on the internal coordinates.
By means of the explicit dictionary between EFT and IIB and $D=11$ supergravity, respectively, this gives rise to
the full reduction ansaetze for the consistent embedding into standard higher-dimensional supergravity.

\bigskip

\underline{Summary of conventions and notation}

The EFT fields are denoted by calligraphic letters, as in (\ref{EFTfields}). 
We keep the same letters for these fields after decomposing the
E$_{6(6)}$ indices down to ${\rm GL}(5)\times {\rm SL}(2)$ in accordance with
the IIB solution of the section constraint (\ref{section0}).
The two-forms require further redefinition which will be denoted by $\tilde{\cal B}_{\mu\nu}$.   

The original type IIB fields and space-time indices in $D=10$ on the other hand
are indicated by hats, and the forms are called $C$:
 \be
  \hat{G}_{\hat{\mu}\hat{\nu}}\;, \qquad \hat{C}_{\hat{\mu}\hat{\nu}\hat{\rho}\hat{\sigma}}\;, \quad {\rm etc.} 
 \ee 
Upon Kaluza-Klein decomposition of the IIB fields, the new variables obtained by a standard procedure of 
flattening and unflattening of indices are denoted by a bar. The presence of Chern-Simons terms in the IIB field 
strengths requires yet another redefinition to bring the gauge structure into canonical form, which
we denote without any hat.
Thus we have the series of field redefinitions 
 \be
  \hat{C}\;\rightarrow \; \overline{C}\;\rightarrow \; C
  \;. 
 \ee
These fields will eventually be identified with the various
components of the EFT fields.

In section~6, we describe Scherk-Schwarz reduction of EFT, parametrizing all EFT
fields in terms of $Y$-dependent E$_{6(6)}$-valued twist matrices and the corresponding $x$-dependent 
fields of five-dimensional supergravity, which we denote by straight letters:
 \be
 \begin{split}
  &g_{\mu\nu}(x,Y) \ \rightarrow \ {\bf g}_{\mu\nu}(x)\;, \qquad
  {\cal M}_{MN}(x,Y) \ \rightarrow \  M_{MN}(x)\;, \\
  &{\cal A}_{\mu}{}^{M}(x,Y) \ \rightarrow \ A_{\mu}{}^{M}(x)\;, \qquad
  {\cal B}_{\mu\nu\, M}(x,Y) \ \rightarrow \ B_{\mu\nu\, M}(x)\;.
 \end{split}
 \ee

\section{Review of E$_{6(6)}$ Exceptional Field Theory}
Here we present a brief review of the E$_{6(6)}$ EFT, starting with the 
generalized Lie derivatives and  their gauge algebra (the `E-bracket'), 
which govern the internal (generalized) diffeomorphisms. We then 
introduce the tensor hierarchy and define the full gauge transformations, 
including generalized external diffeomorphisms, in order to construct 
the complete theory.  

\subsection{Generalized diffeomorphisms and tensor hierarchy}
We start by collecting the relevant facts about E$_{6(6)}$. Its dimension is $78$ 
and we denote the generators by $t_{\boldsymbol{\alpha}}$, with Cartan-Killing form 
$\kappa_{{\boldsymbol{\alpha}}{\boldsymbol{\beta}}}$.  
As recalled in the introduction, 
E$_{6(6)}$ admits two inequivalent fundamental representations of dimension 27, 
denoted by ${\bf 27}$ and $\bar{\bf 27}$ and labelled by indices $M,N=1,\ldots, 27$. 
In these fundamental representations,
there are two cubic E$_{6(6)}$-invariant tensors, the fully symmetric $d$-symbols $d^{MNK}$ and $d_{MNK}$, 
which we normalize as $d_{MPQ}d^{NPQ} = \delta_M{}^N$.
The $d$-symbols define the manifestly E$_{6(6)}$ covariant section constraint~\cite{Coimbra:2011ky}
 \be
  d^{MNK}\,\partial_N \partial_K A \ = \ 0\;, \quad  d^{MNK}\,\partial_NA\, \partial_K B \ = \ 0 \,,  
 \label{sectioncondition}
 \ee  
and also satisfy the following cubic identities 
 \be\label{cubicidentity}
  \begin{split}
   d_{S(MN} \,d_{PQ)T}\, d^{STR} \ &= \ \frac{2}{15}\delta_{(M}{}^{R} \,d_{NPQ)}\;, \\
   d_{STR}\, d^{S(MN} \, d^{PQ)T}  \ &= \ \frac{2}{15}\delta_{R}{}^{(M} \,d^{NPQ)}\;. 
  \end{split}
 \ee  
 
In order to define the generalized Lie derivatives below we need  
the projector onto the adjoint representation in the tensor product 
${\bf 27}\otimes \bar{\bf 27}={\bf 78}+\cdots$, which reads 
\bea\label{adjproj}
\mathbb{P}^M{}_N{}^K{}_L
&\equiv& (t_{\boldsymbol{\alpha}})_N{}^M (t^{\boldsymbol{\alpha}})_L{}^K ~=~
\frac1{18}\,\delta_N{}^M\delta_L{}^K + \frac16\,\delta_N{}^K\delta_L{}^M
-\frac53\,d_{NLR}d^{MKR}\;.
\eea
With respect to a vector like parameter $\Lambda^M$ one would naively define 
the Lie derivative as in standard geometry, acting on, say, a vector as 
 \be\label{standLie} 
  {\cal L}_{\Lambda}V^M \ \equiv \ \Lambda^K \partial_K V^M-\partial_K\Lambda^M V^K\;. 
 \ee 
The problem with applying this definition to EFT is that some fields are subject to further constraints, 
for instance the generalized metric ${\cal M}_{MN}$ is an E$_{6(6)}$-valued matrix, 
and this condition is not preserved under  (\ref{standLie}).  
This is fixed by simply projecting the tensor $\partial_K\Lambda^M$ living 
in ${\bf 27}\otimes \bar{\bf 27}$ onto the adjoint by means of the projector (\ref{adjproj}). 
Gauge transformations w.r.t.~the internal diffeomorphism parameter 
$\Lambda^M$ for a vector with upper or lower indices in terms of the 
\textit{generalized} Lie derivative, denoted by $\mathbb{L}_{\Lambda}$ in the following,  
are thus defined as~\cite{Coimbra:2011ky}
\bea\label{genLie}
\delta V^M &=& \mathbb{L}_{\Lambda} V^M 
\ \equiv \ \Lambda^K \partial_K V^M - 6\, \mathbb{P}^M{}_N{}^K{}_L\,\partial_K \Lambda^L\,V^N
+\lambda\,\partial_P \Lambda^P\,V^M
\;,
\nonumber\\
\delta W_M &=& \mathbb{L}_{\Lambda} W_M 
\ \equiv \ \Lambda^K \partial_K W_M + 6\, \mathbb{P}^N{}_M{}^K{}_L\,\partial_K \Lambda^L\,W_N
+\lambda'\,\partial_P \Lambda^P\,W_M
\;.
\eea
Here we also included a density term proportional to $\lambda \, \partial_P \Lambda^P$. 
The generalized Lie derivatives are consistent for arbitrary density weights $\lambda$, 
and indeed in formulating EFT it is crucial to assign particular non-trivial weights to the fields. 
Writing out the projector (\ref{adjproj}), the gauge transformations are given by 
\bea\label{explicitLie}
   \delta _{\Lambda}V^M &=&   \Lambda^K \partial_K V^M-\partial_K\Lambda^M V^K
  +\Big(\lambda-\frac{1}{3}\Big)\,\partial_P\Lambda^P\, V^M
  +10\, d_{NLR}\, d^{MKR}  \partial_K\Lambda^L V^N\;,
  \nonumber\\
    \delta _{\Lambda}W_M &=&   \Lambda^K \partial_K W_M + \partial_M\Lambda^K W_K
  +\Big(\lambda+\frac{1}{3}\Big)\,\partial_P\Lambda^P\, W_M
  -10\, d_{MLR}\, d^{NKR}\partial_K\Lambda^L W_N .
\eea
The generalized Lie derivatives can similarly be defined for E$_{6(6)}$ tensors with an arbitrary 
number of upper and lower fundamental indices. In particular, the gauge transformations 
for the generalized metric take the form $\delta_{\Lambda}{\cal M}_{MN}=\mathbb{L}_{\Lambda}{\cal M}_{MN}$, 
with the generalized Lie derivative for density weight $\lambda=0$. 
In this form the condition ${\cal M}\in {\rm E}_{6(6)}$ is indeed preserved. 

Given the modified form of generalized Lie derivatives, as opposed to the conventional 
Lie derivatives, it is no longer clear that they are consistent, in particular that they 
satisfy an algebra, i.e., that they lead to gauge transformations that close. 
Closure can, however, be established, but here it is crucial to employ the section constraint (\ref{sectioncondition}). 
An explicit computation then shows that the generalized Lie derivatives close according to 
 \be
  \big[\mathbb{L}_{\Lambda_1},\mathbb{L}_{\Lambda_2}\big] \ = \ \mathbb{L}_{[\Lambda_1,\Lambda_2]_{\rm E}}\;, 
 \ee
with the `E-bracket'   
 \be\label{Ebracket}
   \big[\Lambda_1,\Lambda_2\big]_{\rm E}^M \ \equiv \ 
   2\,\Lambda_{[1}^{K}\partial_K\Lambda_{2]}^{M}
   -10\,d^{MNP}\,d_{KLP}\,\Lambda_{[1}^{K}\partial_N\Lambda_{2]}^{L}\;. 
  \ee 
The first term in here has the same form as the standard Lie bracket governing the 
algebra of standard diffeomorphisms. The second term explicitly involves 
the E$_{6(6)}$ structure in form of the $d$-symbols. Thus, the gauge algebra on this space 
differs from the diffeomorphism algebra. 
In particular, the Lie derivative of a generalized vector w.r.t.~another generalized 
vector (both of weights $\tfrac{1}{3}$) does not coincide with their E-bracket. 
More precisely, the antisymmetric part coincides with the E-bracket, but there 
is a non-trivial symmetric part, given by 
 \be\label{symmLieIdentity}
  \big(\mathbb{L}_{V}W+\mathbb{L}_{W}V\big)^M \ = \ 10\,d^{MNK} d_{PQK}\partial_N\big(V^PW^Q\big)  \;. 
 \ee 
Moreover,  
the E-bracket does not define a Lie algebra in that the Jacobi identity is not satisfied. 
The non-trivial `Jacobiator' as well as the `anomalous' symmetric part in (\ref{symmLieIdentity}) 
are, however, of the form $\Lambda^M=d^{MNK}\partial_N\chi_K$, 
for some explicit function $\chi$, 
and one can verify that due to the section constraint the Lie derivative vanishes for this parameter. 
Hence, the Jacobi identity does hold acting on fields satisfying the strong constraint (see \cite{Berman:2012vc}
for more details), but the non-vanishing Jacobiator has important consequences, 
upon taking into account the external coordinate dependence.

So far we have defined the generalized internal diffeomorphisms by 
generalized Lie derivatives. We will refer to a tensor structure as transforming 
`covariantly' iff its transformation is governed by the generalized Lie derivative (of some weight)
and call such objects generalized tensors. 
Since all fields are functions of internal and external coordinates $Y^M$ and $x^{\mu}$, respectively, 
we now need to set up a calculus that allows us to differentiate w.r.t.~$x^{\mu}$. 
Indeed, as all fields and parameters in the full theory, $\Lambda^M=\Lambda^M(x,Y)$ depends 
on the external $x^{\mu}$ and therefore the derivative $\partial_{\mu}$ of any tensor field
is not covariant in the above sense. 
In order to remedy this we introduce a gauge connection ${\cal A}_{\mu}{}^{M}$, 
of which we can think as taking values in the `E-bracket algebra', and define 
the covariant derivatives 
 \be
 \label{DDcov}
  {D}_{\mu} \ \equiv \ \partial_{\mu}-\mathbb{L}_{{\cal A}_{\mu}}\;. 
 \ee  
The covariant derivative of any generalized tensor then transforms covariantly 
provided the gauge vector transforms as $\delta_{\Lambda}{\cal A}_{\mu}{}^{M}={ D}_{\mu}\Lambda^{M}$, 
where the gauge parameter $\Lambda^M$ carries weight $\lambda_{\Lambda}=\tfrac{1}{3}$. 
Next, we would like to define a field strength for ${\cal A}_{\mu}{}^{M}$. 
Naively, one would write the standard formula for the field strength or curvature 
of a gauge connection, but with the Lie bracket replaced by the E-bracket (\ref{Ebracket}). 
However, since the E-bracket does not satisfy the Jacobi identity the resulting object 
does not transform covariantly and also does not satisfy a Bianchi identity. 
Since the failure of the E-bracket to satisfy the Jacobi identity is of the form $d^{MNK}\partial_N\chi_K$
we can repair this by introducing two-forms ${\cal B}_{\mu\nu \,M}$ with appropriate 
gauge transformations 
and adding the term $d^{MNK}\,\partial_K {\cal B}_{\mu\nu\,N}$ to the field strength. 
This defines (the beginning of) the so-called tensor hierarchy, originally introduced 
in gauged supergravity~\cite{deWit:2005hv,deWit:2008ta}.  
Using (\ref{Ebracket}) we thus obtain the field strength 
\bea\label{YM}
{\cal F}_{\mu\nu}{}^M &=&  
2\, \partial_{[\mu} {\cal A}_{\nu]}{}^M 
-2\,{\cal A}_{[\mu}{}^K \partial_K {\cal A}_{\nu]}{}^M 
+10\, d^{MKR}d_{NLR}\,{\cal A}_{[\mu}{}^N\,\partial_K {\cal A}_{\nu]}{}^L
\nonumber\\
&&{}
 + 10 \, d^{MNK}\,\partial_K {\cal B}_{\mu\nu\,N}
\;.
\eea
This tensor transforms covariantly under the appropriate gauge transformations 
of ${\cal A}$ and ${\cal B}$ given in (\ref{deltaAB}) below. 
The presence of the 2-form in (\ref{YM}) also ensures that this
field strength satisfies a modified covariant Bianchi identity
 \be
  3 \,{D}_{[\mu}{\cal F}_{\nu\rho]}{}^M \ = \ 10\, d^{MNK}\partial_K{\cal H}_{\mu\nu\rho\,N}\;,
  \label{Bianchi}
 \ee 
giving rise to the  3-form curvature of the 2-form. 
The 3-form field strength ${\cal H}_{\mu\nu\rho \,M}$ is defined by this equation as
\bea\label{HCurvature}
\ {\cal H}_{\mu\nu\rho\,M} &=&
3\,{D}_{[\mu} {\cal B}_{\nu\rho]\,M}
-3\,d_{MKL}\, {\cal A}_{[\mu}{}^K\,\partial_{\vphantom{[}\nu} {\cal A}_{\rho]}{}^L 
+ 2\,d_{MKL}\, {\cal A}_{[\mu}{}^K {\cal A}_{\vphantom{[}\nu}{}^P \partial_P {\cal A}_{\rho]}{}^L 
\nonumber\\
&&{}
-10\, d_{MKL}d^{LPR}d_{RNQ}\, {\cal A}_{[\mu}{}^K {\cal A}_{\vphantom{[}\nu}{}^N\,\partial_P {\cal A}_{\rho]}{}^Q+\cdots 
\;,
\eea
where ${\cal B}$ carries weight $\lambda_{{\cal B}}=\frac23$,
up to terms that vanish under the projection with $d^{MNK} \partial_K$.
Now in turn we can establish a Bianchi identity for ${\cal H}$, which reads 
\bea\label{calHBianchi}
4\, {D}_{[\mu} {\cal H}_{\nu\rho\sigma]M} &=& 
-3 \,d_{MPQ} {\cal F}_{[\mu\nu}{}^P {\cal F}_{\rho\sigma]}{}^Q
+\dots\;,
\eea
again up to terms annihilated by the projection with $d^{MNK}\partial_K$.

We close this section by collecting the complete bosonic gauge transformations.  
The external and internal metric $g_{\mu\nu}$ and ${\cal M}_{MN}$ transform 
under internal generalized diffeomorphisms as a scalar density of weight $\tfrac{2}{3}$
and a symmetric 2-tensor of weight zero, respectively. 
Recalling that ${\cal A}$ carries weight $\lambda=\tfrac{1}{3}$ and noting that ${\cal B}$
carries weight $\lambda=\tfrac{2}{3}$, the gauge transformations then read  
 \be
 \begin{split}
   \delta {\cal A}_\mu{}^M \ &= \ D_\mu \Lambda^M  
   -10\, d^{MNK} \partial_K\Xi_{\mu N}\;, \\
   \Delta {\cal B}_{\mu\nu M} \ &= \ 2D_{[\mu}\Xi_{\nu]\,M} 
   +d_{MKL}\Lambda^K{\cal F}_{\mu\nu}{}^{L}+ {\cal O}_{\mu\nu M}\;, 
 \end{split}  
 \label{deltaAB}
 \ee
where we defined 
\bea\label{DELTAB}
\Delta {\cal B}_{\mu\nu\,N} &\equiv& \delta {\cal B}_{\mu\nu\,N} + d_{NKL}\,{\cal A}_{[\mu}{}^K\, \delta {\cal A}_{\nu]}{}^L
\;. 
\eea
Here we also specified the gauge transformations under the new parameter $\Xi_{\mu M}$
of weight $\tfrac{2}{3}$ 
associated to the 2-form, and we note that the gauge transformations are 
so far only determined  
up to yet unspecified terms ${\cal O}_{\mu\nu M}$ satisfying 
 \be
  d^{MNK} \partial_K{\cal O}_{\mu\nu N} \ = \ 0\;.
 \ee 
This corresponds to the gauge redundancy of the next form in the tensor hierarchy, 
but it turns out that this ambiguity drops out of all terms in the action and equations 
of motion.

We finally give the form of the \textit{external} diffeomorphisms of the $x^{\mu}$, which are 
generated by a parameter $\xi^{\mu}=\xi^{\mu}(x,Y)$, 
 \bea
 \delta_{\xi} e_{\mu}{}^{a} &=& \xi^{\nu}{D}_{\nu}e_{\mu}{}^{a}
 + {D}_{\mu}\xi^{\nu} e_{\nu}{}^{a}\;, \nonumber\\
\delta_{\xi} {\cal M}_{MN} &=& \xi^\mu \,{D}_\mu {\cal M}_{MN}\;,\nonumber\\
\delta_{\xi} {\cal A}_{\mu}{}^M &=& \xi^\nu\,{\cal F}_{\nu\mu}{}^M + {\cal M}^{MN}\,g_{\mu\nu} \,\partial_N \xi^\nu
\;,\nonumber\\
\Delta_{\xi} {\cal B}_{\mu\nu\,M} &=& \tfrac1{2\,\sqrt{10}}\,\xi^\rho\,
 e\varepsilon_{\mu\nu\rho\sigma\tau}\, {\cal F}^{\sigma\tau\,N} {\cal M}_{MN} 
 \;. 
 \label{skewD}
\eea
Let us note that they take the same form as standard diffeomorphisms generated 
by conventional Lie derivatives, except that all partial derivatives are replaced by gauge covariant 
derivatives. Moreover, in $\delta{\cal A}_{\mu}$ there is an additional ${\cal M}$-dependent term 
and in $\Delta {\cal B}_{\mu\nu}$ the naively covariant form $\xi^{\rho}{\cal H}_{\mu\nu\rho}$ 
has been replaced according to a duality relation to be discussed momentarily. 
We will discuss these external diffeomorphisms, in particular their gauge algebra, 
in more detail in sec.~2.4 below.

\subsection{ E$_{6(6)}$ covariant dynamics}
Let us now define the dynamics of the E$_{6(6)}$ EFT by giving the unique  
action principle on the extended space, which decomposes into the five terms 
 \be\label{EFTaction}
  S_{\rm EFT} \ = \ S_{\rm EH}+{S}_{\rm sc}+{S}_{\rm VT}+S_{\rm top} -V\;. 
 \ee 
The first term formally takes the same form as the standard Einstein-Hilbert term, 
 \be
 \label{EH}
  S_{\rm EH} \ = \ \int d^5x \,d^{27}Y\,e\,\widehat{R} \ = \ 
  \int d^5x \,d^{27}Y\,e\,e_{a}{}^{\mu}e_{b}{}^{\nu} \widehat{\cal R}_{\mu\nu}{}^{ab}\;, 
 \ee
except that in the definition of the Riemann tensor all partial derivatives 
are replaced by ${\cal A}_{\mu}$ covariant derivatives and one adds 
an additional term to make it properly local Lorentz invariant, 
$\widehat{R}_{\mu\nu}{}^{ab}\equiv R_{\mu\nu}{}^{ab}+{\cal F}_{\mu\nu}{}^{M}
e^{\rho[a}\partial_M e_{\rho}{}^{b]}$. The second term is the `scalar-kinetic' term 
defined by 
\bea
{\cal L}_{\rm sc} &=& \frac1{24}\,e\,g^{\mu\nu}\,{D}_{\mu}{\cal M}_{MN}\,{D}_{\nu}{\cal M}^{MN}
\;,
\label{Lsc}
\eea
with $e\equiv\sqrt{|g|}$.
The third term in (\ref{EFTaction}) is the kinetic term for the gauge-vectors, written in terms of the 
gauge covariant curvature (\ref{YM}), 
 \bea
{\cal L}_{\rm VT}&\equiv &
-\frac14\,e\,{\cal F}_{\mu\nu}{}^M{\cal F}^{\mu\nu\,N}\,{\cal M}_{MN}
\;.
\label{LVT}
\eea
The fourth term is a Chern-Simons-type topological term, which is only gauge invariant 
up to boundary terms. It is most conveniently defined 
by writing it as a manifestly gauge invariant action in one higher dimension, where 
it reduces to a total derivative, reducing it to the boundary integral in one dimension lower. 
Using form notation it reads 
\bea
S_{\rm top} &=& 
\int d^5 x\,d^{27}Y\,{\cal L}_{\rm top} 
\nonumber\\
&=&
\tfrac{1}{6}\sqrt{10}\, \!\int d^{27}Y \int_{{\cal M}_6}\,\left(
d_{MNK}\,{\cal F}^M \wedge  {\cal F}^N \wedge  {\cal F}^K
-40\, d^{MNK}{\cal H}_M\,  \wedge \partial_N{\cal H}_K
\right)
\,.
\label{CSlike}
\eea
Under a general variation of ${\cal A}$ and ${\cal B}$ the topological Lagrangian varies as 
\bea
\delta{\cal L}_{\rm top} &=&
\tfrac18\,\sqrt{10}\,\varepsilon^{\mu\nu\rho\sigma\tau}
\Big(d_{MNK}\,
{\cal F}_{\mu\nu}{}^M {\cal F}_{\rho\sigma}{}^N  \delta {\cal A}_\tau{}^K
+\tfrac{20}{3}\,d^{MNK}\,\partial_N{\cal H}_{\mu\nu\rho \,M} \,
\Delta {\cal B}_{\sigma\tau\,K} \Big)
\;.
\label{vartopo}
\eea
The final term in the action is the 
`scalar potential' that involves only internal derivatives $\partial_M$ and reads 
\be\label{fullpotential}
 \begin{split}
  V \ = \ &-\frac{1}{24}{\cal M}^{MN}\partial_M{\cal M}^{KL}\,\partial_N{\cal M}_{KL}+\frac{1}{2} {\cal M}^{MN}\partial_M{\cal M}^{KL}\partial_L{\cal M}_{NK}\\
  &-\frac{1}{2}g^{-1}\partial_Mg\,\partial_N{\cal M}^{MN}-\frac{1}{4}  {\cal M}^{MN}g^{-1}\partial_Mg\,g^{-1}\partial_Ng
  -\frac{1}{4}{\cal M}^{MN}\partial_Mg^{\mu\nu}\partial_N g_{\mu\nu}\;. 
 \end{split} 
 \ee 
Its form is uniquely determined by the internal generalized diffeomorphism invariance 
(up to the relative coefficient between the last two terms in 
the second line that is, however, universal for all EFTs).

The field equations of the E$_{6(6)}$ EFT follow by varying (\ref{EFTaction}) naively w.r.t.~all 
fields. For now we focus on the field equations for the two-form only, because they will be significant below. 
The 2-form ${\cal B}_{\mu\nu M}$ does not enter with a kinetic term, but appears inside the Yang-Mills-type 
kinetic term, c.f.~the definition (\ref{YM}), and the topological term (\ref{CSlike}). 
Therefore, its field equations are 
first order and read 
\be\label{dualityrel}
d^{MNK}\partial_K  \left(e{\cal M}_{NL} {\cal F}^{\mu\nu L}
 +\frac16\,\sqrt{10}\,  \varepsilon^{\mu\nu\rho\sigma\tau}\,
  {\cal H}_{\rho\sigma\tau N}\right) \ = \ 0
  \,. 
 \ee 
These equations take the same form as the standard duality relations in five dimensions 
between vectors and two-forms. However, here they appear only under a differential 
operator, which thus leads to different sets of duality relations for different solutions of the section constraint.

\subsection{Fermions and Supersymmetry}

The bosonic sector of exceptional field theory is uniquely determined upon imposing invariance
under generalized diffeomorphisms in the internal and external space-time.
Supersymmetry has not been imposed in order to determine the interactions; however, as expected
the bosonic action (\ref{EFTaction}) can be embedded into a supersymmetric theory~\cite{Musaev:2014lna}.
The fermions of the theory are those of the maximal five-dimensional theory~\cite{Cremmer:1980gs}, 
however, living now on the full $(5+27)$-dimensional space-time (subject to the section constraint).
In particular, they are ${\rm SO}(1,4)$ symplectic Majorana spinors spinors and we refer to~\cite{deWit:2004nw} 
for our spinor conventions.\footnote{
Just for the conventions for the Levi-Civita density we follow~\cite{Hohm:2013vpa,Musaev:2014lna},
with the two conventions related by 
$\varepsilon_{\mu\nu\rho\sigma\tau}^{[1312.0614]}=-i\varepsilon_{\mu\nu\rho\sigma\tau}^{[{\rm hep-th}\slash0412173]}$.
Accordingly, $\gamma$-matrices satisfy $\gamma^{abcde} = i\varepsilon^{abcde}$\,.
}
With respect to the R-symmetry group (or generalized internal Lorentz group) ${\rm USp}(8)$, the fermion fields
fall into irreducible representations 
with the gravitino fields $\psi_\mu^i$ transforming in the fundamental 8, and the spin-$\frac12$ 
fermions $\chi^{ijk}$ transforming in the totally anti-symmetric, $\Omega$-traceless 42
\bea
\chi^{ijk}=\chi^{\llbracket ijk\rrbracket}
~\equiv~ \chi^{ijk} - \frac12\,\Omega^{[ij} \chi^{k]mn}\Omega_{mn}
\;,
\eea
where $\Omega_{ij}=\Omega_{[ij]}$ denotes the symplectic invariant tensor. 
Here and in the following we use the notation of double brackets $\llbracket\dots\rrbracket$
to denote the projection of an ${\rm USp}(8)$ tensor onto the $\Omega$-traceless part.
With respect to generalized internal diffeomorphisms (\ref{genLie}) 
the fermionic fields transform as weighted scalars of weight 
$\lambda_{\psi}=\frac16$, $\lambda_{\chi}=-\frac16$\,.

Coupling of the fermions requires the introduction of frame fields underlying the
external and internal metric, 
\bea
g_{\mu\nu}~=~ e_\mu{}^a e_\nu{}^b\eta_{ab}\;,\qquad
{\cal M}_{MN} &=& {\cal V}_M{}^{ij} {\cal V}_{N\,ij}\;,
\label{defgM}
\eea
with the f\"unfbein $e_\mu{}^a$, and the pseudo-real 27-bein 
\bea
\left\{ {\cal V}_{M}{}^{ij}, {\cal V}_{M\,ij}
= ({\cal V}_M{}^{ij})^*={\cal V}_M{}^{kl}\Omega_{ki}\Omega_{lj} \right\}
\;,
\eea
satisfying ${\cal V}_{M}{}^{ij}={\cal V}_{M}{}^{\llbracket ij\rrbracket}$\,.
The inverse 27-bein is defined as
\bea
{\cal V}_M{}^{ij}{\cal V}_{ij}{}^N &=& \delta_M{}^N\;,\qquad
{\cal V}_M{}^{kl}{\cal V}_{ij}{}^M~=~\delta^{kl}_{ij}-\frac18\Omega_{ij}\Omega^{kl}
\;,
\label{invV}
\eea
with conventions $\d^{ij}_{kl}=\frac12(\d^i_k\d^j_l-\d^i_l\d^j_k)$ and $\Omega_{ik}\Omega^{jk}=\delta_i^j$\,.
The 27-bein is an ${\rm E}_{6(6)}$ group-valued matrix, which is encoded in 
the structure of its infinitesimal variation, 
\bea
\delta {\cal V}_M{}^{ij} &=& 
-2 \,\delta q_k{}^{[i}\,{\cal V}_M{}^{j]k}   + \delta p{}^{ijkl}\,{\cal V}_M{\,}_{kl} 
\;,
\label{varVpq}
\eea
with $\delta q_i{}^{j}$ and $\delta p{}^{ijkl}$ spanning the ${\bf 36}$ and ${\bf 42}$ of ${\rm USp}(8)$,
respectively, i.e.\
\bea
\delta q_i{}^{j}   &=& - \delta q_l{}^{k} \Omega_{ik} \Omega^{jl} \;,\qquad
\delta p{}^{ijkl} ~=~ \delta p{}^{\llbracket ijkl\rrbracket}
\;,
\label{pq}
\eea
and corresponding to the compact and non-compact generators of $\mathfrak{e}_{6(6)}$, respectively.

The full ${\rm SO}(1,4) \times {\rm USp}(8)$ covariant derivatives are then defined as 
\bea
{\cal D}_\mu \psi{}^i &\equiv&
\partial_\mu \psi{}^i +\frac14\,\omega_{\mu}{}^{ab} \gamma_{ab} \,\psi^i
- {\cal Q}_{\mu\,j}{}^i \psi{}^j -\mathbb{L}_{{\cal A}_\mu}\,\psi^i\;,
\nonumber\\
{\cal D}_M \psi{}^i &\equiv&
\partial_M \psi{}^i +\frac14\,\omega_{M}{}^{ab} \gamma_{ab} \,\psi^i
- {\cal Q}_{M\,j}{}^i \psi{}^j
\;,
\label{DM}
\eea
with spin connections $\omega$, ${\cal Q}$ defined in terms of the bosonic frame fields
and the Lie derivative ${\mathbb{L}}$ taking care of the weight of $\psi^i$ under
generalized diffeomorphisms.
From the spin connections the Christoffel connections 
$\Gamma_{\mu\nu}{}^\rho$, $\Gamma_{MN}{}^K$, can be defined
by the generalized vielbein postulates
\bea
0&\equiv&\nabla_\mu e_\nu{}^a=
{\cal D}_{\mu} e_{\nu}{}^a - \Gamma_{\mu\nu}{}^\rho\,e_\rho{}^a  
 = D_{\mu}e_{\nu}{}^a+\omega_{\mu}{}^{ab}e_{\nu\,b} - \Gamma_{\mu\nu}{}^\rho\,e_\rho{}^a 
\;,
\label{GVP}\\
0&\equiv&\nabla_M {\cal V}_N{}^{ij}=
{\cal D}_M {\cal V}_N{}^{ij} - \Gamma_{MN}{}^K\,{\cal V}_K{}^{ij}
=
{\partial}_M {\cal V}_N{}^{ij} 
+2\,{\cal Q}_{M\,k}{}^{[i} {\cal V}_N{}^{j]k}
- \Gamma_{MN}{}^K\,{\cal V}_K{}^{ij}
\;,
\nonumber
\eea
declaring covariant constancy of the frame fields. In turn, the spin connections are defined by
properly generalized vanishing torsion conditions.
For the ${\rm SO}(1,4)$ connection $\omega_\mu{}^{ab}$ the absence of torsion takes the familiar form 
\bea
{\cal D}_{[\mu} e_{\nu]}{}^a \equiv D_{[\mu}e_{\nu]}{}^a+\omega_{[\mu}{}^{ab}e_{\nu]b}~\stackrel{!}{=}~0
 \qquad \Longleftrightarrow\qquad \Gamma_{[\mu\nu]}{}^\rho ~=~0\;,
 \label{notorsion}
\eea
describing a deformation of Riemannian geometry by the fact that the derivative $D_\mu$ 
is covariantized w.r.t.\ internal 
generalized diffeomorphisms (\ref{DDcov}), under which the f\"unfbein $e_\mu{}^a$
transforms as a weighted scalar.
For the internal sector on the other hand, vanishing torsion translates into the projection condition~\cite{Coimbra:2011ky}
\bea
\Gamma_{MN}{}^K \Big|_{\bf 351} &=& 0\;,
\label{notorsion351}
\eea
for the generalized Christoffel connection, decomposed into irreducible ${\rm E}_{6(6)}$ representations. 
More precisely, by its definition (\ref{GVP}) the Christoffel connection $\Gamma_{MN}{}^K$ is algebra valued
in its last two indices
\bea
\Gamma_{MN}{}^K = \Gamma_M{}^{\boldsymbol{\alpha}}\,(t_{\boldsymbol{\alpha}})_N{}^K
\;,\qquad
\Gamma_M{}^{\boldsymbol{\alpha}} \sim {\bf 27} \otimes {\bf 78} = {\bf 27}\oplus {\bf 351}\oplus{\bf} {\bf 1728}
\;,
\eea
and (\ref{notorsion351}) indicates that $\Gamma_M{}^{\boldsymbol{\alpha}}$ only has non-vanishing components 
in the ${\bf 27}\oplus {\bf} {\bf 1728}$\,.
Explicitly, parametrizing the ${\rm USp}(8)$ connection as
\bea
 {\cal Q}_M{}_j{}^i &=&  {q}_M{}_j{}^i + {\cal V}_M{}^{kl} \Omega^{im} \,q_{kl,jm}
 \;,
 \label{Q_int}
\eea
with $q_{kl,ij}=q_{\llbracket kl\rrbracket ,(ij)}$, equations (\ref{notorsion351}) 
translate into
\bea
q_{kl,mn} &=&
-p_{M\,klp(m}\, {\cal V}_{n)q}{}^{M} \,\Omega^{pq}
-\frac14\, {\cal V}^{pq\,M} \left(p_{M\,pqk(m}\,\Omega_{n)l}-p_{M\,pql(m}\,\Omega_{n)k} \right)
\nonumber\\
&&{}
+\frac14\,\Gamma_{KM}{}^K\,\left(
{\cal V}_{k(m}{}^M \Omega_{n)l}-{\cal V}_{l(m}{}^M \Omega_{n)k}\right)
+u_{kl,mn}
\;, 
\label{q_int}
\eea
with 
\bea
\label{Cartan}
q_{M\,i}{}^j &\equiv& \frac13\,{\cal V}_{ik}{}^N \partial_M {\cal V}_{N}{}^{jk}
\;, \qquad
p_M{}^{ijkl} ~\equiv~ 
 \partial_M {\cal V}_N{}^{[ij} {\cal V}^{kl]\,N}
\;, 
\eea
and $u_{kl,mn}$ satisfying 
\bea
u_{kl,jm}~=~u_{\llbracket kl\rrbracket ,(jm)}\;,\quad
u_{[kl,m]n}&=& 0\;,\quad
u_{kl,jm}\,\Omega^{lj} ~=~0
\;,
\label{conU}
\eea
dropping out from equations (\ref{notorsion351}).
Vanishing torsion thus determines the ${\rm USp}(8)$ connection (and thereby the Christoffel connection)
up to a block $u_{kl,mn}$ transforming in the ${\bf 594}$ of ${\rm USp}(8)$, which
drops out of all field equations and supersymmetry 
variations~\cite{Coimbra:2011ky,Coimbra:2012af,Cederwall:2013naa,Musaev:2014lna}.
The Christoffel connection gives rise to covariant derivatives
\bea
\nabla_M X_{N} &\equiv&
\partial_M X_{N} -\Gamma_{MN}{}^KX_{K}{}-\frac34\, \lambda_X \Gamma_{KM}{}^K X_{N}\;,
\label{nablaMXN}
\eea
where $\lambda_X$ denotes the weight of $X_N$ under generalized diffeomorphisms,
and the trace part in the Christoffel connection is fixed by demanding 
\bea
\nabla_M e &\stackrel{!}{=}&
0\qquad \Longrightarrow\qquad
\Gamma_{NM}{}^N ~=~ \frac45\,e^{-1}\,\partial_M e
\;.
\eea

The remaining connections
in (\ref{DM})
finally are determined by demanding that the 
$\mathfrak{gl}(5)\oplus \mathfrak{e}_{6(6)}$
algebra-valued currents
\bea
{\cal J}_M{}^{ab} &\equiv& e^{a\,\mu}\, {\cal D}[\omega]_M e_\mu{}^b\;,\qquad
{\cal J}_{\mu\,kl}{}^{ij} ~\equiv~ 
{\cal V}_{kl}{}^M{\cal D}[{\cal A},{\cal Q}]_\mu{\cal V}_M{}^{ij}
\;,
\label{JJ}
\eea
of the frame fields live in the complement of the Lorentz algebra $\mathfrak{so}(1,4)\oplus \mathfrak{usp}(8)$,
specifically
\bea
{\cal J}_M{}^{ab}\Big|_{\mathfrak{so}(1,4)} &=& 0 \;,\qquad
{\cal J}_{\mu\,kl}{}^{ij}\Big|_{\mathfrak{usp}(8)} ~=~ 0 \;.
\label{JJcoset}
\eea
They give the explicit form 
\bea
\omega_M{}^{ab} &=& e^{\mu[a}\, \partial_M e_\mu{}^{b]}
\;,\qquad
{\cal Q}_\mu{\,}_i{}^j~=~\frac13 {\cal V}_{ik}{}^M D_\mu {\cal V}_M{}^{jk}\;,
\eea
of the respective spin connections and give rise to the definition of the coset currents
\bea
{\cal J}_M{}^{ab} \ \equiv \ \pi_M{}^{ab} \ = \ \pi_M{}^{(ab)}\;,\qquad
{\cal J}_{\mu\,mn}{}^{ij} \, \Omega^{km}\Omega^{ln} \ \equiv \
{\cal P}_\mu{}^{ijkl} \ = \ {\cal P}_\mu{}^{\llbracket ijkl\rrbracket}
\;.
\eea
Moreover, it turns out that the Lagrangian and supersymmetry transformation rules
are conveniently given in terms of the modified internal spin connections
\bea
\omega^{\pm}_M{}^{ab} &\equiv&
\omega_M{}^{ab} \pm \frac12\,{\cal M}_{MN}\,{\cal F}_{\mu\nu}{}^N\,e^{\mu a} e^{\nu b}
\;,
\label{omegaHat}
\eea
shifted by the non-abelian field strength (\ref{YM}), and we denote the corresponding 
covariant derivatives by ${\cal D}^\pm$\,.

The different curvatures of these spin connections are the building blocks for the bosonic 
Lagrangian and field equations~\cite{Coimbra:2012af,Aldazabal:2013mya,Musaev:2014lna}, once projected
onto the components such that the undetermined part (\ref{conU}) drops out.
Some of the relevant curvatures are obtained from the commutators
\bea
\left[{\cal D}_\mu, {\cal D}_\nu\right] \epsilon^i &=&
\frac14\,\widehat{\cal R}_{\mu\nu}{}^{ab}\,\gamma_{ab}\,\epsilon^i
+\frac23\,{\cal P}_{[\m jklm}{\cal P}_{\n]}{}^{iklm}\,\epsilon^j
-{\cal F}_{\mu\nu}{}^M\,\nabla_M \epsilon^i
\nonumber\\
&&{}
+ \nabla_{M} {\cal F}_{\mu\nu}{}^N  \left(
{\cal V}_N{}^{jk} {\cal V}_{ik}{}^M-
 {\cal V}_{N\,ik} {\cal V}^{jk\,M} \right)\epsilon^j 
-\frac16\,\nabla_M {\cal F}_{\mu\nu}{}^M\, \epsilon^i
\;,
\nonumber\\[1ex]
{\cal V}_{ij}{}^M \left[{\nabla}^-_M,{\cal D}_\mu \right] \epsilon^j &=&
\frac12\,{\cal V}^{jk}{}^M {\cal D}_M {\cal P}_{\mu}{\,}_{ijkn} \epsilon^n 
+\frac14\,{\cal R}^-_{M\mu}{}^{ab}\,\gamma_{ab}\,\epsilon^j
\;,
\label{commutators}
\eea
\bea
{\cal V}^{ik\,M} {\cal V}_{kj}{}^{N}\left[\nabla_M, \nabla_N\right] \e^j+
\left(4{\cal V}^{ik\,M} {\cal V}_{kj}{}^{N} + \frac12{\cal M}^{MN}\,\delta_j^i \right)
\nabla_{(M} \nabla_{N)} \e^j\nonumber\\
\ = \
\frac14 {\cal V}^{ik\,M} {\cal V}_{kj}{}^{N}\,{\cal R}_{MN}{}^{ab}\,\gamma_{ab}\,\e^j
-\frac1{16}\,{\cal R}\,\e^i
\;.
\nonumber
\eea
Explicitly, the curvature tensors read
\bea
\widehat{\cal R}_{\mu\nu}{}^{ab} &= & 2\,D_{[\mu}\omega_{\nu]}{}^{ab}+2\,\omega_{[\mu}{}^{ac}\,\omega_{\nu]c}{}^{b}
+ {\cal F}_{\mu\nu}{}^M\,\omega_M{}^{ab}
\;,
\nonumber\\
{\cal R}^-_{M\mu}{}^{ab} &\equiv& 
\partial_M \, \omega_\mu{}^{ab}
-{\cal D}_\mu \,\omega^-_M{}^{ab}\;,
\nonumber\\
 {\cal R}_{MN}{}^{ab}
&=& 
  - \frac{1}{2} \, e^{\mu[a}e^{b]\nu}g^{\sigma \tau} \nabla_{M} g_{\mu \sigma} \nabla_{N} g_{\nu \tau} 
  \;,
  \label{Riemanns}
\eea
of which the first two enter the Einstein and the vector
field equations, respectively. The curvature scalar ${\cal R}$ is related to the scalar potential 
from (\ref{fullpotential}) as
\bea
{\cal R} &=& V   +\frac{1}{4}\, {\cal M}^{MN}\,\nabla_Mg_{\mu\nu}\nabla_N g^{\mu\nu} + \nabla_M I^M
\;,
\label{RV}
\eea
up to boundary terms $\nabla_M I^M$\,.

The full supersymmetric extension of the bosonic action (\ref{EFTaction}) can be given 
in very compact form in terms of the above spin connections. It reads
\bea
\label{Lsusy} 
e^{-1}{\cal L}&=&
{\cal L}_{\rm bos}
-\bar{\psi}_{\mu i}\gamma^{\mu\nu\rho}{\cal D}_\nu\psi_\rho^i+2\sqrt{2}\,i\,{\cal V}_{ij}{}^M\Omega^{ik}\bar{\psi}_{\mu k}\gamma^{[\mu}{{\nabla}}^+_M\left(\gamma^{\nu]}\psi_\nu{}^j\right)\nonumber\\ 
&&{}-\frac{4}{3}\,\bar{\chi}_{ijk}\gamma^\mu{\cal D}_\mu\chi^{ijk}+8\sqrt{2}\,i\,{\cal V}_{mn}{}^M\Omega^{np}\bar{\chi}_{pkl} {{\nabla}}^+_M\chi^{mkl}\nonumber\\ 
&&{}+\frac{4i}{3}{\cal P}_\mu{}^{ijkl}\bar{\chi}_{ijk}\gamma^\nu\gamma^\mu\psi_\nu{}^m\Omega_{lm} +4\sqrt{2}\,{\cal V}^{ij\,M}\,\bar{\chi}_{ijk}\gamma^\mu{{\nabla}}^-_M\psi_\mu{}^k\;,
\eea
up to quartic fermion terms. 
The latter are expected to coincide with the quartic terms of the $D=5$ theory~\cite{Cremmer:1980gs}.
The Lagrangian (\ref{Lsusy}) is invariant up to total derivatives under the following set
of supersymmetry transformation rules
\bea
\delta_\epsilon \psi_\mu^i&=&{\cal D}_\mu \epsilon^i
-i\sqrt{2}\,{\cal V}^{ij\,M}\left({\nabla}^-_M(\gamma_\mu\epsilon^k)-\frac13\,\gamma_\mu{{\nabla}}^-_M
\epsilon^k\right)\Omega_{jk}\;,\nonumber\\
\delta_\epsilon\chi^{ijk}&=&\frac{i}2\, {\cal P}_\mu{}^{ijkl}\Omega_{lm}\, \gamma^\mu\epsilon^m
+ \frac{3}{\sqrt{2}} \,{\cal V}^{\llbracket ij\,M} \,{\nabla}^-_M\epsilon^{k\rrbracket}\;,
\label{transF}
\eea
for the fermionic fields, and
\bea 
\delta_\epsilon e^a_\mu&=&\frac12\,\bar{\epsilon}_i\gamma^a\psi_\mu^i\;,\qquad
\delta_\epsilon{\cal V}_M{}^{ij} ~=~4i\,\Omega^{im}\Omega^{jn}\,
{\cal V}_M{}^{kl}\,\Omega_{p\llbracket k}\bar{\chi}_{lmn\rrbracket}\epsilon^p\;,\nonumber\\
\delta_\epsilon {\cal A}_\mu{}^M&=&\sqrt{2}\left(i\,\Omega^{ik}\bar{\epsilon}_k
\psi_\mu{}^j+\bar{\epsilon}_k\gamma_{\mu}\chi^{ijk}\right){\cal V}_{ij}{}^M\;,\nonumber\\
\Delta_\epsilon {\cal B}_{\mu\nu}{}_M&=&-\frac{1}{\sqrt{5}}\,
{\cal V}_{M}{}^{ij}\left(2\,\bar{\psi}_{i[\mu}\gamma_{\nu]}\epsilon^k\Omega_{jk}+i\bar{\chi}_{ijk}\gamma_{\mu\nu}\epsilon^k\right)
\;, 
\label{transB}
\eea
for the bosonic fields. Equations (\ref{transF}) depict the Killing spinor equations of the theory.
It is remarkable, that in the supersymmetry transformation rules all explicit appearance of the 
field strength ${\cal F}_{\mu\nu}{}^M$ can be entirely absorbed into the shifted spin connection 
$\omega^-$ form (\ref{omegaHat}) whereas the Lagrangian (\ref{Lsusy}) carries both
$\omega^+$ and $\omega^-$\,.

\subsection{Algebra of external and internal generalized diffeomorphisms}
The algebra of internal generalized diffeomorphisms is governed by the E-bracket 
and has been discussed extensively in the literature. The algebra of the external 
diffeomorphisms, which acts in a more subtle way due to the field-dependent 
modifications in (\ref{skewD}) compared to standard diffeomorphisms, 
has been determined in \cite{Hohm:2015xna} (for the SL$(3)\times {\rm SL}(2)$ EFT, 
but the results generalize immediately). Here we use the opportunity to complete 
the literature by discussing the off-diagonal part of the total gauge algebra, 
i.e., the algebra of external and internal generalized diffeomorphisms. 
This will be important below, when we show that, upon solving the section 
constraint, the internal and external \textit{conventional} diffeomorphisms 
indeed close according to the 10- or 11-dimensional diffeomorphism 
algebra, implying the full diffeomorphism invariance of the resulting supergravities.

For simplicity, let us  first consider a pure E$_{6(6)}$ tensor $T$ (whose indices we suppress) that is an 
external scalar, i.e., does not carry external indices $\mu,\nu,\ldots$ (an example is 
the generalized metric ${\cal M}_{MN}$). We then compute 
for the gauge algebra
 \be
 \begin{split}
  \big[\delta_{\Lambda},\delta_{\xi}\big]T \ &= \ 
  \delta_{\Lambda}(\xi^{\mu}D_{\mu}T)-\delta_{\xi}(\mathbb{L}_{\Lambda} T)\\
  \ &= \ \xi^{\mu}\mathbb{L}_{\Lambda}(D_{\mu}T)-\mathbb{L}_{\Lambda}(\xi^{\mu}D_{\mu}T)\\
  \ &= \ -\mathbb{L}_{\Lambda}\xi^{\mu}\,D_{\mu}T\;. 
 \end{split}
 \ee 
Here we used in the second line that the covariant derivative transforms covariantly. 
Moreover, recall that the gauge parameter $\xi^{\mu}$ is not to be varied in the 
gauge algebra.  
Thus, we have closure 
 \be\label{closurepara}
   \big[\delta_{\Lambda},\delta_{\xi}\big] \ = \ \delta_{\xi'}\;, \qquad 
   \xi'^{\mu} \ = \  -\mathbb{L}_{\Lambda}\xi^{\mu} \ = \ -\Lambda^N\partial_N\xi^{\mu}\;, 
 \ee
which defines the effective ($\Lambda$-transformed) $\xi^{\mu}$ parameter.    
Next, we inspect the (external) vielbein $e_{\mu}{}^{a}$, which is slightly more 
involved because it carries a vector index. With (\ref{skewD}) we compute 
 \be
  \begin{split}
    \big[\delta_{\Lambda},\delta_{\xi}\big]e_{\mu}{}^{a} \ &= \ 
    \delta_{\Lambda}(\xi^{\nu}{ D}_{\nu}e_{\mu}{}^{a}
    + { D}_{\mu}\xi^{\nu} e_{\nu}{}^{a})
    -\delta_{\xi}(\mathbb{L}_{\Lambda}e_{\mu}{}^{a})\\
    \ &= \
     \delta_{\Lambda}(\xi^{\nu}{ D}_{\nu}e_{\mu}{}^{a}
    + { D}_{\mu}\xi^{\nu} e_{\nu}{}^{a})
    -\mathbb{L}_{\Lambda}(\xi^{\nu}{ D}_{\nu}e_{\mu}{}^{a}
   + { D}_{\mu}\xi^{\nu} e_{\nu}{}^{a}) \\
   \ &= \ 
    -(\mathbb{L}_{\delta_{\Lambda}{\cal A}_{\mu}}\xi^{\nu})  e_{\nu}{}^{a}
     -\mathbb{L}_{\Lambda}\xi^{\nu}\,{ D}_{\nu}e_{\mu}{}^{a}
     -\mathbb{L}_{\Lambda}({ D}_{\mu}\xi^{\nu})e_{\nu}{}^{a} \\
   \ &= \ 
   -{D}_{\mu}\Lambda^N\partial_N\xi^{\nu} e_{\nu}{}^{a}
   -\mathbb{L}_{\Lambda}\xi^{\nu}\,{ D}_{\nu}e_{\mu}{}^{a}
   -\Lambda^N{ D}_{\mu}(\partial_N\xi^{\nu})e_{\nu}{}^{a} \\
   \ &= \ -\mathbb{L}_{\Lambda}\xi^{\nu}\,{D}_{\nu}e_{\mu}{}^{a}
    -{ D}_{\mu}(\Lambda^N\partial_N\xi^{\nu}) e_{\nu}{}^{a} \\
   \ &= \ \xi'^{\nu}{ D}_{\nu}e_{\mu}{}^{a}+{ D}_{\mu}\xi'^{\nu} e_{\nu}{}^{a}\;. 
 \end{split}
\ee 
Here we used again, in the third line, the covariance of the covariant derivative, 
due to which various terms cancelled and that $\xi^{\mu}$ is a scalar with respect to 
internal diffeomorphisms. 
We thus established closure according to the same parameter as in (\ref{closurepara}).

Let us now turn to the gauge vector ${\cal A}_{\mu}{}^{M}$, whose transformation 
in (\ref{skewD}) is ${\cal M}$-dependent. In order to simplify the discussion, we first 
consider the minimal variation without this term, 
 \be
  \delta_{\xi}^0 {\cal A}_{\mu}{}^{M} \ \equiv \ \xi^{\nu}{\cal F}_{\nu\mu}{}^{M}\;. 
 \ee
Although this transformation rule is insufficient for the complete gauge invariance of EFT, 
it does lead to a consistent gauge algebra, as we discuss now.  
In order to prove closure of the gauge algebra we have to compute 
 \be
  \begin{split}
   \big[\delta_{\Lambda},\delta_{\xi}^0\big]{\cal A}_{\mu}{}^{M} \ &= \   
   \delta_{\Lambda}(\xi^{\nu}{\cal F}_{\nu\mu}{}^{M})-\delta_{\xi}^0(D_{\mu}\Lambda^M)\\
   \ &= \ \xi^{\nu}\mathbb{L}_{\Lambda}{\cal F}_{\nu\mu}{}^{M}+\mathbb{L}_{\delta_{\xi}^0{\cal A}_{\mu}}\Lambda^M\;. 
  \end{split}
 \ee 
For the second term we find 
 \be
  \mathbb{L}_{\delta_{\xi}^0{\cal A}_{\mu}}\Lambda^M \ = \ \mathbb{L}_{\xi^{\nu}{\cal F}_{\nu\mu}}\Lambda^M
  \ = \ \xi^{\nu}\mathbb{L}_{{\cal F}_{\nu\mu}}\Lambda^M
  -\partial_K\xi^{\nu} {\cal F}_{\nu\mu}{}^{M}\Lambda^K
  +10\,d_{NLR}\, d^{MKR} \partial_K\xi^{\nu} {\cal F}_{\nu\mu}{}^{L} \Lambda^N\;, 
 \ee 
which follows by writing out the Lie derivative and collecting the terms where the 
derivative $\partial_M$ hits the gauge parameter $\xi^{\mu}$. 
The second term in here is the required $\xi'$
transformation, so that we have shown 
 \be\label{closureSTep}
  \big[\delta_{\Lambda},\delta_{\xi}^0\big]{\cal A}_{\mu}{}^{M} \ = \  \delta_{\xi'}^0 {\cal A}_{\mu}{}^{M}
 + \xi^{\nu}(\mathbb{L}_{\Lambda}{\cal F}_{\nu\mu}{}^{M}+\mathbb{L}_{{\cal F}_{\nu\mu}}\Lambda^M)
 +10\,d_{NLR}\, d^{MKR} \partial_K\xi^{\nu} {\cal F}_{\nu\mu}{}^{L} \Lambda^N\;. 
 \ee 
The second term in here is the \textit{symmetrized} Lie derivative that in turn is  
`trivial' and given by (\ref{symmLieIdentity}), 
 \be
   \mathbb{L}_{\Lambda}{\cal F}_{\nu\mu}{}^{M}+\mathbb{L}_{{\cal F}_{\nu\mu}}\Lambda^M
   \ = \ 10\,d_{LNR}\, d^{MKR}\partial_K({\cal F}_{\nu\mu}{}^{L}\Lambda^N)\;. 
 \ee   
Inserting this in (\ref{closureSTep}) above, we finally obtain 
  \be
    \big[\delta_{\Lambda},\delta_{\xi}^0\big]{\cal A}_{\mu}{}^{M} \ = \  \delta_{\xi'}^0 {\cal A}_{\mu}{}^{M}
    +10\,d_{NLR}\, d^{MKR} \partial_K\big(\xi^{\nu} {\cal F}_{\nu\mu}{}^{L} \Lambda^N\big)\;. 
 \ee 
Comparing with the general gauge transformations  of ${\cal A}_{\mu}{}^{M}$ in (\ref{deltaAB}) 
we infer that the additional term on the right-hand side can be interpreted as a field-dependent 
gauge transformation for the one-form parameter $\Xi_{\mu}$ corresponding to the 
two-form potential in the hierarchy. We thus established closure according to 
 \be
   \big[\delta_{\Lambda},\delta_{\xi}^0\big]{\cal A}_{\mu}{}^{M} \ = \  \big( \delta_{\xi'}^0 
   +\delta_{\Xi'}\big){\cal A}_{\mu}{}^{M}\;, \qquad
   \Xi_{\mu N}' \ = \ -d_{NKL}\xi^{\nu}{\cal F}_{\nu\mu}{}^{K}\Lambda^L\;. 
 \ee
We see once more that the higher forms of the tensor hierarchy and their associated 
gauge symmetries are essential for the consistency of EFT.  
   
Let us now return to the full gauge transformations of  ${\cal A}_{\mu}{}^{M}$ w.r.t.~$\xi^{\mu}$, 
including the extra term that we denote in the following by 
$\delta_{\xi}'{\cal A}_{\mu}{}^{M}\equiv {\cal M}^{MN}g_{\mu\nu}\partial_N\xi^{\nu}$. 
We collect the additional contributions in the gauge algebra 
and find 
 \be\label{gaugealgebraSTEEPP}
 \begin{split}
 \big[\delta_{\Lambda},\delta_{\xi}\big]{\cal A}_{\mu}{}^{M} \ &= \ \cdots 
 +\delta_{\Lambda}({\cal M}^{MN}g_{\mu\nu})\partial_N\xi^{\nu}
 +\mathbb{L}_{\delta'_{\xi}{\cal A}_{\mu}}\Lambda^M \\
 \ &= \  \cdots 
 +\mathbb{L}_{\Lambda}({\cal M}^{MN}g_{\mu\nu})\partial_N\xi^{\nu}
 +\mathbb{L}_{{\cal M}^{\bullet N}g_{\mu\nu}\partial_N\xi^{\nu}}\Lambda^M\;, 
 \end{split}
\ee 
where the dots indicate the terms already computed in the previous paragraph. 
The second term on the right-hand side can be written as 
 \be
  \mathbb{L}_{{\cal M}^{\bullet N}g_{\mu\nu}\partial_N\xi^{\nu}}\Lambda^M \ = \ 
  -\mathbb{L}_{\Lambda}( {\cal M}^{MN}g_{\mu\nu}\partial_N\xi^{\nu})
  +10\, d^{MKR} d_{PLR}\partial_K( {\cal M}^{PQ}g_{\mu\nu}\partial_Q\xi^{\nu}\Lambda^L)\;, 
 \ee 
where we used again the identity (\ref{symmLieIdentity}).  
Using this in (\ref{gaugealgebraSTEEPP}) we obtain 
 \be
   \big[\delta_{\Lambda},\delta_{\xi}\big]{\cal A}_{\mu}{}^{M} \ = \ -{\cal M}^{MN}g_{\mu\nu}\,
   \mathbb{L}_{\Lambda}( \partial_N\xi^{\nu})
   +10\, d^{MKR} d_{PLR}\partial_K( {\cal M}^{PQ}g_{\mu\nu}\partial_Q\xi^{\nu}\Lambda^L)\;. 
 \ee
Recalling that $\xi^{\nu}$ is a scalar,  
$\mathbb{L}_{\Lambda}( \partial_N\xi^{\nu})=\partial_N(\mathbb{L}_{\Lambda}\xi^{\nu})$
and so the first term becomes the $\xi'$ transformation defined in (\ref{closurepara}). 
The second term can be interpreted as an additional field-dependent 
contribution to the effective one-form parameter $\Xi'$. 
Thus, in total we learned 
   \be\label{EFTEichalgebra}
   \big[\delta_{\Lambda},\delta_{\xi}\big] \ = \ \delta_{\xi'}+\delta_{\Xi'}\;, 
   \ee
where    
 \be
   \xi'^{\mu} \ = \  -\Lambda^N\partial_N\xi^{\mu}\;, \quad
    \Xi_{\mu M}' \ = \ -d_{MNK}(\xi^{\nu}{\cal F}_{\nu\mu}{}^{N}+{\cal M}^{KL}g_{\mu\nu}\partial_L\xi^{\nu})\Lambda^K\;. 
 \ee
We leave it as an exercise for the reader to verify closure on the two-form, 
which can only be established up to unknown terms corresponding to the 
gauge symmetry of the three-form. 

For completeness we record here that the algebra of external generalized diffeomorphisms 
is given by 
 \be\label{EFTGAUge}
  \big[\delta_{\xi_1},\delta_{\xi_2}\big]  \ = \ \delta_{\xi_{12}}+\delta_{\Lambda_{12}}+\cdots\;, 
 \ee
with effective parameters  
 \be\label{finalparam}
  \begin{split}
     \xi_{12}^\mu\ &\equiv \ \xi_2^\nu{\cal D}_\nu\xi_1^\mu-\xi_1^\nu{\cal D}_\nu\xi_2^\mu
   \;, \\
    \Lambda_{12}^M\ &\equiv \ \xi_2^\mu\xi_1^\nu{\cal F}_{\mu\nu}{}^{M}
    -2\,{\cal M}^{MN}g_{\mu\nu}\,\xi_{[2}^{\mu}\partial^{}_N\xi_{1]}^{\nu}\;. 
   \end{split}
  \ee  
The dots in (\ref{EFTGAUge}) indicate possible gauge transformations corresponding to higher forms 
entering the tensor hierarchy, see \cite{Hohm:2015xna} for more details.

\section{Type IIB solution and embedding of diffeomorphisms}
In this section we will show, for the type IIB solution of the section constraint,  
how the fields and symmetries of EFT are related to those of the standard formulation 
of supergravity in which ten-dimensional diffeomorphism invariance 
is manifest.\footnote{For the M-theory solution we refer the reader to \cite{Hohm:2013vpa}.} 
To this end we show in the first subsection how, upon solving the section constraint, 
the standard diffeomorphism 
algebra is generically embedded in the gauge algebra of EFT 
(in particular the E-bracket), illustrating this 
with a simple toy model. In the second and third subsection we turn to the specific
solution of the section constraint for type IIB and show how the coordinates and tensor fields 
decompose. In the final subsection we return to the external diffeomorphisms of 
EFT and supergravity (that, we recall, are not manifest symmetries), which in the following 
section will be shown to match precisely, thereby proving that EFT leads to a 10-dimensional 
theory with full diffeomorphism invariance.

\subsection{Embedding of standard diffeomorphisms into E-bracket algebra}
\label{subsec:embedding}

We now discuss how to embed the standard diffeomorphisms into the E-bracket 
algebra of EFT. More precisely, we will show that the external and internal 
diffeomorphisms in EFT close in the same way 
as those of a $D=10$ gravity theory, implying that there is a `hidden' 
10-dimensional diffeomorphism symmetry in EFT upon choosing a $D=10$ solution of the section constraint. 

Before focusing on type IIB supergravity, let us start from a generic theory of Einstein gravity,  
coupled to some matter, and inspect the action of the diffeomorphism group under a Kaluza-Klein-type 
decomposition. 
To this end we split the ten-dimensional world and tangent space indices, here and in the 
following indicated by a hat, according to  
$\hat{\mu}  =  (\mu,m)$ and $\hat{a} =  (a,\alpha)$, respectively,  
where $\mu=0,\ldots n-1$, and $m=1,\dots,d$, with $n+d=10$, and similarly for the flat indices.  
Correspondingly, we decompose the tensor fields and symmetry parameters 
of the theory according to this $n+d$ split. 
For instance, the ten-dimensional frame field encoding the metric is written as 
  \be\label{KKgauge}
  E_{\hat{\mu}}{}^{\hat{a}} \ = \ \left(\begin{array}{cc} \phi^{-\kkappa}e_{\mu}{}^{a} &
  A_{\mu}{}^{m} \phi_{m}{}^{\alpha} \\ 0 & \phi_{m}{}^{\alpha}
  \end{array}\right)\,, 
 \ee 
where $\phi=\det (\phi_{m}{}^{\alpha})$ and $\kkappa  =  \frac{1}{n-2}$. Here we employed  a 
gauge fixing of the ten-dimensional Lorentz group SO$(1,9)$ to SO$(1,n-1)\times {\rm SO}(d)$. 
We next perform an analogous decomposition of the remaining gauge symmetries, i.e., of the ten-dimensional 
diffeomorphisms $x^{\hat{\mu}}\rightarrow x^{\hat{\mu}}-\xi^{\hat{\mu}}$ 
and local Lorentz transformations parametrized by $\lambda^{\hat{a}}{}_{\hat{b}}$, acting on the vielbein as 
 \be\label{fullDiff}
  \delta E_{\hat{\mu}}{}^{\hat{a}} \ = \ \xi^{\hat{\nu}}\partial_{\hat{\nu}}E_{\hat{\mu}}{}^{\hat{a}}
  +\partial_{\hat{\mu}}\xi^{\hat{\nu}} E_{\hat{\nu}}{}^{\hat{a}}+\lambda^{\hat{a}}{}_{\hat{b}} E_{\hat{\mu}}{}^{\hat{b}}\;. 
 \ee 
Specifically, we decompose the diffeomorphism parameter as 
 \be
  \xi^{\hat{\mu}} \ = \ (\xi^{\mu}\,,\;\Lambda^m)\;,  
 \ee 
and refer to the diffeomorphisms generated by $\xi^{\mu}$ as `external' and those 
generated by $\Lambda^m$ as `internal'. 
Inserting (\ref{KKgauge}) into (\ref{fullDiff}) we read off the following action of the internal 
diffeomorphisms, 
  \be\label{Lambdaguge}
  \begin{split}
   \delta_{\Lambda}e_{\mu}{}^{a} \ &= \ \Lambda^m\partial_m e_{\mu}{}^{a}+\kkappa\, \partial_m\Lambda^m\,e_{\mu}{}^{a}\;, \\
   \delta_{\Lambda}\phi_m{}^{\alpha} \ &= \ \Lambda^n\partial_n \phi_m{}^{\alpha}+\partial_m\Lambda^n\,\phi_n{}^{\alpha}\;, \\
   \delta_{\Lambda} A_{\mu}{}^m \ &= \ \partial_{\mu}\Lambda^m-A_{\mu}{}^{n}\partial_n\Lambda^m
   +\Lambda^n\partial_n A_{\mu}{}^{m}\;.
  \end{split}
 \ee  
We will also use the notation ${\cal L}_{\Lambda}$ for the conventional Lie derivative of the 
purely internal space, acting in the standard fashion on tensors (of weight zero). Thus, the 
above transformations read
 \be
   \begin{split}
   \delta_{\Lambda}e_{\mu}{}^{a} \ &= \ {\cal L}_{\Lambda} 
   e_{\mu}{}^{a}+\kkappa\, \partial_m\Lambda^m\,e_{\mu}{}^{a}\;, \qquad 
   \delta_{\Lambda}\phi_m{}^{\alpha} \ = \ {\cal L}_{\Lambda}  \phi_m{}^{\alpha} \;, \\
   \delta_{\Lambda} A_{\mu}{}^m \ &= \ \partial_{\mu}\Lambda^m-{\cal L}_{A_{\mu}}\Lambda^m
   \ \equiv \ \partial_{\mu}\Lambda^m +{\cal L}_{\Lambda}A_{\mu}{}^{m}\;.
  \end{split}
 \ee  
Note that here we employ the convention in which the density term is not part of the Lie derivative. 
Analogously to the discussion in EFT, we can define derivatives and non-abelian field strengths 
that are covariant under these
transformations, 
 \be
  {\cal D}^{\KK}_{\mu} \ \equiv \ \partial_{\mu} -{\cal L}_{A_{\mu}}-\lambda\,\partial_{m}A_{\mu}{}^{m}\;, 
  \qquad F_{\mu\nu} \ \equiv \ 2\,\partial_{[\mu}A_{\nu]}
  -[A_{\mu}, A_{\nu}]\;, 
 \ee
where $\lambda$ is the density weight, e.g., $\lambda=\gamma$ for the external vielbein, 
and $[\;,\;]$ the conventional 
Lie bracket.   
Sometimes we will use the notation $D^{\KK}_{\mu}=\partial_{\mu}-{\cal L}_{A_{\mu}}$ for 
the part of the covariant derivative without the density term.\footnote{We emphasize that this is introduced for 
purely notational convenience. In general, acting with $D^{\KK}_{\mu}$ is not a covariant operation.} 
Specifically, for (\ref{Lambdaguge}) we have 
   \be\label{covderfieldstr}
  \begin{split}
   {\cal D}^{\KK}_{\mu}e_{\nu}{}^{a} \ &= \ \partial_{\mu} e_{\nu}{}^{a}-A_{\mu}{}^{m}\partial_m e_{\nu}{}^{a}-\kkappa\, \partial_n A_{\mu}{}^{n} \,e_{\nu}{}^{a}\;, \\
   {\cal D}^{\KK}_{\mu}\phi_m{}^{\alpha} \ &= \ \partial_{\mu}\phi_m{}^{\alpha}-A_{\mu}{}^{n}\partial_n\phi_m{}^{\alpha}-\partial_mA_{\mu}{}^{n} \phi_n{}^{\alpha}\;, \\
   F_{\mu\nu}{}^{m} \ &= \ \partial_{\mu}A_{\nu}{}^{m}-\partial_{\nu}A_{\mu}{}^{m}-A_{\mu}{}^{n}\partial_n A_{\nu}{}^m
   +A_{\nu}{}^{n}\partial_n A_{\mu}{}^m\;. 
  \end{split}
 \ee  
 
Let us now turn to the external diffeomorphisms. These are obtained from  (\ref{fullDiff})
by inserting (\ref{KKgauge}), switching on only the $\xi^{\mu}$ component, and 
adding the compensating Lorentz transformation with parameter 
$ \lambda^{a}{}_{\beta}  =  -\phi^{\kkappa} \phi_{\beta}{}^{m}\partial_m\xi^{\nu}\,e_{\nu}{}^{a}$, 
which is necessary in order to preserve the gauge choice in  (\ref{KKgauge}).
For instance, on the Kaluza-Klein vectors this yields   
  \be\label{origdelA}
  \delta_{\xi}^{\circ}A_{\mu}{}^{m} \ = \ \xi^{\nu} \partial_{\nu}A_{\mu}{}^{m}
  +\partial_{\mu}\xi^{\nu} A_{\nu}{}^{m}-A_{\mu}{}^{n}\partial_n\xi^{\nu} A_{\nu}{}^{m}
  +\phi^{-\frac{2}{3}}G^{mn}g_{\mu\nu}\partial_n\xi^{\nu}\;, 
 \ee 
where $G^{mn}\equiv \phi_{\alpha}{}^{m}\phi^{\alpha n}$, and we specialized to $n=5$, corresponding 
to the $5+5$ split of type IIB that we will analyze momentarily. 
This gauge transformation 
can more conveniently be written in the form of `improved' or `covariant' diffeomorphisms by adding an
\textit{internal} diffeomorphism  (\ref{Lambdaguge}) with field-dependent parameter 
$\Lambda^m=-\xi^{\nu} A_{\nu}{}^{m}$. The gauge-field-dependent 
terms then organize into the covariant field strength in (\ref{covderfieldstr}), 
 \be\label{COVDAA}
  \delta_{\xi}A_{\mu}{}^{m} \ = \ \xi^{\nu} F_{\nu\mu}{}^{m}+\phi^{-\frac{2}{3}}G^{mn}g_{\mu\nu}\partial_n\xi^{\nu}\;. 
 \ee 
We infer that this is of the same structural form as the external diffeomorphism transformation of the 
EFT gauge vector in (\ref{skewD}), and we will verify below that they can be matched precisely upon 
picking the type IIB solution of the section constraint.  
Similarly, these improved external diffeomorphisms act on the internal and external 
vielbein as 
  \be\label{extdiff}
  \begin{split}
  \delta_{\xi}e_{\mu}{}^{a} \ &= \ \xi^{\nu}{\cal D}^{\KK}_{\nu} e_{\mu}{}^{a}+{\cal D}^{\KK}_{\mu}\xi^{\nu}\,e_{\nu}{}^{a}\;, \\
  \delta_{\xi}\phi_m{}^{\alpha} \ &= \ \xi^{\nu}{\cal D}^{\KK}_{\nu}\phi_m{}^{\alpha}\;, 
  \end{split}
 \ee  
again in structural agreement with the corresponding transformations (\ref{skewD}) in EFT.

\medskip

Next, we inspect the algebra of diffeomorphisms under this decomposition. Since 
the internal diffeomorphisms (five-dimensional in the case we are interested in) 
act on the fields via standard Lie derivatives w.r.t.~the internal space, 
see (\ref{Lambdaguge}),  they close according to the standard Lie bracket, 
 \bea\label{standardLIE}
{}\big[ \delta_{\Lambda_1},\delta_{\Lambda_2}\big] \ = \ \delta_{\Lambda_{12}}
\;,
\qquad
\Lambda_{12}^m ~\equiv~ 
[\Lambda_2,\Lambda_1]^m \ \equiv \ 
\Lambda_{2}^k \partial_k \Lambda_{1}^m-\Lambda_{1}^k \partial_k \Lambda_{2}^m
\;. 
\eea
This is embedded in the E-bracket algebra 
(\ref{Ebracket}) by solving the section constraint and restricting to the five `lowest components'  
of the generalized diffeomorphism parameter. 

The mixed algebra between internal and external diffeomorphisms is straightforwardly  
computed in the form of improved diffeomorphisms (\ref{COVDAA}), (\ref{extdiff}). 
In fact, in this form every term on the right-hand side of the gauge variation is 
covariant w.r.t.~the Lie derivative ${\cal L}_{\Lambda}$, with all derivatives entering via 
covariant derivatives or field strengths.\footnote{The variation of the gauge vectors in (\ref{COVDAA}) 
contains the partial derivative term $\partial_n\xi^{\nu}$, but $\xi^{\nu}$ has to be viewed 
as a scalar w.r.t.~internal diffeomorphisms, hence its partial derivative is a covariant 
vector.} We thus compute, for instance, on the vector 
 \be
 \begin{split}
  \big[\delta_{\Lambda},\delta_{\xi}\big]A_{\mu}{}^{m} \ &= \ \delta_{\Lambda}
  \big(\xi^{\nu} F_{\nu\mu}{}^{m}+\phi^{-\frac{2}{3}}G^{mn}g_{\mu\nu}\partial_n\xi^{\nu}\big)
  -\delta_{\xi}\big(\partial_{\mu}\Lambda^m+{\cal L}_{\Lambda}A_{\mu}{}^{m}\big) \\
  \ &= \ \xi^{\nu}{\cal L}_{\Lambda}F_{\nu\mu}{}^{m}+{\cal L}_{\Lambda}
  \big(\phi^{-\frac{2}{3}}G^{mn}g_{\mu\nu}\big)\partial_n\xi^{\nu}
  -{\cal L}_{\Lambda}\big(\delta_{\xi}A_{\mu}{}^{m}\big)\;.  
 \end{split} 
 \ee
Here we used the covariance of the expressions in $\delta_{\xi}A_{\mu}{}^{m}$. 
Thus, the terms in $\delta_{\Lambda}\delta_{\xi}A$ agree precisely with those in 
$\delta_{\xi}\delta_{\Lambda}A$, except that $\xi$, being a parameter and not a field, 
is not varied in the former but appears under the Lie derivative in the latter. 
These correspond to the left-over terms that do not cancel and that can in turn be interpreted
as \textit{external} diffeomorphisms with a parameter $\xi$ that is `rotated' (with the opposite sign)
by the \textit{internal} diffeomorphisms. Hence, the 
gauge algebra is given by   
 \be\label{LambdaXialgebra}
  \big[ \delta_{\Lambda},\delta_{\xi}\big] \ = \ \delta_{\xi'}\;, \qquad
  \xi'^{\mu} \ = \ -{\cal L}_{\Lambda}\xi^{\mu} \ = \ 
  -\Lambda^m\partial_m\xi^{\mu}\;. 
 \ee 
The same conclusion follows for the  external and internal vielbein. 
This algebra is embedded in the corresponding part of the gauge algebra 
of EFT, see  (\ref{closurepara}).   
 
Finally, we turn to the gauge algebra of external diffeomorphisms with themselves. 
Using again the improved diffeomorphisms (\ref{COVDAA}), (\ref{extdiff}), 
an explicit computation shows 
 \be
  \big[\delta_{\xi_1},\delta_{\xi_2}\big] \ = \ \delta_{\xi_{12}}+\delta_{\Lambda_{12}}\;, 
 \ee
where 
 \be\label{XIXIalgebra}
  \xi_{12}^{\mu} \ = \ 2\,\xi_{[2}^{\nu}{\cal D}^{\KK}_{\nu}\xi_{1]}^{\mu}\;, \qquad
  \Lambda_{12}^{m} \ = \ \xi_{2}^{\mu}\xi_1^{\nu} F_{\mu\nu}{}^{m}
  -2\,\phi^{-\frac{2}{3}} G^{mn} g_{\mu\nu}\,\xi_{[2}^{\mu}\partial^{}_n\xi_{1]}^{\nu}\;. 
 \ee   
This is of the same structural form as the corresponding part of the gauge algebra 
(\ref{finalparam})\footnote{It should be noted that, in general, in EFT there are higher-form transformations 
on the right-hand side of the gauge algebra, corresponding to the higher forms in the tensor hierarchy, 
which are not present here.  
As these are needed because of the anomalous `Jacobiator' of the E-bracket, which vanishes 
on solutions of the section constraint, this is perfectly consistent with the embedding of the 
conventional diffeomorphism algebra.} 
and, together with our results below, implies that the full ten-dimensional diffeomorphism algebra 
is embedded in the gauge algebra of EFT.

\medskip

So far we discussed the decomposition of fields and symmetries for pure (Einstein) gravity,  
but in supergravity there are additional matter fields, typically with associated gauge symmetries, 
which have to be decomposed similarly. Before turning to the specific field content of type IIB, 
let us consider a toy model, which exhibits already all essential features. 
We consider an abelian gauge vector $\hat{B}_{\hat{\mu}}$ (such as the 
RR one-form in type IIA) with gauge symmetries 
 \be
  \delta \hat{B}_{\hat{\mu}} \ = \ 
  \partial_{\hat{\mu}}\chi+
  \xi^{\hat{\nu}}\partial_{\hat{\nu}}\hat{B}_{\hat{\mu}}
  +\partial_{\hat{\mu}}\xi^{\hat{\nu}}\hat{B}_{\hat{\nu}}\;,  
 \ee 
for abelian parameter $\chi$.  
Next we decompose the components as in (\ref{KKgauge}) and 
redefine 
 \be
  \begin{split}
   B_{m} \ &= \ \hat{B}_{m} \;, \\
   B_{\mu} \ &= \ \hat{B}_{\mu}-A_{\mu}{}^{m}\hat{B}_{m}\;. 
  \end{split}
 \ee 
(In terms of the notation introduced in sec.~4 this corresponds to the action with the `bar operator', 
$B \rightarrow \overline{B}$.) 
For these redefined fields the abelian  gauge symmetry becomes  
 \be 
  \begin{split}
   \delta_{\chi}B_{\mu} \ &= \ {\cal D}^{\KK}_{\mu}\chi \ = \ \partial_{\mu}\chi-A_{\mu}{}^{m}\partial_m\chi\;, \\
   \delta_{\chi}B_{m} \ &= \ \partial_m\chi\;, 
   \end{split}
  \ee  
and for the diffeomorphisms 
 \be
  \begin{split}
   \delta B_m \ &= \ \xi^{\nu}\partial_{\nu} B_m+\partial_m\xi^{\nu}\hat{B}_{\nu}+{\cal L}_{\Lambda}B_{m}\;, \\ 
   \delta B_{\mu} \ &= \ {\cal L}_{\Lambda} B_{\mu}+{\cal L}_{\xi}B_{\mu}
   -A_{\mu}{}^{m}\partial_m\xi^{\nu} B_{\nu}-\phi^{-\frac{2}{3}} G^{mn}B_{m} g_{\mu\nu}\partial_n\xi^{\nu}\;, 
  \end{split}
 \ee  
where ${\cal L}_{\xi}$ denotes the standard Lie derivative w.r.t.~$\xi^{\mu}$ (with partial derivatives). 
Adding now field-dependent gauge transformations as above, with $\Lambda^m = -\xi^{\nu}A_{\nu}{}^{m}$ 
and $\chi  =  -\xi^{\nu}B_{\nu}$,   
this can be written more covariantly as 
 \be
  \delta_{\xi}B_{m} \ = \ \xi^{\nu}\widehat{\cal D}_{\nu} B_{m} \ \equiv \ 
  \xi^{\nu}\big(\partial_{\nu}B_m-{\cal L}_{A_{\nu}}B_m-\partial_m B_{\nu}\big)\;, 
 \ee
for the internal components, and as 
  \be\label{deltaxiG}
   \delta_{\xi}B_{\mu} \ = \ \xi^{\nu} G_{\nu\mu}-    \phi^{-\frac{2}{3}} G^{mn}g_{\mu\nu} \partial_m\xi^{\nu}\;, 
  \ee
where 
 \be
  G_{\mu\nu} \ \equiv \ {\cal D}^{\KK}_{\mu}B_{\nu}-{\cal D}^{\KK}_{\nu}B_{\mu}\;. 
 \ee
Note that due to the non-commutativity of covariant derivatives this is \textit{not} an invariant field strength. 
Rather, $G$ transforms as 
 \be\label{noncovvarG}
  \delta_{\Lambda,\chi} G_{\mu\nu} \ = \ \Lambda^m\partial_m G_{\mu\nu}   
  -\partial_m\chi F_{\mu\nu}{}^{m}\;.  
 \ee
 We could define a fully $\chi$-invariant field strength by setting 
 $\bar{G}_{\mu\nu}\equiv G_{\mu\nu}+F_{\mu\nu}{}^{m}B_{m}$, but it turns out that the match 
 with EFT requires the (analogue of the) above form. 
 In fact, in EFT a slightly more general notion of covariance is appropriate: 
 the gauge parameters analogous to $\Lambda^m$ and $\chi$ will be components of the 
 generalized diffeomorphism parameter $\Lambda^M$, and the field strengths 
 $G_{\mu\nu}$ and $F_{\mu\nu}{}^{m}$  correspond to components of the EFT field strength
 ${\cal F}_{\mu\nu}{}^{M}$, so that transformations such as 
 (\ref{noncovvarG}) originate from the covariant transformation 
 governed by the full generalized Lie derivative of EFT, 
 $\delta_{\Lambda}{\cal F}_{\mu\nu}{}^{M}=\mathbb{L}_{\Lambda}{\cal F}_{\mu\nu}{}^{M}$.  
 
We finally note that it is straightforward to verify that 
the transformations (\ref{deltaxiG}) and  (\ref{noncovvarG}) 
close according to the gauge algebras (\ref{LambdaXialgebra}) and (\ref{XIXIalgebra}), 
encoding the full diffeomorphism algebra.  Conversely, starting with component 
fields $B_{\mu}$, $B_m$, and gauge symmetries closing according to the above algebra 
(\ref{LambdaXialgebra}), (\ref{XIXIalgebra}), we can reconstruct the form with 
manifest (say ten-dimensional) diffeomorphism invariance.

\subsection{Type IIB solution of section constraint}

We now turn to the specific solution of the section constraint that will be shown 
to lead to a formulation that is on-shell equivalent to type IIB supergravity. 
To this end we have to break E$_{6(6)}$ to ${\rm GL}(5)\times {\rm SL}(2)$, 
embedding the residual group according to 
\bea
{\rm GL}(5)\times {\rm SL}(2) &\subset& {\rm SL}(6) \times {\rm SL}(2) \;\, \subset \;\, {\rm E}_{6(6)}
\;. 
\eea
In this case, the fundamental  and the adjoint representation of ${\rm E}_{6(6)}$ break as
\bea
\bar{{\bf  27}} &\rightarrow& (5,1)_{+4}+(5',2)_{+1}+(10,1)_{-2}+(1,2)_{-5}
\;,
\label{split27B}
\\
{\bf 78} &\rightarrow&  (5,1)_{-6} + (10',2)_{-3} + \left(1+15+20\right)_{0} + (10,2)_{+3} + (5',1)_{+6}
\;,
\label{split78B}
\eea
with the subscripts referring to the charges under ${\rm GL}(1)\subset {\rm GL}(5)$.
An explicit solution to the section condition (\ref{sectioncondition}) is given by 
restricting the $Y^M$ dependence of all fields to the five coordinates in the $(5,1)_{+4}$. 
Explicitly, splitting the coordinates $Y^M$ and the fundamental indices according to (\ref{split27B}) into
\bea
\left\{Y^M\right\} &\rightarrow&\left\{ \,y^m\,,\; y_{m\,\alpha}\,,\; y^{mn}\,,\;  y_\alpha \,\right\}
\;,
\label{YbreakB}
\eea
with internal indices $m, n = 1, \dots, 5$ and SL$(2)$ indices  $\alpha=1, 2$, 
the non-vanishing components of the $d$-symbol are given by
 \bea
d^{MNK}  &:&
 d^{m}{}_{n\alpha,\beta} = \tfrac1{\sqrt{10}}\, \delta^m_n \varepsilon_{\alpha\beta}\;,\quad
d^{mn}{}_{k\alpha,l\beta} = \tfrac1{\sqrt{5}}\, \delta^{mn}_{kl}\,\varepsilon_{\alpha\beta}\;,\quad
d^{mn,kl,p} = \tfrac1{\sqrt{40}}\,\varepsilon^{mnklp}\;, \nonumber\\
d_{MNK} &:&   d_{m}{}^{n\alpha,\beta} = \tfrac1{\sqrt{10}}\, \delta^n_m \varepsilon^{\alpha\beta}\;,\quad
d_{mn}{}^{k\alpha,l\beta} = \tfrac1{\sqrt{5}}\, \delta_{mn}^{kl}\,\varepsilon^{\alpha\beta}\;,\quad
d_{mn,kl,p} = \tfrac1{\sqrt{40}}\,\varepsilon_{mnklp}\;, 
 \label{dbreakB}
 \eea
and all those related by symmetry, $d^{MNK}=d^{(MNK)}$\,.
In particular, the ${\rm GL}(1)$ grading guarantees that all components $d^{m\,n\,k}$ vanish. 
It follows 
that the section condition (\ref{sectioncondition}) indeed is solved by restricting the coordinate dependence
of all fields according to
\bea
\left\{\partial^{m\,\alpha}A =0\;,\; \partial_{mn} A = 0\;,\; \partial^\alpha A = 0\right\}\qquad
\Longleftrightarrow\qquad
A(x^\mu,Y^M) &\longrightarrow& A(x^\mu,y^m)
\;. 
\label{explicit_sectionB}
\eea
Indeed, the section constraint then reduces to $d^{Mnk}\partial_n\otimes \partial_k=0$, for which all relevant components 
of the $d$-symbol simply vanish.

\subsection{Decomposition of EFT fields}
\label{subsec:decomposition}

In this subsection we analyze various objects of EFT, e.g., the generalized metric and 
the gauge covariant curvatures, in terms of the component fields originating under the above 
decomposition of E$_{6(6)}$, together with their gauge symmetries. This sets the stage for our analysis 
in sec.~4, where we start from type IIB supergravity and perform the complete Kaluza-Klein 
decomposition in order to match it to the fields and symmetries discussed here. 
Thus, here we split tensor fields and indices according to
(\ref{split27B})--(\ref{dbreakB}), 
assuming the explicit solution (\ref{explicit_sectionB}) of the section condition.

To begin, let us consider the $p$-form field content of the ${\rm E}_{6(6)}$ EFT under 
the split (\ref{split27B}). This yields 
\bea
{\cal A}_{\mu}{}^{M}\;: \quad \left\{ {\cal A}_\mu{}^m, {\cal A}_{\mu\,m\,\alpha}, {\cal A}_{\mu\,kmn},  
{\cal A}_{\mu\,\alpha} \right\}\;,\qquad
{\cal B}_{\mu\nu\, M}\;: \quad 
\left\{{\cal B}_{\mu\nu}{}^\alpha ,  {\cal B}_{\mu\nu\,mn}, {\cal B}_{\mu\nu}{}^{m\,\alpha} , {\cal B}_{\mu\nu\,m} \right\}
\;,
\label{AB_B}
\eea
where we have defined ${\cal A}_{\mu\,kmn}\equiv \frac12\varepsilon_{kmnpq}{\cal A}_\mu{}^{pq}$\,.
However, the EFT Lagrangian actually depends on the two-forms only under certain derivatives, 
\bea
\left\{\,\partial_m {\cal B}_{\mu\nu}{}^\alpha \,,\;  \partial_{[k}{\cal B}_{|\mu\nu|\,mn]}\,,\; \partial_m {\cal B}_{\mu\nu}{}^{m\,\alpha}\,  \right\}
\;, 
\label{onlyBB}
\eea
introducing an additional redundancy in the two-form field content, which will be important for the match with type IIB. 
As discussed above,
the vector fields ${\cal A}_\mu{}^m$ will be identified with the IIB Kaluza-Klein vector fields, 
which transform according to (\ref{Lambdaguge}) and in particular close 
according to the standard Lie bracket 
of five-dimensional diffeomorphisms, see (\ref{standardLIE}), embedded into the E-bracket~(\ref{Ebracket}).

Let us now work out the general formulas of the ${\rm E}_{6(6)}$-covariant formulation
with (\ref{dbreakB}) and imposing the explicit solution of the section condition (\ref{explicit_sectionB}) on all fields.
We then obtain, by inserting (\ref{dbreakB}) into (\ref{YM}), the following covariant field strengths
of the different vector fields in (\ref{AB_B}), 
\bea\label{FieldSTrengths}
{\cal F}_{\mu\nu}{}^m &=&  2\partial_{[\mu} {\cal A}_{\nu]}{}^m 
- {\cal A}_{\mu}{}^n\partial_n {\cal A}_{\nu}{}^m
+{\cal A}_{\nu}{}^n\partial_n {\cal A}_{\mu}{}^m
\;,
\nonumber\\
{\cal F}_{\mu\nu\,m\alpha} &=&
2D^{\KK}_{[\mu} {\cal A}^{}_{\nu]\,m\alpha}
+\varepsilon_{\alpha\beta} \,  \partial_m \tilde{\cal B}_{\mu\nu}{}^\beta
\;,
\nonumber\\
{\cal F}_{\mu\nu\,kmn} &=&
2D^{\KK}_{[\mu} {\cal A}^{}_{\nu]\,kmn}
-3\,\sqrt{2} \, \varepsilon^{\alpha\beta}  {\cal A}_{[\mu\,[k|\alpha|} \partial_m {\cal A}_{\nu]}{}_{n]\beta} 
+3\,\partial_{[k} \tilde{\cal B}_{|\mu\nu|\,mn]} 
\;,
\nonumber\\
{\cal F}_{\mu\nu\,\alpha} &=&
2 D^{\KK}_{[\mu} {\cal A}^{}_{\nu]\,\alpha}
-2(\partial_k {\cal A}_{[\mu}{}^k) \, {\cal A}_{\nu]}{}_\alpha 
-\sqrt{2}\,{\cal A}_{[\mu}{}^{mn} \partial_n {\cal A}_{\nu]\,m\alpha}\nonumber\\
&&{}
-\sqrt{2}\,{\cal A}_{[\mu|m\alpha|}{} \partial_n {\cal A}_{\nu]}{}^{mn}
-\varepsilon_{\alpha\beta} \,\partial_k \tilde{\cal B}_{\mu\nu}{}^{k\beta}
\;,
\eea
with the redefined  two-forms
\bea
\tilde{\cal B}_{\mu\nu}{}^\alpha &\equiv&
\sqrt{10}\, {\cal B}_{\mu\nu}{}^\alpha-\varepsilon^{\alpha\beta}\,{\cal A}_{[\mu}{}^n {\cal A}_{\nu]}{}_{n\beta}
\;,
\quad\nonumber\\
\tilde {\cal B}_{\mu\nu\,mn} &\equiv& \sqrt{10}\, {\cal B}_{\mu\nu\,mn} + {\cal A}_{[\mu}{}^{k}  {\cal A}_{\nu]\,kmn}
\;,\nonumber\\
\tilde{\cal B}_{\mu\nu}{}^{k\alpha} &\equiv& 
\sqrt{10}\, {\cal B}_{\mu\nu}{}^{k\alpha}+\varepsilon^{\alpha\beta}\,{\cal A}_{[\mu}{}^k  {\cal A}_{\nu]}{}_\beta
\;.
\eea
Here all covariant derivatives are $D^{\KK}_\mu \equiv \partial_\mu - {\cal L}_{{\cal A}_\mu}$, 
covariantized w.r.t.\  to the action of the five-dimensional internal diffeomorphisms reviewed above.
The corresponding vector gauge transformations, obtained from (\ref{deltaAB}),  are given by
\bea
\delta {\cal A}_\mu{}^m &=& D^{\KK}_\mu\Lambda^m\;,
\nonumber\\
\delta {\cal A}_{\mu\,m\alpha} &=&
D^{\KK}_\mu \Lambda_{m\alpha}
+{\cal L}_\Lambda {\cal A}_{\mu\,m\alpha}
-  \varepsilon_{\alpha\beta} \, \partial_m 
\tilde\Xi_\mu{}^\beta 
\;,
\nonumber\\
 \delta {\cal A}_{\mu\,kmn} &=&
D^{\KK}_\mu \Lambda_{kmn}
+{\cal L}_\Lambda {\cal A}_{\mu\,kmn}
-3\sqrt{2} \, \varepsilon^{\alpha\beta} \, \partial_{[k} {\cal A}_{|\mu|}{}_{m|\alpha|} \Lambda_{n]\beta}
-3\,  \partial_{[k}\tilde\Xi_{|\mu|\,mn]} 
\label{deltaAAA}
\;,
\eea
with
\bea
\tilde{\Xi}_\mu{}^\alpha &\equiv&
\sqrt{10}\, \Xi_\mu{}^\alpha -\varepsilon^{\alpha\beta}\,\Lambda^n {\cal A}_{\mu\,n\beta}
\;,\qquad
\tilde{\Xi}_{\mu\,mn} ~\equiv~ \sqrt{10}\,\Xi_{\mu\,mn} + \Lambda^{k} {\cal A}_{\mu\,kmn}
\;. 
\eea
For the vector fields ${\cal A}_{\mu\,\alpha}$ we observe that
its gauge variation contains the contribution 
\bea
\delta {\cal A}_{\mu\,\alpha} &=& \dots + \varepsilon_{\alpha\beta}\,\partial_k \tilde{\Xi}_{\mu}{}^{k\beta}
\;. 
\label{dAa}
\eea
This implies  that it can entirely be gauged away by the tensor gauge symmetry 
associated with the two-forms ${\cal B}_{\mu\nu}{}^{k\beta}$.
Consequently, it will automatically disappear from the Lagrangian upon integrating out 
$\partial_k {\cal B}_{\mu\nu}{}^{k\beta}$.
The remaining two-form field strengths in turn come with gauge transformations
\bea
\delta \tilde{\cal B}_{\mu\nu}{}^\alpha &=&
2 D^{\KK}_{[\mu} \tilde \Xi^{}_{\nu]}{}^\alpha 
+{\cal L}_\Lambda \tilde{\cal B}_{\mu\nu}{}^\alpha
-\varepsilon^{\alpha\beta}\, \Lambda_{n\beta} {\cal F}_{\mu\nu}{}^n+\tilde{\mathcal{O}}_{\mu \nu}{}^{\alpha}
\;,\nonumber\\
\delta \tilde{\cal B}_{\mu\nu\,mn} &=& 
2  D^{\KK}_\mu \left(\tilde{\Xi}_{\nu\,mn} +\frac1{\sqrt{2}}\,\varepsilon^{\alpha\beta} \, {\cal A}_{\nu\,m\alpha} \,\Lambda_{n\beta}\right)
+\sqrt{2}  \,\partial_m {\cal A}_{\mu \,n\alpha} \, \tilde{\Xi}_\nu{}^\alpha 
\nonumber\\
&&{}
+{\cal L}_\Lambda \tilde{\cal B}_{\mu\nu\,mn} 
-\frac1{\sqrt{2}} \,  \Lambda_{[m|\alpha|} \,\partial_{n]} \tilde{\cal B}_{\mu\nu}{}^\alpha
+\Lambda_{mnk} \,{\cal F}_{\mu\nu}{}^k\nonumber\\
&&{}
+\frac{1}{\sqrt{2}}\,\varepsilon^{\alpha\beta} \,{\cal F}_{\mu\nu \, m\alpha} \,\Lambda_{n\beta}+\tilde{\cal{O}}_{\mu \nu m n}
\;,
\label{deltaBB}
\eea
where 
\bea
\tilde{\mathcal{O}}_{\mu \nu}{}^{\alpha} &\equiv& \sqrt{10} \,\mathcal{O}_{\mu \nu}{}^{\alpha}
\;,\\
\tilde{\mathcal{O}}_{\mu \nu \, m n} &\equiv& \sqrt{10} \, \mathcal{O}_{\mu \nu \, m n}+\partial_m \left(2\Lambda^k\,\tilde{\cal B}_{\mu\nu \, n k}+\sqrt{2} {\cal A}_{\mu \, n \alpha} \, \Xi_{\nu}{}^{\alpha}+\sqrt{2}\,\varepsilon^{\alpha\beta} \, {\cal A}_{\mu \, n \alpha}{\cal A}_{\nu \, k \beta}\right)\;. 
\nonumber
\eea
Finally, the associated three-form field strengths are obtained from (\ref{HCurvature}) and read 
\bea
\tilde{\cal H}_{\mu\nu\rho}{}^\alpha &\equiv&
\sqrt{10}\,{\cal H}_{\mu\nu\rho}{}^\alpha 
~=~
3\,D^{\KK}_{[\mu} \tilde{\cal B}^{}_{\nu\rho]}{}^\alpha 
+3\,\varepsilon^{\alpha\beta}\,{\cal F}_{[\mu\nu}{}^n {\cal A}_{\rho]\,n\beta}
\;,
\label{HB}\\
\tilde{\cal H}_{\mu\nu\rho\,mn} &\equiv&
\sqrt{10}\,{\cal H}_{\mu\nu\rho\,mn}
\nonumber
\\
&=&
3\, D^{\KK}_\mu \tilde {\cal B}_{\nu\rho\,mn}
-3\,{\cal F}_{\mu\nu}{}^k {\cal A}_{\rho\,kmn}
-3\sqrt{2}\, \varepsilon^{\alpha\beta}\, {\cal A}_\mu{}_{m\alpha} D_\nu {\cal A}_\rho{}_{n\beta}
+3\sqrt{2}\, {\cal A}_\mu{}_{m\alpha} \partial_n \tilde{\cal B}_{\nu\rho}{}^\alpha
\;.\nonumber
\eea
More precisely, this holds 
up to terms that are projected out from the Lagrangian under $y$-derivatives.
The expressions on the r.h.s.\ in (\ref{deltaBB})--(\ref{HB}) are understood to
be projected onto the corresponding antisymmetrizations in their parameters, i.e. $[mn]$,
$[\mu\nu]$, $[\mu\nu\rho]$, etc.

It is also instructive to give the component form of the Bianchi identities originating from (\ref{Bianchi}) 
and (\ref{calHBianchi}). From the latter we obtain the components 
\bea
4\,D^{\KK}_{[\mu} \tilde{\cal H}^{}_{\nu\rho\sigma]}{}^\alpha
&=& 
6\,\varepsilon^{\alpha\beta}\,{\cal F}_{[\mu\nu}{}^n {\cal F}_{\rho\sigma]\,n\beta}\;.  
\label{BianchiA}
\eea
After a straightforward but somewhat tedious computation one finds 
\bea
4\,D^{\KK}_{[\mu} \tilde{\cal H}_{\nu\rho\sigma]\,mn} 
+4\sqrt{2}\, {\cal A}_\mu{}_{m\alpha} \partial_n \tilde{\cal H}_{\nu\rho\sigma}{}^\alpha
&=&
-6\,{\cal F}_{[\mu\nu}{}^k {\cal F}_{\rho\sigma]\,kmn}
-3\sqrt{2}\, \varepsilon^{\alpha\beta}\, {\cal F}_{[\mu\nu\,|m\alpha|} {\cal F}_{\rho\sigma]\,n\beta}
\nonumber\\
&&{}\hspace{-1cm}
+3\sqrt{2}\,
\partial_m \left(  \varepsilon_{\alpha\beta}\,\tilde{\cal B}_{\mu\nu}{}^\alpha\partial_n \tilde{\cal B}_{\rho\sigma}{}^\beta \right)
-12\, \partial_m \left({\cal F}_{\mu\nu}{}^k   \tilde {\cal B}_{\rho\sigma\,kn}\right)
\nonumber\\
&&{}\hspace{-1cm}
-6\sqrt{2}\, \partial_m \left( {\cal A}_\mu{}_{n\alpha}  \varepsilon^{\alpha\beta}
{\cal F}_{\nu\rho}{}^k {\cal A}_{\sigma\,k\beta} \right)
\;.  
\label{BianchiB}
\eea
Again, the indices $m,n$ and $\mu,\nu , \rho,\sigma$ in here are totally antisymmetrized, 
which we did not indicate explicitly in order not to clutter the notation. 

\medskip

Let us now move to the scalar field content of the theory. In the EFT formulation,
they parametrize the symmetric matrix ${\cal M}_{MN}$\,. We now 
need to choose a parametrization of this matrix in accordance with the decomposition (\ref{split78B}).
In standard fashion~\cite{Cremmer:1997ct}, we build the matrix as ${\cal M}_{MN}=({\cal V}{\cal V}^T)_{MN}$ from a `vielbein' ${\cal V}\in {\rm E}_{6(6)}$
in triangular gauge 
\bea
{\cal V} &\equiv& 
{\rm exp}\left[\varepsilon^{klmnp}\,c_{klmn} \, t_{(+6)\,p}\right] \,
{\rm exp}\left[b_{mn}{}^\alpha\,t_{(+3)}{}_{\alpha}^{mn}\right]\,
{\cal V}_5\,{\cal V}_2\,
{\rm exp} \left[\Phi\, t_{(0)}\right]
\;.
\label{V27B}
\eea
Here, $t_{(0)}$ is the E$_{6(6)}$ generator associated to the GL(1) grading of (\ref{split78B}), ${\cal V}_2$, ${\cal V}_5$ 
denote matrices in the SL(2) and SL(5) subgroup, respectively,
parametrized by vielbeins $\nu_2$, $\nu_5$.
The $t_{(+n)}$ refer to the E$_{6(6)}$ generators of positive grading in (\ref{split78B}), with
non-trivial commutator
\bea
{}
\left[t_{(+3)}{}_\alpha^{kl}, t_{(+3)}{}_\beta^{mn}\right] &=& \varepsilon_{\alpha\beta}\,\varepsilon^{klmnp}\,t_{(+6)\,p}
\;.
\eea
All generators are evaluated in the
fundamental ${\bf 27}$ representation (\ref{split27B}), such that the symmetric matrix ${\cal M}_{MN}$ takes the block form
\bea
{\cal M}_{KM} &=& \left(
\begin{array}{cccc}
{\cal M}_{km}&{\cal M}_k{}^{m\beta}&{\cal M}_{k,mn}&{\cal M}_{k}{}^{\beta}\\
{\cal M}^{k\alpha}{}_{m}&{\cal M}^{k\alpha,}{}^{m\beta}&{\cal M}^{k\alpha}{}_{mn}&{\cal M}^{k\alpha,}{}^{\beta}\\
{\cal M}_{kl,m}&{\cal M}_{kl}{}^{m\beta}&{\cal M}_{kl,mn}&{\cal M}_{kl}{}^{\beta}\\
{\cal M}^\alpha{}_{m}&{\cal M}^{\alpha,m\beta}&{\cal M}^\alpha{}_{mn}&{\cal M}^\alpha{}^{\beta}
\end{array}
\right)
\;.
\label{M2B}
\eea
Explicit evaluation of (\ref{V27B}) determines the various blocks in (\ref{M2B}). For instance, \
\bea
{\cal M}_{mn,kl} &=& e^{2\Phi/3}\,m_{m[k}m_{l]n}
+2 e^{5\Phi/3} \, b_{mn}{}^\alpha b_{kl}{}^\beta\,
m_{\alpha\beta}\;, 
\eea
while the components in the last line are  given by\footnote{
The explicit expressions (\ref{calMcomponents}) and (\ref{tildeMB}) 
for the matrix components of ${\cal M}_{MN}$ and $\tilde{\cal M}_{MN}$
correct some typos in equations (5.22) and (5.24),
respectively, in the published version of~\cite{Hohm:2013vpa}.
}
\bea\label{calMcomponents}
{\cal M}^{\alpha\beta} &=& e^{5\Phi/3}\,m^{\alpha\beta}\;,\qquad
{\cal M}^{\alpha}{}_{mn} = \sqrt{2} \,e^{5\Phi/3}\,m^{\alpha\beta} \varepsilon_{\beta\gamma} \,b_{mn}{}^\gamma
\;,
\nonumber\\
{\cal M}^{\alpha,m\beta} &=& 
\frac12\,e^{5\Phi/3}\,m^{\alpha\gamma}\varepsilon_{\gamma\delta}\,\varepsilon^{mklpq}\,b_{kl}{}^\beta b_{pq}{}^\delta
-\frac1{24}\,e^{5\Phi/3}\,m^{\alpha\beta}\,\varepsilon^{mklpq}\,c_{klpq}
\;,
\nonumber\\
{\cal M}^\alpha{}_m &=& 
\frac23\,e^{5\Phi/3}\, m_{\beta\gamma}  \,\varepsilon^{kpqrs}
 \left(
b_{mk}{}^{[\alpha} b_{pq}{}^{\beta]} b_{rs}{}^{\gamma}
+\frac1{8}\,\varepsilon^{\alpha\beta}\,b_{mk}{}^\gamma\,c_{pqrs} \right)
\;,
\eea
with the symmetric matrix $m^{\alpha\beta}=(\nu_2)^\alpha{}_u (\nu_2)^{\beta\,u}$
built from the ${\rm SL}(2)$ vielbein from (\ref{V27B}).
We will also need the following combinations of the matrix entries of ${\cal M}_{MN}$ 
(that emerge after integrating out some of the fields), 
\bea
\tilde{\cal M}_{MN} &\equiv& {\cal M}_{MN}- {\cal M}_{M}{}^{\alpha} ({\cal M}^{\alpha\beta})^{-1} {\cal M}_{N}{}^\beta
\;, 
\label{tildeM}
\eea
for which we find 
\bea
\tilde{\cal M}_{mn,kl} &=& e^{2\Phi/3}\,m_{m[k}m_{l]n}
\;,\nonumber\\
\tilde{\cal M}_{mn}{}^{k\alpha} &=& \frac1{\sqrt{2}}\,e^{2\Phi/3}\,
\varepsilon_{mnpqr} m^{kp} m^{qu}m^{rv} b_{uv}{}^\alpha
\;,\nonumber\\
\tilde{\cal M}_{mn,k} &=&
-\frac1{6\sqrt{2}}\,e^{2\Phi/3}\,\varepsilon^{uvpqr}\,m_{mu}m_{nv} \left( c_{kpqr}-6\varepsilon_{\alpha\beta}\, b_{kp}{}^\alpha b_{qr}{}^\beta\right)\;, 
\nonumber\\
\tilde{\cal M}^{m\alpha,n\beta} &=&
e^{-\Phi/3}\,m^{mn}m^{\alpha\beta}
+2\, e^{2\Phi/3}\,\,m^{kp}
\left(
m^{mn} m^{lq} -2\, m^{ml} m^{nq} 
\right) 
b_{kl}{}^\alpha b_{pq}{}^\beta 
\;,
\label{tildeMB}
\eea
etc., with $m_{mn}=(\nu_5)_m{}^a (\nu_5)_n{}^a$.

Next, we can work out the covariant derivatives of the various `scalar components' of 
the generalized metric.  
Using (\ref{dbreakB}) we find for the covariant derivatives of the matrix parameters in (\ref{M2B}) 
\bea\label{covderMcomp}
{\cal D}_\mu \Phi &=& D^{\KK}_\mu\Phi + \tfrac45\,\partial_k {\cal A}_\mu{}^k \;,\nonumber\\
{\cal D}_\mu m_{mn} &=& D^{\KK}_\mu m_{mn}+ \tfrac{2}{5}\,\partial_k {\cal A}_\mu{}^k\, m_{mn}\;,\nonumber\\
{\cal D}_\mu b_{mn}{}^\alpha &=& D^{\KK}_\mu b_{mn}{}^\alpha -\varepsilon^{\alpha\beta} \partial_{[m} {\cal A}_{n]\beta\,\mu}\;,\nonumber\\
{\cal D}_\mu c_{klmn} &=& D^{\KK}_\mu c_{klmn} 
+4 \sqrt{2}\, \partial_{[k} {\cal A}_{lmn]\mu}
+12\,b_{[kl}{}^\alpha\,\partial_{m} {\cal A}_{n]\,\alpha\,\mu}
\;, 
\eea
where we recall that $D^{\KK}_{\mu}$ denotes the covariant derivatives w.r.t.~${\cal A}_{\mu}{}^{m}$ 
(that below will be identified with the Kaluza-Klein vector $A_{\mu}{}^{m}$)  
without the density terms, 
which here have been indicated explicitly, thereby defining the weight of all fields. 
The form of these covariant derivatives implies in particular that we have the following gauge symmetries 
on these fields, 
\bea
\delta \Phi &=& {\cal L}_\Lambda \Phi - \tfrac45\,\partial_k \Lambda{}^k \;,\nonumber\\
\delta m_{mn} &=& {\cal L}_\Lambda m_{mn}- \tfrac{2}{5}\,\partial_k \Lambda{}^k\, m_{mn}\;,\nonumber\\
\delta b_{mn}{}^\alpha &=& {\cal L}_\Lambda b_{mn}{}^\alpha +\varepsilon^{\alpha\beta} \partial_{[m} \Lambda_{n]\beta}\;,\nonumber\\
\delta c_{klmn} &=& {\cal L}_\Lambda c_{klmn} 
-4 \sqrt{2}\, \partial_{[k} \Lambda_{lmn]}
-12\,b_{[kl}{}^\alpha\,\partial_{m} \Lambda_{n]\,\alpha}
\;.
\label{deltaMMM}
\eea

We close this section by giving some relevant formulas for the decompositions of various terms 
in the action upon putting the solution of the section constraint. 
The scalar kinetic term (\ref{Lsc}) yields
\bea
\frac1{24}\,D_{\mu} {\cal M}_{MN}D^{\mu} {\cal M}^{MN} &=&
-\frac56\, {\cal D}_\mu\Phi {\cal D}^\mu\Phi 
+\frac14\,{\cal D}_\mu m_{\alpha\beta}{\cal D}^\mu m^{\alpha\beta}
+\frac14\,{\cal D}_\mu m_{mn}{\cal D}^\mu m^{mn}
\nonumber\\
&&{}
- e^\Phi\,{\cal D}_\mu b_{mn}{}^\alpha {\cal D}^\mu b_{kl}{}^\beta\,m^{mk} m^{nl} m_{\alpha\beta}
\nonumber\\
&&{}
-\frac1{48}\,e^{2\Phi}\, \widehat{\cal D}_\mu c_{klmn} \widehat{\cal D}^\mu c_{pqrs} m^{kp}m^{lq}m^{mr}m^{ns}\;, 
\label{Lscal}
\eea
where we defined 
\bea
\widehat{\cal D}_\mu c_{klmn} &\equiv&
{\cal D}_\mu c_{klmn} + 12 \varepsilon_{\alpha\beta}\,b_{kl}{}^\alpha {\cal D}_\mu b_{mn}{}^\beta
\;.
\label{Dhatc}
\eea
The `scalar potential' (\ref{fullpotential})
takes the form
\bea
V & = &
3\,e^{7\Phi/3}\, \partial_{[k} b_{mn]}{}^\alpha \partial_{l} b_{pq}{}^\beta\,
m^{kl}m^{mp}m^{nq}m_{\alpha\beta}
\nonumber\\
&&{}
+
\frac{5}{48}\,e^{10\Phi/3}\,
X_{klmnp} X_{qrstu} m^{kq}m^{lr}m^{ms}m^{nt}m^{pu} + V_\Phi(\partial_k \Phi, \partial_k m_{mn})
\;,
\label{potXX}
\eea
where the last term combines all contributions with the internal derivative acting on $\Phi$ and $m_{mn}$,
and
\bea
X_{klmnp} &\equiv & 
\partial_{[k} c_{lmnp]} + 12\,\varepsilon_{\alpha\beta}\,b_{[kl}{}^\alpha \partial_m b_{np]}{}^\beta
\;.
\label{potX}
\eea
Finally, we give the topological term (\ref{CSlike}) in this parametrization, 
\bea
  {\cal L}_{\rm top} &=&
\frac1{8}\,\varepsilon^{\mu\nu\rho\sigma\tau} \varepsilon^{klmnp}\,\Big(
\frac{\sqrt{2}}{6}\,
\varepsilon^{\alpha\beta} \,
{\cal F}_{\mu\nu\,m\alpha} {\cal F}_{\rho\sigma\,n\beta}\,A_{\tau\,pkl}
+\frac16\,{\cal F}_{\mu\nu}{}_{mnq} {F}_{\rho\sigma}{}^{q}\, A_\tau{}_{klp}
\nonumber\\
&&{}
\qquad
-\frac{\sqrt{2}}{2}\, \varepsilon^{\alpha\beta}\, A_{\mu\,m\alpha}\partial_n A_{\nu\,p\beta}  {F}_{\rho\sigma}{}^{q}\,   A_\tau{}_{klq}
+\frac12\,
 \partial_p \tilde{B}_{\mu\nu\,mn}   {F}_{\rho\sigma}{}^{q}\,    A_\tau{}_{klq}
\nonumber\\
&&{}
\qquad
+\sqrt{2}\,  \varepsilon^{\alpha\beta}\, 
A_\mu{}_{m\alpha} D_\nu A_\rho{}_{n\beta} \,\partial_p \tilde{B}_{\sigma\tau\,kl}
-\sqrt{2}\, 
A_\mu{}_{m\alpha} \partial_n \tilde{B}_{\nu\rho}{}^\alpha\,\partial_p \tilde{B}_{\sigma\tau\,kl}
\nonumber\\
&&{}
\qquad
 +\frac23\, \varepsilon^{\alpha\beta}\, 
A_{\mu\,m\alpha} \partial_n A_{\nu\,k\beta} A_{\rho\,l\gamma} \partial_p \tilde{B}_{\sigma\tau}{}^\gamma
-
 \varepsilon^{\alpha\beta}\, \varepsilon^{\gamma\delta}
A_{\mu\,m\alpha} \partial_n A_{\nu\,k\beta} A_{\rho\,l\gamma} D_\sigma A_{\tau\,p\delta}
\nonumber\\
&&{}
\qquad
 +\frac{\sqrt{2}}9\,
 \,\partial_m \tilde{\cal H}_{\mu\nu\rho}{}^\alpha
\,A_{\sigma\, n\alpha}  A_{\tau}{}_{klp}
-  D_\mu \tilde B_{\nu\rho\,mn} \partial_p \tilde{B}_{\sigma\tau\,kl}
-\frac{2}3\,
\varepsilon_{\alpha\beta} \,\tilde{\cal H}_{\mu\nu\rho}{}^\beta \partial_k \tilde{B}_{\sigma\tau}{}^{k\alpha}
\nonumber\\
&&{}
\qquad
+{\cal O}(A_{\mu\,\alpha})
\Big)
\;.
\label{LtopexpB}
\eea

\subsection{External diffeomorphisms}
Let us finally turn to the action of the external diffeomorphisms (\ref{skewD}) 
under the type IIB decomposition. 
On the external vielbein $e_{\mu}{}^{a}$ this symmetry reduces to that found in the 
Kaluza-Klein decomposition in (\ref{extdiff}), because on scalar-densities such as 
$e_{\mu}{}^{a}$ and $\xi^{\mu}$ the gauge-covariant derivative of EFT simply reduces to the 
Kaluza-Klein covariant derivative w.r.t.~${\cal A}_{\mu}{}^{m}$.  
For the internal generalized metric ${\cal M}_{MN}$ the external diffeomorphism 
transformations on the various components in (\ref{M2B}) are read off from 
(\ref{skewD}), with the EFT covariant derivatives written out in (\ref{covderMcomp}). 

Next, we consider the external diffeomorphism transformations of the vector fields, 
which are more subtle due to the presence of the term involving the inverse of  
the generalized metric ${\cal M}$. From (\ref{calMcomponents}) we determine 
the relevant components of the matrix ${\cal M}^{MN}$, 
\bea
{\cal M}^{m,n} &=& e^{4\Phi/3}\,m^{mn} \;,
\nonumber\\
{\cal M}_{m\alpha,}{}^n &=& 2\,e^{4\Phi/3}\, \varepsilon_{\alpha\beta}\,m^{nk} b_{km}{}^\beta\;,
\nonumber\\
{\cal M}^{mn,k} &=&
-\frac{\sqrt{2}}{12}\,e^{4\Phi/3}\,\varepsilon^{mnpqr}\,m^{ks}\left(
c_{pqrs} - 6 \,\varepsilon_{\alpha\beta}\,b_{pq}{}^\alpha b_{rs}{}^\beta\right)
\;.
\label{Minvcomp}
\eea
This in turn determines the following gauge variations of the vector field components 
in (\ref{AB_B}), 
 \be
  \begin{split}
   \delta_{\xi}{\cal A}_{\mu}{}^{m} \ &= \ \xi^{\nu}{\cal F}_{\nu\mu}{}^{m}+{\cal M}^{m,n}g_{\mu\nu}\partial_n\xi^{\nu}\;, \\
   \delta_{\xi}{\cal A}_{\mu \, m\alpha} \ &= \ \xi^{\nu}{\cal F}_{\nu\mu m\alpha}
   +{\cal M}_{m\alpha,}{}^{n}g_{\mu\nu}\partial_n\xi^{\nu}\;, \\
   \delta_{\xi}{\cal A}_{\mu \, mnk} \ &= \ \tfrac{1}{2} \epsilon_{mnkpq}\xi^{\nu}{\cal F}_{\nu\mu}{}^{pq}
   +\tfrac{1}{2}\epsilon_{mnkpq}{\cal M}^{pq,n}\partial_n\xi^{\nu}\;, 
  \end{split}
 \ee  
with the field strengths given in  (\ref{FieldSTrengths}). 
As a first check that EFT subjected to this solution of the section constraint is equivalent to 
type IIB supergravity, we infer from the first variation in here that ${\cal A}_{\mu}{}^{m}$ has 
the same external diffeomorphism variation as the Kaluza-Klein vector, c.f.~(\ref{COVDAA}), 
 \be
  \delta_{\xi}{\cal A}_{\mu}{}^{m} \ = \ \xi^{\nu} {\cal F}_{\nu\mu}{}^{m}+\phi^{-\frac{2}{3}}G^{mn}g_{\mu\nu}\partial_n\xi^{\nu}\;, 
 \ee 
therefore justifying the identification of both fields. Indeed, the fields strength components 
${\cal F}_{\mu\nu}{}^{m}$ reduce to the Kaluza-Klein components  $F_{\mu\nu}{}^{m}$, 
see (\ref{FieldSTrengths}) and (\ref{covderfieldstr}), and the metric-dependent 
terms coincide upon identifying 
 \be
  e^{4\Phi/3}\,m^{mn} \ = \ \phi^{-2/3} G^{mn}\;, 
 \ee 
which relates the matrix $m^{mn}\in{\rm SL}(5)$ and the scale factor $\Phi$ 
to the metric $G^{mn}$ with dynamical 
determinant $\phi^2$. (This relation can be fixed, for instance, by noting with (\ref{deltaMMM}) that both sides 
transform in the same way under internal diffeomorphisms.)  
The precise match for the remaining vector field components will 
be the subject of the following sections.


\section{Type IIB supergravity and its Kaluza-Klein decomposition}


In this section, we review ten-dimensional IIB supergravity and bring it into the form that
allows for a convenient translation of its field content into the various components of the EFT fields identified above.

\subsection{Type IIB supergravity}

Denoting ten-dimensional curved 
indices by $\hat{\mu},\hat{\nu},\ldots$, the type IIB field content 
is given by 
 \be
  E_{\hat{\mu}}{}{}^{\hat{a}}\;, \quad 
  m_{\alpha\beta}\;,  \quad
  \hat{C}_{\hat{\mu}\hat{\nu}}{}^{\alpha}\;, \quad
  \hat{C}_{\hat{\mu}\hat{\nu}\hat{\rho}\hat{\sigma}}\;, \quad\alpha,\beta=1,2\;,
  \label{IIBfields}
 \ee
i.e., the zehnbein, the two ${\rm SL}(2)/{\rm SO}(2)$ coset scalars
parametrizing the symmetric ${\rm SL}(2)$ matrix $m_{\alpha\beta}$, a doublet of 2-forms  
and a 4-form. The 2-forms combine RR 2-form and the NS B-field,  
with the abelian field strengths given by
\be
  \hat{F}_{\hat{\mu}\hat{\nu}\hat{\rho}}{}^{\alpha} \ = \ 
  3\,\partial_{[\hat{\mu}}\hat{C}_{\hat{\nu}\hat{\rho}]}{}^{\alpha}\;. 
 \ee 
 The Chern-Simons (CS)-modified curvature of the 4-form is given in components by 
 \be
   \hat{F}_{\hat{\mu}_1\ldots\hat{\mu}_5} \ \equiv \ 5\,\partial_{[\hat{\mu}_1}\hat{C}_{\hat{\mu}_2\ldots \hat{\mu}_5]}
  -\frac54 \, 
  \varepsilon_{\alpha\beta}\,\hat{C}_{[\hat{\mu}_1\hat{\mu}_2}{}^{\alpha}\hat{F}_{\hat{\mu}_3\hat{\mu}_4\hat{\mu}_5]}{}^{\beta}\;, 
 \label{F5}
 \ee 
 such that they satisfy the Bianchi identities
 \be
6\,\partial_{[\hat{\mu}_1}\hat{F}_{\hat{\mu}_2\hat{\mu}_3\hat{\mu}_4\hat{\mu}_5\hat{\mu}_6]}
=-\frac52\,\varepsilon_{\alpha\beta}\hat{F}_{[\hat{\mu}_1\hat{\mu}_2\hat{\mu}_3}{}^{\alpha}\hat{F}_{\hat{\mu}_4\hat{\mu}_5\hat{\mu}_6]}{}^{\beta}
\;,
\label{BianchiIIB}
\ee
and transform as
\be
\begin{split}
\delta \hat{C}_{\hat{\mu} \hat{\nu}}{}^{\alpha}&\ = \ 2\,\partial_{[\hat{\mu}}\hat{\lambda}_{\hat{\nu}]}{}^{\alpha}\;,\\
\delta \hat{C}_{\hat{\mu}\hat{\nu}\hat{\rho}\hat{\sigma}}& \ = \ 4\,\partial_{[\hat{\mu}}\hat{\lambda}_{\hat{\nu}\hat{\rho}\hat{\sigma}]}+\frac12\, \varepsilon_{\alpha \beta} \hat{\lambda}_{[\hat{\mu}}{}^{\alpha}\hat{F}_{\hat{\nu}\hat{\rho}\hat{\sigma}]}{}^{\beta}
\;, 
\end{split}
\label{Ctensor}
\ee
under tensor gauge transformations.
The IIB field equations have been constructed in~\cite{Schwarz:1983wa,Schwarz:1983qr,Howe:1983sra}.
They can be described by a pseudo-action which in our conventions is given by
\be
\begin{split}
S \ = \ \int d^{10}x  \sqrt{|G|}\;\Big(&\hat{R}+\frac{1}{4} \partial_{\hat{\mu}}m_{\alpha \beta}\partial^{\hat{\mu}}m^{\alpha \beta}-\frac{1}{12}\hat{F}_{\hat{\mu}_1\hat{\mu}_2\hat{\mu}_3}{}^{\alpha}\hat{F}^{\hat{\mu}_1\hat{\mu}_2\hat{\mu}_3}{}^{\beta}m_{\alpha \beta}\\
&-\frac{1}{30}\hat{F}_{\hat{\mu}_1\hat{\mu}_2\hat{\mu}_3\hat{\mu}_4\hat{\mu}_5}\hat{F}^{\hat{\mu}_1\hat{\mu}_2\hat{\mu}_3\hat{\mu}_4\hat{\mu}_5}\Big)\\
&-\frac{1}{864}\,
\int d^{10}\hat{x}\,\varepsilon_{\alpha\beta}\,\varepsilon^{\hat{\mu}_1\ldots \hat{\mu}_{10}}C_{\hat{\mu}_1\hat{\mu}_2\hat{\mu}_3\hat{\mu}_4}\hat{F}_{\hat{\mu}_6\hat{\mu}_7\hat{\mu}_8}{}^{\alpha}\hat{F}_{\hat{\mu}_8\hat{\mu}_9\hat{\mu}_{10}}{}^{\beta}\;,
\end{split}
\label{actionIIB}
\ee
and which after variation of the fields has to be supplemented with the standard 
self-duality equations for the 5-form field strength 
\bea
\hat{F}_{\hat{\mu}\hat{\nu}\hat{\rho}\hat\sigma\hat\tau}&=&
\frac1{5!}\,\sqrt{|G|}\,\varepsilon_{\hat{\mu}\hat{\nu}\hat{\rho}\hat\sigma\hat\tau
\hat\mu_1\hat\mu_2\hat\mu_3\hat\mu_4\hat\mu_5}\,
\hat{F}^{\hat\mu_1\hat\mu_2\hat\mu_3\hat\mu_4\hat\mu_5}
\;,
\label{FF55}
\eea
with $|G|\equiv |{\rm det}\,G_{\hat\mu\hat\nu}|= |{\rm det}\,E_{\hat\mu}{}^{\hat a}|^2$.
It is straightforward to verify that the integrability conditions of the self-duality equations together with the Bianchi identities (\ref{BianchiIIB})
coincide with the second-order field equations obtained by variation of (\ref{actionIIB}).
Our ${\rm SL}(2)$ conventions can be translated into the ${\rm SU}(1,1)/{\rm U}(1)$
conventions of \cite{Schwarz:1983qr}, by combining the real components of the doublet 
$\hat{F}_{\hat\mu\hat\nu\hat\rho}{}^\alpha$ into a complex ${F}$
\bea
{F}_{\hat\mu\hat\nu\hat\rho}{} &\equiv &
\hat{F}_{\hat\mu\hat\nu\hat\rho}{}^1+i\,\hat{F}_{\hat\mu\hat\nu\hat\rho}{}^2
\;,
\eea
and parametrizing the symmetric ${\rm SL}(2)$ matrix $m_{\alpha\beta}$
in terms of a complex scalar $B$ as
\bea
m_{\alpha\beta} &\equiv& (1-BB^*)^{-1} \left(
\begin{array}{cc}
(1-B)(1-B^*) & i (B-B^*) \\
 i (B-B^*)  & (1+B)(1+B^*)
 \end{array}
\right)
\;.
\eea
In terms of the complex combinations 
\bea
G_{\hat\mu\hat\nu\hat\rho}\ \equiv \ f(F_{\hat\mu\hat\nu\hat\rho}-B F_{\hat\mu\hat\nu\hat\rho}^*)\;,\quad
P_{\hat\mu} \ \equiv \ f^2 \partial_{\hat\mu} B\;,
\qquad\mbox{with}\;\;f=(1-BB^*)^{-1/2}\;, 
\eea 
charged under the ${\rm U}(1)\subset{\rm SU}(1,1)$,
the kinetic terms of (\ref{actionIIB}) translate into those of \cite{Schwarz:1983qr} with
\bea
 m_{\alpha\beta} \hat{F}_{\hat{\mu}\hat{\nu}\hat{\rho}}{}^{\alpha}\hat{F}^{\hat{\mu}\hat{\nu}\hat{\rho}\,\beta}
 &=& G^*_{\hat{\mu}\hat{\nu}\hat{\rho}} G^{\hat{\mu}\hat{\nu}\hat{\rho}}  
\;,\nonumber\\
\frac14\,\partial_{\hat{\mu}} m_{\alpha\beta}\partial^{\hat{\mu}} m^{\alpha\beta}
&=& -2\,P^*_{\hat{\mu}} P^{\hat{\mu}}
\;.
\eea
In the following, we will perform the standard $5+5$ 
Kaluza-Klein redefinitions of the IIB fields
but keeping the dependence on all ten coordinates.

\subsection{Kaluza-Klein decomposition and field redefinitions} 
\label{subsec:redefinitions}

We now split the the coordinates according to a $5+5$ Kaluza-Klein decomposition
into
\be
x^{\hat{\mu}}=(x^\mu,y^m) \;,
\ee
and similarly for the flat indices $\hat{a}=(\underline{a},\underline{\alpha})$\,.
The $\mu$ and $\underline{a}$ indices range from $0,\ldots ,4$ and respectively represent the curved and flat indices of 
what we will refer to as the external space. 
Similarly, the indices $m$ and $\underline{\alpha}$ range from $1,\ldots,5$ and are associated with the internal space.
After partial fixation of the Lorentz gauge symmetry, the vielbein may be brought into triangular form (\ref{KKgauge})
\be
E_{\hat{\mu}}{}^{\hat{a}}=\left(\begin{array}{cc}
\phi^{-1/3}\;e_{\mu}{}^{\underline{a}} & 
A_{\mu}{}^{m}\phi_{m}{}^{\underline{\alpha}} \\
 0  & \phi_{m}{}^{\underline{\alpha}}
 \end{array}\right)\;, 
 \label{E10}
 \ee
 parametrized in terms of two 5 by 5 matrices $e_\mu{}^{\underline{a}}$ and $\phi_m{}^{\underline{\alpha}}$
with $\phi\equiv {\rm det}(\phi_m{}^{\underline{\alpha}}$), and the Kaluza-Klein vectors $A_\mu{}^m$.
We stress again that all fields depend on all ten coordinates, such that we are still describing the full
IIB theory.
The result of the ten-dimensional Einstein-Hilbert term in the parametrization (\ref{E10}) has been
given in~\cite{Hohm:2013vpa} and in particular features the non-abelian Kaluza-Klein field strength
\bea
{F}_{\mu\nu}{}^m &\equiv& 2\,\partial_{[\mu} A_{\nu]}{}^m 
-A_{\mu}{}^n\partial_n A_\nu{}^m+A_{\nu}{}^n\partial_n A_\mu{}^m
\;.
\label{FYM}
\eea

In order to describe the Kaluza-Klein decomposition of the $p$-forms, we introduce 
in standard Kaluza-Klein manner the projector 
$P_{\mu}{}^{\hat{\nu}}=E_{\mu}{}^{\underline{a}}E_{\underline{a}}{}^{\hat{\nu}}$.
It converts 10-dimensional curved indices into 5-dimensional ones such that
the resulting fields transform covariantly (i.e.\ according to the structure of their internal indices)
under internal diffeomorphisms. 
We denote its action by a bar on the corresponding $p$-form components,
\bea
\overline{C}_\mu &\equiv& P_{\mu}{}^{\hat{\nu}}\,\hat{C}_{\hat{\nu}}\;,\qquad
\mbox{etc.}\;,
\eea
such that the IIB two- and four-form give rise to the components 
\be
\begin{split}
\overline{C}_{mn}{}^{\alpha} \ &= \ \hat{C}_{mn}{}^{\alpha}\;,\\
\overline{C}_{\mu\, m}{}^{\alpha} \ &= \ \hat{C}_{\mu m}{}^{\alpha}-A_{\mu}{}^p\hat{C}_{pm}{}^{\alpha}\;,\\
\overline{C}_{\mu\nu}{}^{\alpha} \ &= \ \hat{C}_{\mu \nu}{}^{\alpha}-2A_{[\mu}{}^{p}\hat{C}_{|p|\nu]}{}^{\alpha}+A_{\mu}{}^{p}A_{\nu}{}^{q}\hat{C}_{pq}{}^{\alpha}\;,\\[1ex]
\overline{C}_{mnkl} \ &= \ \hat{C}_{mnkl}\;,\\
\overline{C}_{\mu \, nkl} \ &= \ \hat{C}_{\mu nkl}-A_{\mu}{}^{p}\hat{C}_{pnkl}\;,\\
\overline{C}_{\mu \nu\, kl} \ &= \ \hat{C}_{\mu \nu kl}-2A_{[\mu}{}^{p}\hat{C}_{|p|\nu]kl}+A_{\mu}{}^{p}A_{\nu}{}^{q}\hat{C}_{pqkl}\;,\\
\overline{C}_{\mu \nu \rho \,l} \ &= \ \hat{C}_{\mu \nu \rho\, l}-3A_{[\mu}{}^{p}\hat{C}_{|p|\nu \rho]\,l}+3A_{[\mu}{}^{p}A_{\nu}{}^{q}\hat{C}_{|pq|\rho ]\,l}-A_{\mu}{}^{p}A_{\nu}{}^{q}A_{\rho}{}^{r}\hat{C}_{pqrl}\;,\\
\overline{C}_{\mu \nu \rho \sigma} \ &= \ \hat{C}_{\mu \nu \rho \sigma}-4A_{[\mu}{}^{p}\hat{C}_{|p|\nu \rho \sigma]}+6A_{[\mu}{}^{p}A_{\nu}{}^{q}\hat{C}_{|pq|\rho \sigma]}
-4A_{[\mu}{}^{p}A_{\nu}{}^{q}A_{\rho}{}^{r}\hat{C}_{|pqr| \sigma]}
\\
&\quad
+A_{\mu}{}^{p}A_{\nu}{}^{q}A_{\rho}{}^{r}A_{\sigma}{}^{s}\hat{C}_{pqrs}
\;.
\end{split}
\label{C2C4}
\ee
The same redefinition applies to field strengths and gauge parameters.
The redefined fields now transform covariantly under internal diffeomorphisms. 
Indeed, separating ten-dimensional diffeomorphisms into 
$\xi^{\hat{\mu}}=(\xi^{\mu},\Lambda^m)$, we find together with (\ref{Ctensor}) 
\be
\begin{split}\label{transC2}
\delta \overline{C}_{mn}{}^{\alpha}&=2\partial_{[m}\overline{\lambda}_{n]}{}^{\alpha}+\mathcal{L}_{\Lambda}\overline{C}_{mn}{}^{\alpha}\;,\\
\delta \overline{C}_{\mu \,m}{}^{\alpha} \ &= \ D^{\KK}_{\mu}\overline{\lambda}_m{}^{\alpha}-\partial_m \overline{\lambda}_{\mu}{}^{\alpha}+\mathcal{L}_{\Lambda}\overline{C}_{\mu \,m}{}^{\alpha}
\;, \\
\delta \overline{C}_{\mu \nu}{}^{\alpha} \ &= \ 2 \, D^{\KK}_{[\mu}\overline{\lambda}^{}_{\nu]}{}^{\alpha}+{F}_{\mu \nu}{}^k\overline{\lambda}_k{}^{\alpha}+\mathcal{L}_{\Lambda}\overline{C}_{\mu \nu}{}^{\alpha}\;,
\end{split}
\ee
for the transformation behaviour 
of the redefined 2-forms under gauge transformations and internal diffeomorphisms. 
As in the previous section, derivatives $D^{\KK}_\mu$
are covariantized w.r.t.\ the action of internal diffeomorphisms, i.e.
\bea
D^{\KK}_{\mu}\overline{\lambda}_m{}^{\alpha} &\equiv&
\partial_{\mu}\overline{\lambda}_m{}^{\alpha} -A_\mu{}^n\partial_n  \overline{\lambda}_m{}^{\alpha}
-\partial_m A_\mu{}^n \overline{\lambda}_n{}^{\alpha}\;,
\qquad\mbox{etc.}\,.
\eea
In contrast to $D=11$ supergravity for which these redefinitions and covariant gauge transformations have
been explicitly worked out  in~\cite{Hohm:2013vpa}, the presence of Chern-Simons terms in the IIB
field strengths (\ref{F5}) requires a further redefinition for the components of the 4-form in order to
establish the dictionary to the fields of EFT.
This is related to the fact that tensor gauge transformations for the EFT $p$-forms that we have discussed
in the previous section do not mix these forms with the scalar fields of the theory.
This motivates the following and final field redefinition\footnote{
Similar redefinitions have been discussed in \cite{Ciceri:2014wya}
in order to recover part of the ${\rm E}_{6(6)}$ tensor hierarchy structure 
from the IIB supersymmetry variations.
}
\be
\begin{split}\label{redefC4}
C_{klmn}\ &\equiv \ \overline{C}_{klmn}\;,\\
C_{\mu\, kmn}\ &\equiv \ \overline{C}_{\mu\, kmn}-\frac38\,\varepsilon_{\alpha \beta}\overline{C}_{\mu \,[k}{}^{\alpha}\overline{C}_{mn]}{}^{\beta}\;,\\
C_{\mu \nu\, mn}\ &\equiv \ \overline{C}_{\mu \nu \,mn}-\frac18\,\varepsilon_{\alpha \beta}\overline{C}_{\mu\nu}{}^{\alpha}\overline{C}_{mn}{}^{\beta}\;,\\
C_{\mu \nu \rho\, m}\ &\equiv \ \overline{C}_{\mu \nu \rho\, m}-\frac38\,\varepsilon_{\alpha \beta}
\overline{C}_{[\mu \nu}{}^{\alpha}\overline{C}_{\rho]\,m}{}^{\beta}\;,\\
C_{\mu\nu\rho\sigma}\ &\equiv \ \overline{C}_{\mu \nu \rho \sigma}\;.
\end{split}
\ee
For the components of the two-form $\overline{C}_{\mu\nu}{}^\alpha$, etc., there is no further
redefinition, so for simplicity of the notation, we will simply drop their bars in the following
\bea
C_{mn}{}^\alpha \equiv  \overline{C}_{mn}{}^\alpha\;,\qquad
C_{\mu m}{}^\alpha \equiv  \overline{C}_{\mu m}{}^\alpha\;,\qquad
C_{\mu\nu}{}^\alpha \equiv  \overline{C}_{\mu\nu}{}^\alpha\;.
\eea

Although we have not seen the 3-form and the 4-form in the tensor hierarchy of the ${\rm E}_{6(6)}$ EFT, 
we will show later that it is possible to test their expressions by comparing the reduced $D=10$ self duality equations (\ref{FF55})
to the first order duality equations \eqref{dualityrel} from EFT.
The redefined 4-forms (\ref{redefC4}) continue to transform covariantly under internal diffeomorphisms
with their total gauge transformations given by
\be
\begin{split}\label{transC4}
\delta C_{mnkl} \ &= 4\partial_{[m}\overline\lambda_{nkl]}+\frac32\, \epsilon_{\alpha \beta} \partial_{[m}\overline{\lambda}_{n}C_{kl]}{}^{\beta}+\mathcal{L}_{\Lambda}C_{mnkl}\;,\\
\delta C_{\mu \,kmn} \ &= \ D^{\KK}_{\mu}\overline\lambda_{kmn}-3\partial_{[k}\overline\lambda_{|\mu|mn]} +\mathcal{L}_{\Lambda}C_{\mu \,kmn}
\\ & \quad +\frac34\, \varepsilon_{\alpha \beta} \left(\overline{\lambda}_{[k}{}^{\alpha}\partial_m{C}_{|\mu|n]}{}^{\beta}-\partial_{[m}\overline{\lambda}_k{}^{\alpha}{C}_{|\mu|n]}{}^{\beta}\right)\;,\\
\delta C_{\mu \nu\, mn} \ &= \ 2\, D^{\KK}_{[\mu} \overline\lambda_{\nu]\, mn}+2\partial_{[m}\overline\lambda_{n] \mu \nu}+{F}_{\mu \nu}{}^k\overline\lambda_{kmn}+\mathcal{L}_{\Lambda}C_{\mu \nu mn}
\\&\quad+\frac14\, \varepsilon_{\alpha\beta}\left(-2\partial_{[m}{C}_{|\mu| n]}{}^{\alpha}\overline{\lambda}_{\nu}{}^{\beta}+F_{\mu \nu\, [m}{}^{\alpha}\overline{\lambda}_{n]}{}^{\beta}-\overline{\lambda}_{[m}{}^{\alpha}\partial_{n]}{C}_{\mu \nu}{}^{\beta}\right)\;.
\end{split}
\ee
We see that after the redefinitions (\ref{redefC4}), the variation of $\delta {C}_{\mu \,kmn}$
and $\delta C_{\mu\nu \,mn}$ no longer carry any scalar fields $\overline{C}_{mn}{}^\alpha$
and are thus of the form to be matched with the fields and transformations of EFT.
The field strengths appearing on the r.h.s.\ of (\ref{transC4}) are the Kaluza-Klein field strength (\ref{FYM}) and 
the modified three-form field strength
\bea
F_{\mu\nu\, n}{}^{\alpha}&\equiv&\overline{F}_{\mu \nu\, n}{}^{\alpha}-{F}_{\mu \nu}{}^k {C}_{k n}{}^{\alpha}\;,\nonumber\\
&=&2D_{[\mu}{C}_{\nu] m}{}^{\alpha}+\partial_{m}{C}_{\mu \nu}{}^{\alpha}\;,
\label{redefF2}
\eea
again redefined such that the scalar contribution is split off.
For completeness we also give the remaining components of the three-form field strength
\be
\begin{split}
{F}_{kmn}{}^{\alpha}\ &\equiv \ \overline{F}_{kmn}{}^{\alpha}=3\partial_{[k}{C}_{mn]}{}^{\alpha}\;,\\
{F}_{\mu\,mn}{}^\alpha \ &\equiv \ \overline{F}_{\mu\,mn}{}^\alpha \ = \ D^{\KK}_\mu {C}_{mn}{}^\alpha - 2 \partial_{[m} {C}_{|\mu|\,n]}{}^\alpha \;,\\
{F}_{\mu\nu \rho}{}^{\alpha}\ &\equiv \ \overline{F}_{\mu\nu \rho}{}^{\alpha}\ =\ 3\,D^{\KK}_{[\mu}{C}_{\nu\rho]}{}^{\alpha}-3\,{F}_{[\mu\nu}{}^k{C}_{\rho]k}{}^{\alpha}\;,
\end{split}
\ee
as well as the properly redefined 
components of the five-form field strength,
expressed in terms of the components (\ref{redefC4})
according to
\bea
{F}_{mpqrs}&\equiv&\overline{F}_{mpqrs}=5\,\partial_{[m} {C}_{pqrs]} 
-\frac54\,\varepsilon_{\alpha\beta}\,{C}_{[mp}{}^{\alpha} \overline{F}_{qrs]}{}^\beta
\;,\nonumber\\
{F}_{\mu pqrs}&\equiv& \overline{F}_{\mu pqrs}\nonumber\\
&=& D^{\KK}_{\mu}{C}_{pqrs}-4\partial_{[p} C_{|\mu| qrs]}-\frac34\,\varepsilon_{\alpha \beta}{C}_{[pq}{}^{\alpha}{F}_{|\mu|rs]}{}^{\beta}+\frac32\,\varepsilon_{\alpha \beta}{C}_{[p q}{}^{\alpha}\partial_r{C}_{|\mu|s]}{}^{\beta}\;,\nonumber\\
F_{\mu \nu kmn}&\equiv&\overline{F}_{\mu \nu kmn}-\frac34\,\varepsilon_{\alpha \beta}F _{\mu \nu [k}{}^{\alpha}{C}_{mn]}{}^{\beta}-{F}_{\mu \nu}{}^p(\overline{C}_{pkmn}-\frac38\,\varepsilon_{\alpha \beta}{C}_{[km}{}^{\alpha}{C}_{|p|n]}{}^{\beta})\nonumber\\
&=&2\,D^{\KK}_{[\mu} C_{\nu] kmn}+3\partial_{[k} C_{|\mu \nu| mn]}-\frac32\,\varepsilon_{\alpha \beta} {C}_{\mu [k}{}^{\alpha}\partial_{m}{C}_{|\nu|n]}{}^{\beta}\;,\nonumber\\
F_{\mu \nu \rho mn}&\equiv&\overline{F}_{\mu \nu \rho mn}-\frac14 \varepsilon_{\alpha \beta} \overline{F}_{\mu \nu \rho}{}^{\alpha}{C}_{mn}{}^{\beta}\nonumber\\
&=&3D^{\KK}_{[\mu} C_{\nu\rho]mn}-2\partial_{[m}C_{|\mu \nu \rho| n]}-3{F}_{[\mu \nu}{}^k C_{\rho] k m n}\nonumber\\
&&{}-\frac32\,\varepsilon_{\alpha \beta}(\partial_{[m}{C}_{[\mu\nu}{}^{\alpha}{C}_{\rho]n]}{}^{\beta}+ {C}_{[\mu |m|}{}^{\alpha}D_{\nu}{C}_{\rho]n}{}^{\beta})\;,\nonumber\\
{F}_{\mu \nu \rho \sigma m}&\equiv&
\overline{F}_{\mu \nu \rho \sigma m}\nonumber\nonumber\\
&=&4D^{\KK}_{[\mu} C_{\nu \rho \sigma]m}+\partial_{m}{C}_{\mu \nu \rho \sigma}+6 {F}_{[\mu \nu}{}^p C_{\rho \sigma] p m}\nonumber\\
&&{}
+\frac32\,\varepsilon_{\alpha \beta}{F}_{[\mu\nu}{}^k  {C}_{\rho |m|}{}^{\alpha} {C}_{\sigma]k}{}^{\beta}
-\frac34\,\varepsilon_{\alpha \beta}{C}_{[\mu \nu}{}^{\alpha}\partial_{|m|}{C}_{\rho \sigma]}{}^{\beta}
+\varepsilon_{\alpha\beta}\,{C}_{\mu\,m}{}^\alpha {\cal F}_{\nu\rho\sigma}{}^\beta
\;,\nonumber\\
{F}_{\mu\nu\rho\sigma\tau}&\equiv&\overline{F}_{\mu\nu\rho\sigma\tau}=
5D^{\KK}_{[\mu}{C}_{\nu\rho\sigma\tau]}-10{F}_{[\mu\nu}{}^m C_{\rho\sigma\tau]m}-\frac{15}{4}\,\varepsilon_{\alpha \beta}{C}_{[\mu\nu}{}^{\alpha}D^{\KK}_{\rho}{C}_{\sigma \tau]}{}^{\beta}\;.
\label{FFFFF}
\eea

\subsection{External diffeomorphisms}

In the previous subsection we have decomposed the IIB fields according to a 5+5 Kaluza-Klein split
(without giving up the dependence on the 5 internal coordinates) and spelled out their transformations
under internal diffeomorphisms and tensor gauge transformations after suitable redefinitions of the
various components. Before fully establish the dictionary to the fields in the EFT basis, we will in
this section compute the behaviour of the redefined IIB fields under external diffeomorphisms $\xi^\mu$,
whose parameter may in general also depend on all 10 coordinates.

Above, we have already discussed the transformation of the KK vector fields
under external diffeomorphisms
 \be\label{COVDAA2}
  \delta^{\rm cov}_{\xi}A_{\mu}{}^{m} \ = \ \xi^{\nu} F_{\nu\mu}{}^{m}
  +\phi^{-\frac{2}{3}}G^{mn}g_{\mu\nu}\partial_n\xi^{\nu}\;,
 \ee 
c.f.\ (\ref{COVDAA}), which is in agreement
with the EFT gauge vector transformations reduced to this component. 
Let us now test the remaining vector components from the IIB $p$-forms.
For ${C}_{\mu m}{}^{\alpha}$, as redefined in (\ref{C2C4}), a straightforward
calculation gives
 \bea
  \delta_{\xi}{C}_{\mu m}{}^{\alpha} &=& 
  {\cal L}_{\xi}{C}_{\mu m}{}^{\alpha}
   -\phi^{-\frac{2}{3}}G^{nk}C_{nm}{}^{\alpha} g_{\mu\nu}\partial_k\xi^{\nu}
  \nonumber\\
  &&{}
   +\partial_m\xi^{\nu} A_{\nu}{}^{n} {C}_{\mu n}{}^{\alpha}
  -A_{\mu}{}^{n}\partial_n\xi^{\nu}{C}_{\nu m}{}^{\alpha}
  +\partial_m\xi^{\nu}{C}_{\mu \nu}{}^{\alpha}
 \;,
 \eea  
under external diffeomorphisms. 
The origin of the second term is the corresponding variation of the Kaluza-Klein
vector (\ref{COVDAA2}) which enters the redefined fields in (\ref{C2C4}).
As for the Kaluza-Klein vector field, it follows
that the last three terms are eliminated by field dependent gauge transformations 
with parameters (parameter redefinition)
 \be\label{Fieldparam}
  \Lambda^m \ = \ -\xi^{\nu} A_{\nu}{}^{m}\;, \quad
  \overline{\lambda}_{m}{}^{\alpha} \ = \ -\xi^{\nu}{C}_{\nu m}{}^{\alpha}\;, \quad
  \overline{\lambda}_{\mu}{}^{\alpha} \ = \ -\xi^{\nu}{C}_{\nu\mu}{}^{\alpha}
  \;,
 \ee
which render the action of the diffeomorphism manifestly gauge covariant. 
Together, the variation takes the form
\be\label{FINALdeltaxiC}
     \delta^{\rm cov}_{\xi}{C}_{\mu m}{}^{\alpha} \ = \ \xi^{\nu}F_{\nu\mu m}{}^{\alpha}
     -\phi^{-\frac{2}{3}}G^{nk}{C}_{nm}{}^{\alpha} g_{\mu\nu}\partial_k\xi^{\nu}
     \;.
 \ee
Note in particular that the field strength entering this formula is the one 
defined in (\ref{redefF2}) which does not carry any scalar contributions.
This is the form of the variation that we will be able to match with the corresponding 
variation for the fields in the EFT basis.

Next let us consider the variation of the 4-form component ${C}_{\mu mnk}$. 
After standard Kaluza-Klein redefinition (\ref{C2C4}), some straightforward calculation
yields
 \bea
   \delta_{\xi}\overline{C}_{\mu mnk} &=& \, \xi^{\nu}
   \left(2D^{\KK}_{[\nu}\overline{C}_{\mu]\,mnk}+3\partial_{[m}\overline{C}_{|\nu\mu|nk]}\right)
   +{\cal L}_{\xi^{\nu}A_{\nu}}\overline{C}_{\mu mnk}\nonumber\\
   &&+D^{\KK}_{\mu}(\xi^{\nu}\overline{C}_{\nu mnk})-3\,\partial_{[m}(\xi^{\nu}\overline{C}_{|\nu\mu|nk]})
   +\phi^{-\frac{2}{3}}G^{lp}\,C_{mnkl}\,g_{\mu\nu}\partial_p\xi^{\nu}\;,
 \eea
 for the variation under external diffeomorphisms in terms of the redefined fields.
 In the first term we recognize the covariant field strength $F_{\nu\mu mnk}$ from 
 (\ref{FFFFF}) up to its bilinear contributions. 
 These will be completed once we consider the variation of the redefined four form 
  \be\label{doubletildevar}
 \begin{split}
   \delta_{\xi}C_{\mu mnk} \ = \ &\,\delta_{\xi}\overline{C}_{\mu mnk} 
   -\frac{3}{8}\, \varepsilon_{\alpha\beta} 
   \delta_{\xi}{C}_{\mu[m}{}^{\alpha}{C}_{nk]}{}^{\beta}
     -\frac{3}{8}\, \varepsilon_{\alpha\beta} 
  {C}_{\mu[m}{}^{\alpha}\delta_{\xi}{C}_{nk]}{}^{\beta}
  \;, 
 \end{split} 
 \ee 
with the second term obtained via (\ref{FINALdeltaxiC}), and the third 
term carrying 
 \bea 
  \delta_{\xi} {C}_{mn}{}^{\alpha}
  &=&\xi^{\nu}{F}_{\nu mn}{}^{\alpha}+2\partial_{[m}(\xi^{\nu}{C}_{|\nu| n]}{}^{\alpha})+\mathcal{L}_{\xi^{\nu}A_{\nu}}{C}_{mn}{}^{\alpha}\;.
 \eea
Combining all these contributions and supplementing the variation by the
gauge transformations with parameters (\ref{Fieldparam}), we arrive at the
final form
\bea
   \delta^{\rm cov}_{\xi}C_{\mu mnk} &=&\,\xi^{\nu}\,F_{\nu\mu mnk}+
    \phi^{-\frac{2}{3}}G^{lp}\left(C_{mnkl}+\frac{3}{8}\, \varepsilon_{\alpha\beta}
    {C}_{l[m}{}^{\alpha} {C}_{nk]}{}^{\beta}\right)
    g_{\mu\nu}\partial_p\xi^{\nu}
\;.
\label{finaldelC4}
\eea

In the next section, we will provide the complete dictionary between the Kaluza-Klein redefined 
fields of type IIB supergravity and the fundamental fields in the $E_{6(6)}$ EFT. 
In particular, matching the EFT equations against the IIB self-duality equations (\ref{FF55}),
we will explicitly reconstruct the remaining 4-form components $C_{\mu\nu\rho m}$, $C_{\mu\nu\rho\sigma}$\,.

\section{Embedding of type IIB into E$_{6(6)}$ Exceptional Field Theory}
\label{sec:embedding}

In this section, we provide an explicit dictionary between the Kaluza-Klein redefined fields of 
type IIB supergravity and those of the $E_{6(6)}$ exceptional field theory after picking solution
(\ref{explicit_sectionB}) of the section constraint. 
We first show that the fundamental EFT fields can be identified among the redefined IIB fields 
on a pure kinematical level by comparing the transformation behaviour under diffeomorphisms
and gauge transformations. We then show that the equivalence also holds on the 
dynamical level by reproducing the IIB self-duality equations (\ref{FF55}) from the 
EFT field equations. In particular, this will allow us to obtain explicit expressions for
the remaining 4-form components $C_{\mu\nu\rho m}$, $C_{\mu\nu\rho\sigma}$ which do
not show up among the fundamental EFT fields, but whose existence follows
 from the EFT dynamics.

\subsection{Kinematics}

Before identifying the details of the IIB embedding, let us first 
revisit the resulting field content of EFT after picking solution
(\ref{explicit_sectionB}) of the section constraint. 
With the split (\ref{split27B}), (\ref{split78B}),
the full $p$-form field content of the ${\rm E}_{6(6)}$ Lagrangian in this basis
is given by (\ref{AB_B})
\bea
\left\{ {\cal A}_\mu{}^m, {\cal A}_{\mu\,m\,\alpha}, {\cal A}_{\mu\,kmn},  {\cal A}_{\mu\,\alpha} \right\}\;,\qquad
\left\{{\cal B}_{\mu\nu}{}^\alpha ,  {\cal B}_{\mu\nu\,mn}, {\cal B}_{\mu\nu}{}^{m\,\alpha}  \right\}
\;,
\label{f1}
\eea
where, more precisely, the Lagrangian depends on the 2-forms
only under certain contractions with internal derivatives, c.f.~(\ref{onlyBB}).
The EFT scalar sector is described by the fields parametrizing the
${\rm E}_{6(6)}$ generalized metric ${\cal M}_{MN}$ (\ref{M2B})
\bea
\{ \Phi,  m_{mn}, m_{\alpha\beta}, b_{mn}{}^{\alpha}, c_{klmn} \}
\;.\label{f2}
\eea
Comparing the index structure of these fields to the field content of the Kaluza-Klein 
decomposition of IIB supergravity given in the previous section allows to give a
first qualitative correspondence between the two formulations. With the discussion
of section~\ref{subsec:embedding} in mind, it appears natural to relate the field
${\cal A}_\mu{}^m$ to the IIB Kaluza-Klein vector field $A_\mu{}^m$, and the
scalars $\Phi$, $m_{mn}$, to the remaining components of the internal IIB metric
(\ref{E10}).

According to their index structure, the fields $\{b_{mn}{}^{\alpha}, {\cal A}_{\mu\,m\,\alpha}, 
{\cal B}_{\mu\nu}{}^\alpha\}$ from (\ref{f1}), (\ref{f2})
will relate to the different components of the ${\rm SL}(2)$ doublet of ten-dimensional two-forms.
Similarly the fields $c_{klmn}, {\cal A}_{\mu\,kmn}, {\cal B}_{\mu\nu\,mn}$ will translate
into the components of the (self-dual) IIB four-form. 
The remaining fields ${\cal A}_{\mu\,\alpha}, {\cal B}_{\mu\nu}{}^{m\,\alpha}$
descend from components of the doublet of dual six-forms.
The two-form tensors ${\cal B}_{\mu\nu\,m}$ that complete the two-forms in (\ref{f1})
into the full ${\bf 27}$ ${\cal B}_{\mu\nu\,M}$ of ${\rm E}_{6(6)}$ do not figure in the ${\rm E}_{6(6)}$ 
covariant Lagrangian. They represent the degrees of freedom on-shell dual to the Kaluza-Klein vector fields, 
i.e.\ descending from the ten-dimensional dual graviton. 

Recall that in the EFT formulation, all vector fields in (\ref{f1}) appear with a Yang-Mills kinetic term 
whereas the two-forms couple via a topological term and are on-shell dual to the vector fields. 
In order to match the structure of IIB supergravity, we will thus have to trade the 
Yang-Mills vector fields ${\cal A}_{\mu\,\alpha}$ for a propagating two-form ${\cal B}_{\mu\nu}{}^\alpha$.
Let us make this more explicit. The $\alpha$-component of the EFT duality equations~(\ref{dualityrel})
yields
\bea
e\,{\cal M}^{\alpha\beta}\,{\cal F}^{\mu\nu}{}_{\beta} &=& -\frac16\,\varepsilon^{\mu\nu\rho\sigma\tau}\,
\tilde{\cal H}_{\rho\sigma\tau}{}^\alpha - e\,{\cal M}^\alpha{}_{\underline{M}}\,{\cal F}^{\mu\nu\,\underline{M}}
\;,
\label{dualityA}
\eea
where we have introduced the index split 
\bea
\{X^M\} &\longrightarrow& \{X^{\underline{M}}, X_\alpha \}
\;.
\eea
With the two-form fields $\tilde{\cal B}_{\mu\nu}{}^{k\beta}$ entering
${\cal F}_{\mu\nu\,\beta}$ on the l.h.s.\ of (\ref{dualityA}), this
duality equation then allows to eliminate all $\tilde{\cal B}_{\mu\nu}{}^{k\beta}$ from the Lagrangian. 
The gauge symmetry (\ref{dAa}) shows that in the process, the vector fields ${\cal A}_{\mu\,\alpha}$ also disappear
from the Lagrangian.\footnote{
Strictly speaking, equation (\ref{dualityA}) only holds 
up to an $x$-dependent `integration constant' ${\cal C}^{\mu\nu\,\alpha}(x)$,
since it enters under $y$-derivative. To fix this freedom,
we have to combine the equation with the vector field equations, 
\bea
D_\nu\left(e\,{\cal M}^{\alpha}{}_M\,{\cal F}^{\nu\mu\,M}\right) &=& 
\frac14\,\varepsilon^{\mu\nu\rho\sigma\tau}\,\varepsilon^{\alpha\beta}\,
 {\cal F}_{\nu\rho}{}^k {\cal F}_{\sigma\tau\,k\beta} 
\;,
\eea
and the Bianchi identity (\ref{BianchiA}), leaving us with $D_\mu{\cal C}^{\mu\nu\,\alpha}=0$\,.
In the following we will directly set ${\cal C}^{\mu\nu\,\alpha}=0$.
}
We infer from (\ref{dualityA}) that the kinetic term for the remaining vector fields
changes into the form 
\bea
e^{-1}\,{\cal L}_{\rm kin,1} &=&
-\frac14\, {\cal F}_{\mu\nu}{}^M {\cal F}^{\mu\nu}{}^N \tilde{\cal M}_{MN}
\;,
\label{FFM}
\eea
with $\tilde{\cal M}_{MN}$ from (\ref{tildeM}).
At the same time, the two-forms $\tilde{B}_{\mu\nu}{}^\alpha$ are 
promoted into propagating fields with kinetic term 
\bea
e^{-1}\,{\cal L}_{\rm kin,2} &=&
-\frac1{12}\,e^{-5\Phi/3}\,m_{\alpha\beta} \,
\tilde{\cal{H}}_{\mu\nu\rho}{}^\alpha \tilde{\cal{H}}^{\mu\nu\rho\,\beta}\;. 
\label{HHM}
\eea
After this dualization, the remaining field content thus is given by
\bea
\left\{ \Phi,  m_{mn}, b_{mn}{}^{\alpha}, c_{klmn}, 
 {\cal A}_\mu{}^m, {\cal A}_{\mu\,m\,\alpha}, {\cal A}_{\mu\,kmn},  
 {\cal B}_{\mu\nu}{}^\alpha ,  {\cal B}_{\mu\nu\,mn}  \right\}
\;,
\label{fff}
\eea
with all except for the last field representing propagating degrees of freedom.
In contrast, the two-form ${\cal B}_{\mu\nu\,mn}$ is related by a first order
duality equation~(\ref{dualityrel}) to ${\cal A}_{\mu\,kmn}$, remnant of the IIB
self-duality equations (\ref{FF55}).
In the following, we will make the dictionary fully explicit.

\subsection{Dictionary and match of gauge symmetries}
\label{subsec:dictionary}

Having established the match of degrees of freedom between IIB supergravity and
EFT upon choosing the IIB solution of the section condition, we can now make the map more precise
by inspecting the gauge and diffeomorphism transformations on both sides.
After Kaluza-Klein decomposition and redefinition of the IIB fields,
as described in section~\ref{subsec:redefinitions}, the resulting components turn out to be
proportional to the EFT fields in their decomposition given in section~\ref{subsec:decomposition} above.
Specifically, comparing the variation of the EFT vector and two-form fields
(\ref{deltaAAA}), (\ref{deltaBB}), to the corresponding transformations in 
\eqref{transC2}, \eqref{transC4}, allows us to establish the dictionary
\bea
&&A_\mu{}^m = {\cal A}_\mu{}^m\;,\qquad
{C}_{\mu m}{}^{\alpha}=-\varepsilon^{\alpha \beta}\mathcal{A}_{\mu \, m \beta}
\;,\qquad{C}_{\mu \nu}{}^{\alpha}=\tilde{\mathcal{B}}_{\mu \nu}{}^{\alpha}\;,
\nonumber\\
&&
C_{\mu \nu \,mn}=\frac{\sqrt{2}}{4}\,\tilde{\mathcal{B}}_{\mu \nu \,mn}\;,
\qquad
C_{\mu kmn}=\frac{\sqrt{2}}{4}\,\mathcal{A}_{\mu \,kmn}
=\frac{\sqrt{2}}{8}\, \varepsilon_{mnkpq} \, \mathcal{A}_{\mu}{}^{pq}
\;,
\label{dictAB}
\eea
respectively. The corresponding gauge parameters
translate with the same proportionality factors, and also the redefined
IIB field strengths (\ref{redefF2}), (\ref{FFFFF}) precisely translate into the
EFT analogues
\bea
&&F_{\mu\nu}{}^m = {\cal F}_{\mu\nu}{}^m\;,\qquad
F_{\mu\nu\, m}{}^{\alpha}=-\varepsilon^{\alpha \beta}\mathcal{F}_{\mu\nu \, m \beta}
\;,\qquad
F_{\mu\nu\, kmn}=\frac{\sqrt{2}}{4}\,\mathcal{F}_{\mu\nu \,kmn}
\;.
\label{dictFF}
\eea

This dictionary may be further confirmed upon comparing the action of 
external diffeomorphisms on both sides.
Indeed, the variations calculated in \eqref{COVDAA2}, \eqref{FINALdeltaxiC}, 
\eqref{finaldelC4} above,
precisely reproduce the EFT transformation law~\eqref{skewD} 
for the vectors ${\cal A}_\mu{}^M$,
provided we identify the components of the scalar matrix ${\cal M}^{MN}$
\eqref{Minvcomp} with the IIB fields according to
 \bea
  \phi^{-\frac{2}{3}}G^{mn} \ = \ e^{4\Phi/3}m^{mn}\;, \qquad
  C_{mn}{}^{\alpha} \ = \ -2 b_{mn}{}^{\alpha}\;,\qquad
C_{mnkl}=-\frac14\,c_{mnkl}\;.
\label{dictM}
\eea
This last identification is precisely compatible with the
gauge transformation behaviour \eqref{deltaMMM} as compared to the 
scalar components of \eqref{transC2}, \eqref{transC4}.
Let us also note, that with this dictionary the EFT covariant derivatives (\ref{covderMcomp})
for the scalar fields precisely translate into the components of the IIB field strengths 
\bea
{\cal D}_\mu b_{mn}{}^\alpha &=& -\frac12\,\overline{F}_{\mu mn}{}^\alpha\;,
\nonumber\\
\widehat{\cal D}_\mu c_{klmn} &=& -4\,\overline{F}_{\mu klmn}\;,
\label{dictFF2}
\eea
with $\widehat{\cal D}_\mu c_{klmn}$ from (\ref{Dhatc}).
Similarly, we have the identification
\bea
\partial_{[k} c_{lmnp]} + 12\,\varepsilon_{\alpha\beta}\,b_{[kl}{}^\alpha \partial_m b_{np]}{}^\beta~=~
X_{klmnp} &=& -\frac4{5}\,\overline{F}_{klmnp}
\;,
\eea
with $X_{klmnp}$ from (\ref{potX}).

We have thus identified the elementary EFT fields among the 
Kaluza-Klein components of the IIB fields.
So far, the identification has been solely based on the matching
of gauge symmetries on both sides. We will in the following show that the embedding of IIB into EFT
also holds dynamically on the level of the equations of motion.

\subsection{Dynamics and reconstruction of 3- and 4-forms}

In this section, we will show how the full IIB self-duality equations (\ref{FF55}) follow from 
the EFT dynamics. Along the way, we will establish explicit expressions for the remaining
components of the ten-dimensional 4-form, thereby completing the explicit embedding 
of the IIB theory.
To begin with, it is useful to first rewrite the various components 
of the IIB self-duality equations in terms of the Kaluza-Klein decomposed fields introduced
in section~\ref{subsec:redefinitions} above. With the IIB metric~\eqref{E10} given
in term of the EFT fields as
\bea
G_{\hat\mu\hat\nu} &=& \left(
\begin{array}{cc}
e^{5\Phi/6}\, g_{\mu\nu} + {\cal A}_\mu{}^m {\cal A}_\nu{}^n\,\phi_{mn}&  e^{-\Phi/2}\, m_{kn}\,{\cal A}_\mu{}^k\\
 e^{-\Phi/2}\, m_{mk} \,{\cal A}_\nu{}^k & e^{-\Phi/2}\, m_{mn} 
\end{array}
\right)
\;,
\label{G10}
\eea
the IIB self-duality equations (\ref{FF55}) split into the following three components
\bea
\overline{F}_{\mu\nu\rho\,mn} &=& 
\frac1{12}\,e^{2\Phi/3} \, 
\sqrt{-g}\, \varepsilon_{\mu\nu\rho\sigma\tau} \varepsilon_{mnklp}\,
\overline{F}{}^{\sigma\tau}{}_{qrs}\,m^{kq}m^{lr}m^{pq} 
\;,
\label{FF1}\\
\overline{F}_{\mu\nu\rho\sigma\,m} 
&=& 
-\frac1{24}\, e^{2\Phi} \, 
\sqrt{-g}  \,\varepsilon_{\mu\nu\rho\sigma\tau} m_{mn} \varepsilon^{nklpq}\,\overline{F}{}^{\tau}{}_{klpq}\, 
\;,
\label{FF2}\\
\overline{F}_{\mu\nu\rho\sigma\tau} &=& 
\frac1{120}\,e^{10\Phi/3}  \, 
\sqrt{-g}  \,\varepsilon_{\mu\nu\rho\sigma\tau} \varepsilon^{mnklp}\,\overline{F}{}_{mnklp} 
\;.
\label{FF3}
\eea
On the r.h.s.\ all external indices are raised and lowered with the metric $g_{\mu\nu}$,
and both $\varepsilon$-symbols denote the numerical tensor densities.
All explicit appearance of Kaluza-Klein vectors ${\cal A}_\mu{}^m$ from (\ref{G10})
is absorbed in the redefined $\overline{F}$'s.
We will now reproduce these equations one by one from the EFT dynamics.

Let us start from the $[mn]$ component of the EFT duality equations (\ref{dualityrel})
which can be integrated to
\bea
\tilde{\cal H}_{\mu\nu\rho\,mn} +{\cal O}_{mn\,\mu\nu\rho}
&=& \frac1{2}\,
e\varepsilon_{\mu\nu\rho\sigma\tau} \,
{\cal M}_{mn,M}\,
{\cal F}^{\sigma\tau\,M}
\;,
\label{explicit0}
\eea
where the ${\cal O}_{mn\,\mu\nu\rho}$ keeps track of the integration ambiguity and satisfies
\bea
\partial_{[k}{\cal O}_{mn]\,\mu\nu\rho}=0\qquad
\Longrightarrow\qquad
{\cal O}_{mn\,\mu\nu\rho} &\equiv& \partial_{[m} \xi_{n]\,\mu\nu\rho}\quad
\mbox{(locally)}\;.
\eea
Eliminating ${\cal F}_{\mu\nu\,\alpha}$ on the r.h.s.\ of (\ref{explicit0}) by means of (\ref{dualityA}) 
turns ${\cal M}_{MN}$ into $\tilde{\cal M}_{MN}$, such that
upon using the explicit expressions (\ref{tildeMB}), we obtain
\bea
 \partial_{[m} \xi_{n]\,\mu\nu\rho} 
&=& \frac1{12}\,e^{2\Phi/3}\,e\,\varepsilon_{\mu\nu\rho\sigma\tau} \varepsilon_{mnklp} m^{kq}m^{lr}m^{ps}\,
\widehat{\cal F}^{\sigma\tau}{}_{qrs}
\nonumber\\
&&{}
-\tilde{\cal H}_{\mu\nu\rho\,mn} 
- \sqrt{2}\,\varepsilon_{\alpha\beta}\,b_{mn}{}^\alpha \tilde{\cal H}_{\mu\nu\rho}{}^\beta
\;,
\label{explicit1}
\eea
with
\bea
\widehat{\cal F}_{\mu\nu\,klm} &\equiv&
{\cal F}_{\mu\nu\,klm}
+3\sqrt{2}\,b_{[kl}{}^\alpha {\cal F}_{|\mu\nu| m]\alpha} 
+ 3\sqrt{2}\,\varepsilon_{\alpha\beta}\,b_{n[k}{}^\alpha b_{lm]}{}^\beta {\cal F}_{\mu\nu}{}^n
+\frac12\,\sqrt{2}\,c_{klmn} \,{\cal F}_{\mu\nu}{}^n
\;.\nonumber\\
&=& 2\,\sqrt{2}\,\overline{F}_{\mu\nu\,klm}
\;,
\eea
where the last identity is easily confirmed upon using 
the dictionary of field strengths (\ref{FFFFF}), (\ref{dictFF}) and scalars (\ref{dictM}).
Together, the relation (\ref{explicit1}) then gives rise to
\bea
F_{\mu\nu\rho\,mn} -\frac14\,\varepsilon_{\alpha\beta}\,C_{mn}{}^\alpha \, \overline{F}_{\mu\nu\rho}{}^\beta
&=& \frac1{12}\,e^{2\Phi/3}\,e\,\varepsilon_{\mu\nu\rho\sigma\tau} \varepsilon_{mnklp} m^{kq}m^{lr}m^{ps}\,
\overline{F}^{\sigma\tau}{}_{qrs}
\;,\label{explicit2}
\eea
and thus precisely reproduces (\ref{FF1}) if we identify the 3-form component $C_{\mu\nu\rho\,m}$ from (\ref{redefC4}) as
\bea
C_{\mu\nu\rho\,m} &=& -\frac18\,\sqrt{2}\,\xi_{m\,\mu\nu\rho}
\;.
\label{defC3}
\eea
We have thus reproduced the first of the components of the IIB self-duality equations and 
along the way identified one of the missing components (\ref{defC3}) of the IIB four-form, that is not among
the fundamental EFT fields. It is defined by the first order differential equations (\ref{explicit2}) 
in terms of the EFT fields up to a gradient 
\bea
C_{\mu\nu\rho\,m}&\longrightarrow& 
C_{\mu\nu\rho\,m}+ \partial_m \lambda_{\mu\nu\rho}
\;,
\label{gaugexi}
\eea
corresponding to a gauge transformation in the IIB theory.

Let us continue towards the other components (\ref{FF2}), (\ref{FF3}),
of the self-duality relations. Consider the external curl of (\ref{explicit0}), which reads
\bea
4D_{[\mu} \tilde{\cal H}_{\nu\rho\sigma]\,mn} +4D^{\KK}_{[\mu} {\cal O}^{}_{\nu\rho\sigma]\,mn} &=& 
2\,e\varepsilon_{\tau\lambda[\nu\rho\sigma}\,D^{\KK}_{\mu]}\,\left({\cal M}_{mn,N}\,{\cal F}^{\tau\lambda\,N}\right)
\;, 
\eea
and use the Bianchi identity (\ref{BianchiB}) to find
\bea
4\,\partial_m \left( D^{\KK}_{[\mu} \xi^{}_{\nu\rho\sigma]\,n} \right)
&=&
6\,{\cal F}_{[\mu\nu}{}^k {\cal F}_{\rho\sigma]\,kmn}
+3\sqrt{2}\, \varepsilon^{\alpha\beta}\, {\cal F}_{[\mu\nu\,|m\alpha|} {\cal F}_{\rho\sigma]\,n\beta}
\nonumber\\
&&{}
+4\sqrt{2}\,  \partial_m \tilde{\cal H}_{[\mu\nu\rho}{}^\alpha\,{\cal A}_{\sigma]}{}_{n\alpha}
-e\varepsilon_{\mu\nu\rho\sigma\lambda}\,D^{\KK}_\tau\left({\cal M}_{mn,N}\,{\cal F}^{\tau\lambda\,N}\right)
\nonumber\\
&&{}
-3\sqrt{2}\,
\partial_m \left(  \varepsilon_{\alpha\beta}\,\tilde{\cal B}_{[\mu\nu}{}^\alpha\partial_{|n|} \tilde{\cal B}_{\rho\sigma]}{}^\beta \right)
+12\, \partial_m \left({\cal F}_{[\mu\nu}{}^k   \tilde{\cal B}_{\rho\sigma]\,kn}\right)
\nonumber\\
&&{}
+6\sqrt{2}\, \partial_m \left( \varepsilon^{\alpha\beta}
{\cal A}_{[\mu\,|n\alpha|}  {\cal F}_{\nu\rho}{}^k {\cal A}_{\sigma]\,k\beta} \right)
\;,
\label{interFA}
\eea
where both, left and right hand side are supposed to be explicitly projected
onto their part antisymmetric in $[mn]$\,.

In order to simplify the second line, we make use of the equations of motion
obtained by varying the Lagrangian~(\ref{EFTaction}) w.r.t.\ the vector fields $ {\cal A}_\mu{}^{mn}$
and using the duality equation (\ref{dualityA}) in order to eliminate ${\cal F}_{\mu\nu\,\alpha}$, 
\bea
0 &=&
-\frac1{24}\,\sqrt{2}\,   
\partial_{[m} \left( e^{2\Phi}m_{n]k}\,\widehat{\cal D}^\mu c_{pqrs}  \varepsilon^{kpqrs} \right)
+ D^{\KK}_\nu \left({\cal M}_{mn,M} {\cal F}^{\nu\mu\,M} \right) 
\nonumber\\
&&{}
+\frac1{6}\,\sqrt{2}\,\varepsilon^{\mu\nu\rho\sigma\tau}\,
 \partial_{[m} {\cal A}_{|\nu|n]\alpha} 
\tilde{\cal H}_{\rho\sigma\tau}{}^\alpha
-\frac1{12}\,\sqrt{2}\,\varepsilon^{\mu\nu\rho\sigma\tau}\,
  {\cal A}_{\nu\,[m|\alpha|} \partial_{n]} \tilde{\cal H}_{\rho\sigma\tau}{}^\alpha 
\nonumber\\
&&{}
+\frac3{4}\,\varepsilon^{\mu\nu\rho\sigma\tau} \,\Big(
\frac{\sqrt{2}}{6}\,
\varepsilon^{\alpha\beta} \,
{\cal F}_{\nu\rho\,m\alpha} {\cal F}_{\sigma\tau\,n\beta}
+\frac13\,{\cal F}_{\nu\rho\,mnp} {\cal F}_{\sigma\tau}{}^{p}
 +\frac{\sqrt{2}}9\,
 {\cal A}_{\nu [m|\alpha|}\,\partial_{n]} \tilde{\cal H}_{\rho\sigma\tau}{}^\alpha
\, 
\Big)
.
\eea
Together we find for (\ref{interFA})
\bea
4\,\partial_m \left( D^{\KK}_\mu \xi_{\nu\rho\sigma\,n} \right)
&=&
-\frac1{24}\,\sqrt{2}\, e\varepsilon_{\mu\nu\rho\sigma\lambda}\,\partial_m 
\left( e^{2\Phi}m_{nk}\,\widehat{\cal D}^\lambda c_{pqrs}  \varepsilon^{kpqrs} \right)
\nonumber\\
&&{}
-3\sqrt{2}\,
\partial_m \left(  \varepsilon_{\alpha\beta}\,\tilde{\cal B}_{\mu\nu}{}^\alpha\partial_n \tilde{\cal B}_{\rho\sigma}{}^\beta \right)
+12\, \partial_m \left({\cal F}_{\mu\nu}{}^k   \tilde{\cal B}_{\rho\sigma\,kn}\right)
\nonumber\\
&&{}
+6\sqrt{2}\, \partial_m \left( {\cal A}_\mu{}_{n\alpha}  \varepsilon^{\alpha\beta}
{\cal F}_{\nu\rho}{}^k {\cal A}_{\sigma\,k\beta} \right)
-4\sqrt{2}\,\partial_m\left(
{\cal A}_{\mu\,n\alpha} \tilde{\cal H}_{\nu\rho\sigma}{}^\alpha
\right)
\;,
\eea
again, projected onto the antisymmetric part $[mn]$\,.
The entire equation thus takes the form of an internal curl and can be integrated to
\bea
-\frac1{24}\,\sqrt{2}\, e\varepsilon_{\mu\nu\rho\sigma\lambda}\,  e^{2\Phi}m_{nk}\,\widehat{\cal D}^\lambda c_{pqrs}  \varepsilon^{kpqrs}&=&
4\, D^{\KK}_{[\mu} \xi_{\nu\rho\sigma]\,n} 
+3\sqrt{2}\,
 \varepsilon_{\alpha\beta}\,\tilde{\cal B}_{[\mu\nu}{}^\alpha\partial_{|n|} \tilde{\cal B}_{\rho\sigma]}{}^\beta 
\nonumber\\
&&{}
-12\,F_{[\mu\nu}{}^k   \tilde {\cal B}_{\rho\sigma]\,kn}
-6\sqrt{2}\,  \varepsilon^{\alpha\beta}\, {\cal F}_{[\mu\nu}{}^k  \,{\cal A}_{\rho\,|n\alpha|} {\cal A}_{\sigma]\,k\beta} 
\nonumber\\
&&{}
+4\sqrt{2}\,
{\cal A}_{\mu\,n\alpha} \tilde{\cal H}_{\nu\rho\sigma}{}^\alpha
+\partial_n \xi_{\mu\nu\rho\sigma}
\;,
\label{dualityC}
\eea
up to an internal gradient $\partial_n \xi_{\mu\nu\rho\sigma}$.
Applying the dictionary (\ref{dictAB}), (\ref{dictFF}) to translate all fields into the IIB
components, this equation becomes
\bea
-\frac1{24}\, e\varepsilon_{\mu\nu\rho\sigma\lambda}\varepsilon^{kpqrs}\,  
e^{2\Phi}m_{nk}\,\overline{F}^\lambda{}_{pqrs}  &=&
\overline{F}_{\mu\nu\rho\sigma\,n} 
-\partial_n \left({C}_{\mu\nu\rho\sigma}+\frac18 \sqrt{2}\, \xi_{\mu\nu\rho\sigma}\right)
\;,
\eea
i.e.\ reproduces equation (\ref{FF2}), provided we
identify the last missing component of the 4-form as
\bea
{C}_{\mu\nu\rho\sigma}&=&-\frac18 \sqrt{2}\, \xi_{\mu\nu\rho\sigma}
\;.
\label{defC4}
\eea
We have thus also reproduced the second component of the IIB self-duality equations and 
along the way identified the last missing components (\ref{defC4}) of the IIB four-form, that is not among
the fundamental EFT fields. It is defined by the first order differential equations (\ref{dualityC}) 
in terms of the EFT fields up to an additive function
\bea
C_{\mu\nu\rho\sigma}&\longrightarrow& C_{\mu\nu\rho\sigma}
+\Lambda_{\mu\nu\rho\sigma}(x)
\;,
\label{xix}
\eea
which we will fix in the following.
In order to find the last component (\ref{FF3}) of the self-duality equations,
we take the external curl of (\ref{dualityC})
\bea
-\partial_n D^{\KK}_{[\mu} \xi_{\nu\rho\sigma\tau]}
&=&
-\frac1{120}\,\sqrt{2}\, e\varepsilon_{\mu\nu\rho\sigma\tau}\,  
D^{\KK}_{\lambda}  \left(e^{2\Phi}m_{nk}\,\widehat{\cal D}^\lambda c_{pqrs}  \varepsilon^{kpqrs}\right) 
+2\sqrt{2}\,
{\cal F}_{[\mu\nu\,|n\alpha|} \tilde{\cal H}_{\rho\sigma\tau]}{}^\alpha
\nonumber\\
&&{}
+4\,{\cal F}_{[\mu\nu}{}^k   \left(
\tilde{\cal H}_{\rho\sigma\tau]\,kn} + \partial_{[k}  \xi_{\rho\sigma\tau]\,n]} \right) 
+2\sqrt{2}\,\varepsilon_{\alpha\beta} \partial_n \tilde{\cal B}_{[\mu\nu}{}^\beta
 \tilde{\cal H}_{\rho\sigma\tau]}{}^\alpha
\nonumber\\
&&{}
-2\sqrt{2}\,
 \varepsilon_{\alpha\beta}\,\tilde{\cal H}_{[\mu\nu\rho}{}^\alpha \partial_{|n|} \tilde{\cal B}_{\sigma\tau]}{}^\beta 
-6\,\sqrt{2} \varepsilon^{\alpha\beta}
{\cal F}_{[\mu\nu}{}^k  {\cal A}_{\rho\,|n \alpha|} {\cal F}_{\sigma\tau]\,k\beta} 
\nonumber\\
&&{}
+6\sqrt{2}\,\varepsilon^{\alpha\beta}
{\cal A}_{[\mu\,|n\alpha|} {\cal F}_{\nu\rho}{}^k {\cal F}_{\sigma\tau] k\beta} 
-3\sqrt{2}\,\partial_n\left(
 \varepsilon_{\alpha\beta}\,\tilde{\cal B}_{[\mu\nu}{}^\alpha D_\rho \tilde{\cal B}_{\sigma\tau]}{}^\beta \right)
 \nonumber\\
 &&{}
 +2\partial_n \left( {\cal F}_{[\mu\nu}{}^k   \xi_{\rho\sigma\tau]\,k}  \right)
\;,
\eea
which after using the equations of motion for $c_{klmn}$
turns into a full internal gradient and can be integrated to the equation
\bea
D^{\KK}_{[\mu} \xi_{\nu\rho\sigma\tau]}
+3\sqrt{2}\,
 \varepsilon_{\alpha\beta}\,\tilde{\cal B}_{[\nu\rho}{}^\alpha D_\mu \tilde{\cal B}_{\sigma\tau]}{}^\beta 
-2  {\cal F}_{[\mu\nu}{}^k   \xi_{\rho\sigma\tau]\,k} 
&=&
\frac{\sqrt{2}}{120}\, e\varepsilon_{\mu\nu\rho\sigma\tau}\varepsilon^{klmnp}\,e^{10\Phi/3}\,X_{klmnp}
\;,
\nonumber\\
\label{explicit7}
\eea
with $X$ from (\ref{potX}), up to some $y$-independent function. The latter can be set to zero by
properly fixing the freedom (\ref{xix}).
After translating (\ref{explicit7}) into the IIB fields, we thus find
\bea
5 D^{\KK}_{[\mu} {{C}}_{\nu\rho\sigma\tau]}
-\frac{15}{4}\,
 \varepsilon_{\alpha\beta}\,\overline{C}_{[\nu\rho}{}^\alpha D^{\KK}_\mu \overline{C}_{\sigma\tau]}{}^\beta 
-10  {\cal F}_{[\mu\nu}{}^k   {{C}}_{\rho\sigma\tau]\,k} 
&=&
\frac{1}{120}\, e\varepsilon_{\mu\nu\rho\sigma\tau}\varepsilon^{klmnp}\,e^{10\Phi/3}\,\overline{F}_{klmnp}
\;.
\nonumber\\
\eea
Thereby we find the last missing component (\ref{FF3})
of the IIB self-duality equation. We have thus shown that the full IIB
self-duality equations (\ref{FF55}) follow from the EFT dynamics,
provided we identify by (\ref{defC3}), (\ref{defC4}) the remaining components 
of the IIB 4-form. Together with the dictionary established in section (\ref{subsec:dictionary}),
this defines all the IIB fields in terms of the fundamental fields from EFT.

\subsection{Complementary checks}

We have in the preceding sections established the full dictionary between the IIB theory and
the EFT fields upon choosing the explicit solution (\ref{explicit_sectionB}) of the section constraint. In particular,
we have defined all the components of the IIB fields (\ref{IIBfields}) in terms of the fundamental 
EFT fields and shown that the EFT dynamics implies the full IIB self-duality equations (\ref{FF55}).
Via integrability this also implies the IIB second order field equations for the 4-form.
The remaining equations of motion of the IIB theory can be verified in a more straightforward 
manner, similar to the analogous discussion for the embedding of $D=11$ supergravity \cite{Hohm:2013vpa},
by using the explicit dictionary. 

As an example, let us collect the contributions to the kinetic terms for the IIB two-form doublet $\hat{C}_{\hat\mu\hat\nu}{}^\alpha$. 
According to their Kaluza-Klein decomposition, these contributions descend from different terms of the EFT Lagrangian:
the kinetic terms (\ref{Lscal}), (\ref{FFM}), (\ref{HHM}), and the scalar potential (\ref{potXX}),
giving rise to
\bea
e^{-1}\,{\cal L}_{{\rm 2-form}} &=&
-e^\Phi\,{\cal D}_\mu b_{mn}{}^\alpha {\cal D}^\mu b_{kl}{}^\beta\,m^{km}m^{ln}\,m_{\alpha\beta} 
-\frac14\,e^{-\Phi/3}\,m^{mn}\,m_{\alpha\beta}\,{\cal F}_{\mu\nu\,m}{}^\alpha {\cal F}^{\mu\nu}{}_n{}^\beta
\nonumber\\
&&{}
-\frac1{12}\,e^{-5\Phi/3}\,m_{\alpha\beta} \,
\tilde{\cal{H}}_{\mu\nu\rho}{}^\alpha \tilde{\cal{H}}^{\mu\nu\rho\,\beta}
-3\,e^{7\Phi/3}\,\partial_{[k} b_{mn]}{}^\alpha \partial_{l} b_{pq}{}^\beta\,m_{\alpha\beta}\,m^{kl}m^{mp}m^{nq}
\;.
\nonumber\\
\eea
Upon translating these fields into the IIB components via (\ref{dictFF})--(\ref{dictFF2}),
the Lagrangian takes the form
\bea
\begin{split}
{\cal L}_{{\rm 2-form}} \ = \ 
-\frac1{12}\,\sqrt{|G|} \Big(\, & 3\,{F}_{\mu mn}{}^\alpha {F}^{\mu mn\,\beta}+3\,{F}_{\mu\nu m}{}^\alpha {F}^{\mu\nu m\,\beta}\\
&+{F}_{\mu\nu\rho}{}^\alpha {F}^{\mu\nu\rho\,\beta}
+{F}_{kmn}{}^\alpha {F}^{kmn\,\beta}
\Big)m_{\alpha\beta}
\;,
\nonumber\\
\end{split}
\eea
where now all indices on the r.h.s.\ are raised and lowered with the full IIB metric (\ref{G10}).
The result thus precisely agrees with the corresponding kinetic term of the IIB (pseudo-)action (\ref{actionIIB}).
Similarly, we find from collecting all the EFT contributions to the 5-form kinetic term
\bea
{\cal L}_{{\rm 5-form}} &=&
-\frac1{15}\,\sqrt{|G|} \left( {F}_{klmnp} {F}^{klmnp}+5\,{F}_{\mu klmn} {F}^{\mu klmn}
+10\,{F}_{\mu\nu klm} {F}^{\mu\nu klm}\right)
\;,
\eea
which reproduces half of the components of the corresponding term in the pseudo-action (\ref{actionIIB}), 
with the other half doubling the contribution due to the self-duality equations (\ref{FF55}).\footnote{
Again, it is important that the self-duality equation (\ref{FF55}) is to be used in the pseudo-action (\ref{actionIIB})
only after deriving the field equations by variation. Strictly speaking, our proof of equivalence holds
on the level of the field equations.}

\section{Generalized Scherk-Schwarz compactification}

The manifestly covariant formulation of EFT described in the previous sections
has proven a rather powerful tool in order to describe consistent truncations by means of
a generalization of the Scherk-Schwarz ansatz~\cite{Scherk:1979zr} to the
exceptional space-time \cite{Hohm:2014qga}. 
This relates to gauged supergravity theories in lower dimensions
(in this case to $D=5$ supergravities), formulated in the embedding tensor formalism.
Via the explicit dictionary of EFT to $D=11$ and type IIB supergravity, this ansatz
then provides the full Kaluza-Klein embedding of various consistent truncations. 

The generalized Scherk-Schwarz ansatz in EFT is governed by a group-valued twist matrix $U\in {\rm E}_{6(6)}$,  depending on 
the internal coordinates, which rotates each fundamental 
group index.  
For instance, for the generalized metric the ansatz reads 
\bea
{\cal M}_{MN}(x,Y) &=& U_{M}{}^{\underline{K}}(Y)\,U_{N}{}^{\underline{L}}(Y)\,{M}_{\underline{K}\underline{L}}(x)\;, 
\label{genSS}
\eea
where $M_{\underline{M}\underline{N}}$ becomes the ${\rm E}_{6(6)}$-valued scalar matrix 
of five-dimensional gauged supergravity. 
This ansatz is invariant under a global ${\rm E}_{6(6)}$ symmetry acting on the underlined indices. 
Indeed, gauged supergravity in the embedding tensor formalism is covariant w.r.t.~a global duality 
group (${\rm E}_{6(6)}$ in the present case), 
although this is not a physical symmetry but rather relates  
different gauged supergravities to each other. 
In addition to the group valued twist matrix, consistency requires that we also
introduce a scale factor $\rho$, depending only on the internal coordinates, for fields 
carrying a non-zero density weight $\lambda$, for which the ansatz contains $\rho^{-3\lambda}$. 
We thus write the 
general reduction ansatz for all bosonic fields of the ${\rm E}_{6(6)}$ EFT (\ref{EFTfields})
as~\cite{Hohm:2014qga}
\bea\label{SSembedding}
{\cal M}_{MN}(x,Y) &=& U_{M}{}^{\underline{K}}(Y)\,U_{N}{}^{\underline{L}}(Y)\,{M}_{\underline{K}\underline{L}}(x)\;, 
\nonumber\\
 g_{\mu\nu}(x,Y) &=& \rho^{-2}(Y)\,{\bf g}_{\mu\nu}(x)\;,\nonumber\\
  {\cal A}_{\mu}{}^{M}(x,Y) &=& \rho^{-1}(Y) A_{\mu}{}^{\underline{N}}(x)(U^{-1})_{\underline{N}}{}^{M}(Y) \;, 
  \nonumber\\
  {\cal B}_{\mu\nu\,M}(x,Y) &=& \,\rho^{-2}(Y) U_M{}^{\underline{N}}(Y)\,B_{\mu\nu\,\underline{N}}(x)
  \;.
 \eea

We will call the above ansatz consistent if the twist matrix $U$ and the function $\rho$ 
factor out of all covariant expressions in the action, the gauge transformations or the equations of motion. 
If this is established, it follows that the reduction is consistent in the strong Kaluza-Klein sense 
that any solution of the lower-dimensional theory can be uplifted to a solution of the full theory, 
with the uplift formulas being (\ref{SSembedding}). Let us explain the required consistency conditions 
for the gauge transformations under internal generalized diffeomorphisms, 
for which the gauge parameter is subject to the same ansatz as the one-form gauge field, 
 \be\label{LambdaAnsatz}
  \Lambda^{M}(x,Y) \ = \ \rho^{-1}(Y)(U^{-1})_{\underline{N}}{}^{M}(Y)\,{\bf \Lambda}^{\underline{N}}(x)\;. 
 \ee 
We start with the field $g_{\mu\nu}$ that transforms as a scalar density of weight $\lambda=\tfrac{2}{3}$.  
Consistency of the ansatz (\ref{SSembedding}) requires that under gauge transformations 
we have 
 \be\label{consisg}
  \delta_{\Lambda} g_{\mu\nu}(x,Y) \ = \ \rho^{-2}(Y)\delta_{\Lambda}{\bf g}_{\mu\nu}(x)\;, 
 \ee
where the expression for $\delta_{\Lambda}{\bf g}_{\mu\nu}$ is $Y$-independent 
and can hence consistently be interpreted as  the gauge transformation for the 
lower-dimensional metric. The variation on the left-hand side yields, upon insertion of (\ref{LambdaAnsatz}), 
 \be
 \begin{split}
  \delta_{\Lambda} g_{\mu\nu} \ &= \ \Lambda^N\partial_N g_{\mu\nu}
  +\tfrac{2}{3}\,\partial_N\Lambda^N g_{\mu\nu}\\
   \ &= \ \rho^{-1}(U^{-1})_{\underline{K}}{}^{N}{\bf \Lambda}^{\underline{K}}\,\partial_N(\rho^{-2}{\bf g}_{\mu\nu})
  +\tfrac{2}{3}\,\partial_N(\rho^{-1}(U^{-1})_{\underline{K}}{}^{N}){\bf \Lambda}^{\underline{K}}\, \rho^{-2}{\bf g}_{\mu\nu}\\
  \ &= \ \tfrac{2}{3}\,\rho^{-3}\Big[\partial_N(U^{-1})_{\underline{K}}{}^{N}-4\,(U^{-1})_{\underline{K}}{}^{N}\,
  \rho^{-1}\partial_N \rho\Big]
  {\bf \Lambda}^{\underline{K}}\, {\bf g}_{\mu\nu}\;. 
 \end{split} 
 \ee 
If we now demand that 
 \be\label{consistencyeq}
  \partial_N(U^{-1})_{\underline{K}}{}^{N}-4\,(U^{-1})_{\underline{K}}{}^{N}\,\rho^{-1}\partial_N \rho 
  \ = \ 3\,\rho\,\vartheta_{\underline{K}}\;, 
 \ee 
where $\vartheta_{\underline{K}}$ is \textit{constant}, then the ansatz (\ref{consisg}) is established with 
 \be\label{DELg}
  \delta_{\Lambda}{\bf g}_{\mu\nu} \ = \ 2\,{\bf \Lambda}^{\underline{M}}\,\vartheta_{\underline{M}}\, {\bf g}_{\mu\nu}\;. 
 \ee 
This corresponds to a gauging of the so-called trombone symmetry that rescales the metric and the other tensor 
fields of the theory with specific weights. Here, $\vartheta_{\underline{K}}$ is the embedding tensor component 
for the trombone gauging, as introduced in \cite{LeDiffon:2008sh}. 
An important consistency condition is that (\ref{consistencyeq}) is a covariant equation under 
internal generalized diffeomorphisms. Treating the (inverse) twist matrix as a vector of weight zero, 
its divergence $\partial_N(U^{-1})_{\underline{M}}{}^{N}$ (recalling that  
the underlined index is inert) is not a scalar.  Indeed, a quick computation with (\ref{explicitLie}) 
using the section constraint shows that it transforms as a scalar density of weight $\lambda=-\tfrac{1}{3}$, 
except for the following anomalous term in the transformation 
 \be
  \Delta_{\Lambda}^{\rm nc}(\partial_N(U^{-1})_{\underline{M}}{}^{N}) \ = \ 
  -\tfrac{4}{3}\, \partial_N(\partial\cdot \Lambda)(U^{-1})_{\underline{M}}{}^{N}\;. 
 \ee 
This contribution is precisely cancelled by the anomalous variation of the second term 
in (\ref{consistencyeq}), provided $\rho$ is a scalar density of weight $\lambda(\rho)=-\tfrac{1}{3}$.  
Then both sides of (\ref{consistencyeq}) are scalar densities of weight $\lambda=-\tfrac{1}{3}$
and the equation is gauge covariant. 

Let us now turn to the consistency conditions required for fields with a non-trivial tensor structure
under internal generalized diffeomorphisms, as the generalized metric. 
In parallel to the above discussion we require that the twist matrices consistently factor out, i.e.\
 \be
  \delta_{\Lambda}{\cal M}_{MN}(x,Y) \ = \ U_M{}^{\underline{K}}(Y) U_{N}{}^{\underline{L}}(Y)
  \delta_{\Lambda}M_{\underline{K}\underline{L}}(x)\;.  
 \ee 
Using the explicit form of the gauge transformations given by generalized Lie derivatives (\ref{explicitLie}) 
one may verify by direct computation that this leads to consistent gauge transformations
\be
\delta_\Lambda {M}_{\underline{M}\underline{N}}(x)  \ = \  
2\,{\bf \Lambda}^{\underline{L}}(x)\left(
\Theta_{\underline{L}}{}^{\boldsymbol{\alpha}}
+\tfrac{9}{2}\,\vartheta_{\underline{R}}\,(t^{\boldsymbol{\alpha}})_{\underline{L}}{}^{\underline{R}}  \right) 
(t_{\boldsymbol{\alpha}})_{(\underline{M}}{}^{\underline{P}}\,{M}_{\underline{N}){\underline{P}}}(x)
\;,
\label{delM}
\ee
provided we assume the consistency conditions
 \be\label{2ndconsistency}
  \big[(U^{-1})_{\underline{M}}{}^{K}(U^{-1})_{\underline{N}}{}^{L}\partial_{K}U_{L}{}^{P}\big]_{\bf 351} \ = \ 
  \tfrac{1}{5}\, \rho\,
  \Theta_{\underline{M}}{}^{{\boldsymbol{\alpha}}} (t_{{\boldsymbol{\alpha}}})_{\underline{N}}{}^{\underline{P}}\;, 
 \ee
where the \textit{constant} $ \Theta_{\underline{M}}{}^{\boldsymbol{\alpha}}$ is the embedding tensor encoding 
conventional (i.e.~non-trombone) gaugings, and the left-hand side is projected onto the ${\bf 351}$ 
sub-representation. Specifically, writing the derivatives of $U$ in terms of 
 \be\label{calX}
  {\cal X}_{ \underline{M}\underline{N}}{}^{\underline{K}} \ \equiv \ 
  (U^{-1})_{\underline{M}}{}^{K}(U^{-1})_{\underline{N}}{}^{L}\partial_{K}U_{L}{}^{\underline{K}}
  \ \equiv \ {\cal X}_{\underline{M}}{}^{{\boldsymbol{\alpha}}}(t_{{\boldsymbol{\alpha}}})_{\underline{N}}{}^{\underline{K}}\;, 
 \ee 
where we used that since $U$ is group valued, $U^{-1}\partial U$ 
is Lie algebra valued (in the indices  $\underline{N}$, $\underline{K}$), so that we can expand 
it in terms of generators as done in the second equality,  
the projector acts as (c.f.~eq.~(4.13) in \cite{deWit:2002vt}), 
 \be\label{351projector}
 \begin{split}
  \big[\,{\cal X}_{\underline{M}}{}^{{\boldsymbol{\alpha}}}\,\big]_{\bf 351} \ &\equiv  \
  (\mathbb{P}_{\bf 351})_{\underline{M}}{}^{{\boldsymbol{\alpha}}\,\underline{N}}{}_{{\boldsymbol{\beta}}}\,
  {\cal X}_{\underline{N}}{}^{{\boldsymbol{\beta}}}\\
   \ &= \ \tfrac{1}{5}\Big(\,{\cal X}_{\underline{M}}{}^{{\boldsymbol{\alpha}}}
  -6\,(t^{{\boldsymbol{\alpha}}})_{\underline{P}}{}^{\underline{N}}\,(t_{{\boldsymbol{\beta}}})_{\underline{M}}{}^{\underline{P}}\,
  {\cal X}_{\underline{N}}{}^{{\boldsymbol{\beta}}}
  +\tfrac{3}{2}\,(t^{{\boldsymbol{\alpha}}})_{\underline{M}}{}^{\underline{P}}\,(t_{{\boldsymbol{\beta}}})_{\underline{P}}{}^{\underline{N}}\,
  {\cal X}_{\underline{N}}{}^{{\boldsymbol{\beta}}}\Big)\;. 
 \end{split} 
 \ee  
Also the condition (\ref{2ndconsistency}) is covariant under internal diffeomorphisms. 
This can be explicitly verified in the same way as the covariance of the torsion tensor (\ref{notorsion351}),  
which lives in the same representation. 
Let us emphasize that solving the consistency equations (\ref{consistencyeq}) 
and (\ref{2ndconsistency}) for $U$ and $\rho$ in general is a rather non-trivial problem. 
It would be important to develop a general theory 
for doing this, which plausibly may require a better understanding of large generalized diffeomorphisms, as in 
\cite{Hohm:2012gk,Hohm:2013bwa,Berman:2014jba,Naseer:2015tia}.

The consistency conditions (\ref{consistencyeq}) and (\ref{2ndconsistency}) can equivalently be encoded
in the structure of a `generalized  parallelization', see~\cite{Grana:2008yw}. To this end, the twist matrix 
$U$ and the scale factor $\rho$ are combined into a  vector of non-zero weight, 
 \be\label{UHat}
  (\widehat{U}^{-1})_{\underline{M}}{}^{N} \ \equiv \ \rho^{-1}\,(U^{-1})_{\underline{M}}{}^{N}\;. 
 \ee  
Since $\rho$ carries weight $-\tfrac{1}{3}$ this is a generalized vector of weight $\tfrac{1}{3}$, 
the same as for the gauge parameter, so that the generalized Lie derivative w.r.t.~$\widehat{U}^{-1}$
is well-defined. 
Both consistency conditions 
(\ref{consistencyeq}) and (\ref{2ndconsistency}) can then be encoded in the single 
manifestly covariant equation
  \be\label{FUllcond}
  \mathbb{L}_{\,\widehat{U}^{-1}_{\underline{M}}}\,\widehat{U}^{-1}_{\underline{N}} \ \equiv  \ 
  -X_{\underline{M}\underline{N}}{}^{\underline{K}}\,\widehat{U}^{-1}_{\underline{K}}\;, 
 \ee
with $X_{\underline{M}\underline{N}}{}^{\underline{K}}$ constant
and related to the $D=5$ embedding tensor as
 \be\label{GaugeStructure}
  X_{\underline{M}\underline{N}}{}^{\underline{K}} \ = \ \big(\Theta_{\underline{M}}{}^{{\boldsymbol{\alpha}}}
  +\tfrac{9}{2}\,\vartheta_{\underline{L}}(t^{{\boldsymbol{\alpha}}})_{\underline{M}}{}^{\underline{L}}\big)
   (t_{{\boldsymbol{\alpha}}})_{\underline{N}}{}^{\underline{K}} - \delta_{\underline{N}}{}^{\underline{K}}\,\vartheta_{\underline{M}}\;, 
 \ee 
 as we briefly verify in the following. In particular, equation (\ref{FUllcond}) implies that
 \be\label{rhoCOND}
  \mathbb{L}_{\,\widehat{U}^{-1}_{\underline{M}}}\,\rho \ = \ -\vartheta_{\underline{M}}\,\rho\;. 
 \ee 
The left-hand side of (\ref{FUllcond}) reads  
 \be\label{GENLieU}
  \begin{split}
    \big(\mathbb{L}_{\,\widehat{U}^{-1}_{\underline{M}}}\,
    \widehat{U}^{-1}_{\underline{N}}\big)^{K} \ = \ \;
    &(\widehat{U}^{-1})_{\underline{M}}{}^{N}\partial_N
    (\widehat{U}^{-1})_{\underline{N}}{}^{K}-6\,(t^{{\boldsymbol{\alpha}}})_{L}{}^{K}(t_{{\boldsymbol{\alpha}}})_{Q}{}^{P}\,
    \partial_P(\widehat{U}^{-1})_{\underline{M}}{}^{Q}\,(\widehat{U}^{-1})_{\underline{N}}{}^{L}\\
    &
    +\tfrac{1}{3}\,\partial_P(\widehat{U}^{-1})_{\underline{M}}{}^{P}\,(\widehat{U}^{-1})_{\underline{N}}{}^{K}\;. 
  \end{split}
 \ee    
Expressing this in terms of $U$ and $\rho$, writing the derivatives of $U$ in terms of (\ref{calX}), 
and multiplying both sides by $\widehat{U}_{K}{}^{\underline{K}}$, a quick computation yields 
 \be\label{UUUSTep}
 \begin{split}
  \widehat{U}_{K}{}^{\underline{K}}\,\big(\mathbb{L}_{\,\widehat{U}^{-1}_{\underline{M}}}\,
    \widehat{U}^{-1}_{\underline{N}}\big)^{K} \ = \ &\,
  -\rho^{-1}(t_{{\boldsymbol{\alpha}}})_{\underline{N}}{}^{\underline{K}}\,\Big({\cal X}_{\underline{M}}{}^{{\boldsymbol{\alpha}}}
  -6\,(t^{{\boldsymbol{\alpha}}})_{\underline{P}}{}^{\underline{Q}}\, (t_{{\boldsymbol{\beta}}})_{\underline{M}}{}^{\underline{P}}
   \, {\cal X}_{\underline{Q}}{}^{{\boldsymbol{\beta}}}\Big)
   -\tfrac{1}{3}\,\rho^{-1}\,{\cal X}_{\underline{P}\underline{M}}{}^{\underline{P}}\,\delta_{\underline{N}}{}^{\underline{K}}\\
  &\,+\Big(6\,(t^{{\boldsymbol{\alpha}}})_{\underline{N}}{}^{\underline{K}}\,(t_{{\boldsymbol{\alpha}}})_{\underline{M}}{}^{\underline{Q}}\,
  (U^{-1})_{\underline{Q}}{}^{P}
   -\tfrac{4}{3}\,(U^{-1})_{\underline{M}}{}^{P}\,\delta_{\underline{N}}{}^{\underline{K}}\Big)\rho^{-2}\partial_P\rho\;.  
 \end{split}
 \ee   
Next, the form of the projector (\ref{351projector}) 
onto the ${\bf 351}$ allows us to rewrite the terms in parenthesis in the first line of (\ref{UUUSTep}). One finds 
 \be
  \begin{split}
   \widehat{U}_{K}{}^{\underline{K}}\,\big(\mathbb{L}_{\,\widehat{U}^{-1}_{\underline{M}}}\,
    \widehat{U}^{-1}_{\underline{N}}\big)^{K} \ = \ &\,
   -5\,\rho^{-1}\big[\,{\cal X}_{\underline{M}\underline{N}}{}^{\underline{K}}\,\big]_{\bf 351}
   +\tfrac{1}{3}\,\rho^{-1}\,\delta_{\underline{N}}{}^{\underline{K}}\,\big(\partial_P(U^{-1})_{\underline{M}}{}^{P}
   -4\,(U^{-1})_{\underline{M}}{}^{P}\,\rho^{-1}\partial_P\rho\big)\\
   &\,-\tfrac{3}{2}\,\rho^{-1}(t_{\boldsymbol{\alpha}})_{\underline{N}}{}^{\underline{K}}\,(t^{\boldsymbol{\alpha}})_{\underline{M}}{}^{\underline{Q}}
   \big(\partial_P(U^{-1})_{\underline{Q}}{}^{P}
   -4\,(U^{-1})_{\underline{Q}}{}^{P}\,\rho^{-1}\partial_P\rho\big)\;. 
  \end{split}
 \ee   
Finally inserting  (\ref{consistencyeq}) and (\ref{2ndconsistency}), we obtain
 \be
  \widehat{U}_{K}{}^{\underline{K}}\,\big(\mathbb{L}_{\,\widehat{U}^{-1}_{\underline{M}}}\,
    \widehat{U}^{-1}_{\underline{N}}\big)^{K} \ = \ 
    -\Theta_{\underline{M}}{}^{{\boldsymbol{\alpha}}}(t_{{\boldsymbol{\alpha}}})_{\underline{N}}{}^{\underline{K}}
    +\delta_{\underline{N}}{}^{\underline{K}}\,\vartheta_{\underline{M}}
    -\tfrac{9}{2}\,(t_{{\boldsymbol{\alpha}}})_{\underline{N}}{}^{\underline{K}}\,(t^{{\boldsymbol{\alpha}}})_{\underline{M}}{}^{\underline{Q}}
    \,\vartheta_{\underline{Q}}\;, 
  \ee
which implies (\ref{GaugeStructure}) for the 
structure constants defined in (\ref{FUllcond}),  
thereby verifying the equivalence with (\ref{consistencyeq}), (\ref{2ndconsistency}). 

It is straightforward to verify that subject to (\ref{FUllcond}), 
the gauge transformations of all bosonic fields in (\ref{SSembedding}) reduce to the correct gauge transformations 
in gauged supergravity. 
Let us illustrate this for a vector of generic weight $\lambda$, for which the Scherk-Schwarz ansatz reads 
 \be
  V^M(x,Y) \ = \ \rho^{-3\lambda}(U^{-1})_{\underline{N}}{}^{M}(Y)\,V^{\underline{N}}(x)
  \ = \ \rho^{-3\lambda+1}(\widehat{U}^{-1})_{\underline{N}}{}^{M}(Y)\,V^{\underline{N}}(x)\;. 
 \ee 
Using (\ref{FUllcond}) and (\ref{rhoCOND}), its gauge transformation then takes the form
 \be
  \begin{split}
   \delta_{\Lambda}V^M \ &= \ \mathbb{L}_{\,{\bf \Lambda}^{\underline{K}}\,\widehat{U}^{-1}_{\underline{K}}}
   \big(\rho^{-3\lambda+1}(\widehat{U}^{-1})_{\underline{N}}{}^{M}\big)\,V^{\underline{N}} \\
   \ &= \ {\bf \Lambda}^{\underline{K}}\,\Big((-3\lambda+1)\big(\mathbb{L}_{\widehat{U}^{-1}_{\underline{K}}}\,\rho\big)
   \rho^{-3\lambda}\,(\widehat{U}^{-1})_{\underline{N}}{}^{M}+\rho^{-3\lambda+1}
   \mathbb{L}_{\widehat{U}^{-1}_{\underline{K}}}(\widehat{U}^{-1})_{\underline{N}}{}^{M}\Big)\,V^{\underline{N}}\\
   \ &= \ \rho^{-3\lambda+1}(\widehat{U}^{-1})_{\underline{N}}{}^{M}\,\Big(
   (3\lambda-1){\bf \Lambda}^{\underline{K}}\,\vartheta_{\underline{K}}\,V^{\underline{N}}
   -{\bf \Lambda}^{\underline{K}} \,X_{\underline{K}\underline{L}}{}^{\underline{N}}\,
   V^{\underline{L}}\Big)\;, 
  \end{split}
 \ee  
from which we read off, inserting (\ref{GaugeStructure}), 
 \be
  \delta_{{\bf \Lambda}}V^{\underline{N}} \ = \ -{\bf \Lambda}^{\underline{K}}\,
  \big(\Theta_{\underline{K}}{}^{{\boldsymbol{\alpha}}}+\tfrac{9}{2}\,\vartheta_{\underline{P}}(t^{{\boldsymbol{\alpha}}})_{\underline{K}}{}^{\underline{P}}
  \big)(t_{{\boldsymbol{\alpha}}})_{\underline{L}}{}^{\underline{N}}\,V^{\underline{L}}
  +3\,\lambda\,{\bf \Lambda}^{\underline{K}}\,\vartheta_{\underline{K}}\,V^{\underline{N}}\;. 
 \ee 
This is the expected transformation in gauged supergravity with general trombone gauging   
and in particular is compatible with (\ref{delM}) and (\ref{DELg}) for $\lambda=0$ and $\lambda=\tfrac{2}{3}$, 
respectively. 
As the covariant derivatives and field strengths are defined in terms of generalized Lie derivatives 
(or its antisymmetrization, the E-bracket), it follows immediately that also these objects reduce 
`covariantly' under Scherk-Schwarz, e.g.,  
\bea
{\cal D}_\mu g_{\nu\rho}(x,Y) &=&
\rho^{-2}\left(
\partial_\mu - A_{\mu}{}^{N}  {\vartheta}_{N}
\right) {\bf g}_{\nu\rho}
\;,\\
{\cal D}_\mu  {\cal M}_{MN}(x,Y) &=& 
U_M{}^{\underline{P}}U_N{}^{\underline{Q}}\,\Big(
\partial_\mu {M}_{\underline{PQ}} - 2 
A_\mu{}^{\underline{L}}\left(
\Theta_{\underline{L}}{}^{\boldsymbol{\alpha}}\,
+\tfrac{9}{2}\,\vartheta_{\underline{R}}\,(t^{\boldsymbol{\alpha}})_{\underline{L}}{}^{\underline{R}}\right) (t_{\boldsymbol{\alpha}})_{(\underline{M}}{}^{\underline{P}} {M}_{\underline{N})\underline{P}} \Big)
\;.
\nonumber
\eea 
In addition, the covariant two-form field strength reduces consistently,   
\bea
{\cal F}_{\mu\nu}{}^M(x,Y) \ = \ \rho^{-1}\,(U^{-1})_{\underline{N}}{}^M\, {F}_{\mu\nu}{}^{\underline{N}}(x)\;,  
\eea
with the $D=5$ covariant field strength $F_{\mu\nu}{}^N$ given by
\bea
F_{\mu\nu}{}^{\underline{M}} &\equiv& 
2\partial_{[\mu} A_{\nu]}{}^{\underline{M}} +  
X_{\underline{KL}}{}^{\underline{M}}\,A_{[\mu}{}^{\underline{K}} A_{\nu]}{}^{\underline{L}}
+d^{\underline{MKL} } X_{\underline{K}\underline{L}}{}^{\underline{N}}B_{\mu\nu\,\underline{N}} 
\;,
\eea
and similarly for the three-form curvature. 
Finally, one can verify that \textit{internal} covariant derivatives $\nabla_M$, 
whose connection components are only partially determined in terms of 
the physical fields, reduce covariantly under Scherk-Schwarz reduction for 
those contractions/projections that are fully determined. 
To this end one may start from the vielbein postulate that relates the
Christoffel-type connections to the USp(8) valued `spin-connections' 
and use the covariant constraints that determine projections of the 
Christoffel connection, e.g., the generalized torsion constraint (\ref{notorsion351}). 
The latter then determines, via (\ref{2ndconsistency}), 
the corresponding projections of the spin connection in terms of the embedding tensor. 
The general analysis proceedes in complete parallel to the discussion in \cite{Hohm:2014qga}. 
In particular, with the geometric definition (\ref{commutators}) 
of the curvature scalar, which is independent of undetermined connections,  
it follows that the potential reduces consistently and thus yields 
the scalar potential of five-dimensional 
gauged supergravity, whose form is uniquely determined by supersymmetry.  

Let us finally discuss the fermions $\psi_{\mu}{}^{i}$ and $\chi^{ijk}$, 
which  transform under the local Lorentz group USp$(8)$ and are scalar densities 
of weight $\tfrac{1}{6}$ and $-\frac{1}{6}$, respectively. 
Accordingly, the Scherk-Schwarz ansatz simply reads 
 \be
  \psi_{\mu}{}^{i}(x,Y) \ = \ \rho^{-\frac{1}{2}}(Y)\,\psi_{\mu}{}^{i}(x)\;, \qquad
  \chi^{ijk}(x,Y) \ = \ \rho^{\frac{1}{2}}(Y)\, \chi^{ijk}(x)\;. 
  \label{SSfermions}
 \ee 
Note in particular that the ansatz does not involve a `rotation' of the USp$(8)$ indices 
by Killing spinors, in contrast to conventional Kaluza-Klein compactifications. 
This is in accord with the fact that such a rotation is a USp$(8)$ 
transformation, which in the context of EFT is a gauge symmetry, and so would correspond to a  
deformation that is pure gauge and hence irrelevant. 
By the above discussion, the supersymmetry variations (\ref{transF}), (\ref{transB}) reduce consistently under Scherk-Schwarz.
In particular, the terms in the fermion variations of (\ref{transF}) depending on the 
internal covariant derivatives $\nabla_M$, whose connection components are fully determined, 
reduce to the projections of the embedding tensor (more precisely, the `flattened' embedding tensor 
often referred to as the `T-tensor') that determine the tensors $A_1$ and $A_2$ defining 
the fermion shifts in gauged supergravity.

To summarize, the reduction ansatz (\ref{SSembedding}), (\ref{SSfermions}) describes a consistent
truncation of ${\rm E}_{6(6)}$ EFT to a $D=5$ maximal gauged supergravity, provided the twist
matrices satisfy the consistency conditions (\ref{consistencyeq}) and (\ref{2ndconsistency}).
It is intriguing, that the match with lower-dimensional gauged supergravity,  
does in fact not explicitly use the section constraint
(provided the initial scalar potential is written in an appropriate form)
\cite{Berman:2012uy,Aldazabal:2013mya,Hohm:2014qga}.
Formally this allows to reproduce all $D=5$ maximal gauged supergravities, 
and it is intriguing to speculate about their possible higher-dimensional embedding 
upon a possible relaxation of the section constraints that would 
define a genuine extension of the original supergravity theories. 
For the moment it is probably fair to say that our understanding 
of a consistent extension of the framework is still limited.
If on the other hand the twist matrices $U$ do obey the section constraint 
(\ref{sectioncondition}), the reduction ansatz (\ref{SSembedding}), (\ref{SSfermions})
translates into a consistent truncation of the original $D=11$ or type IIB supergravity,
respectively, depending on to which solution of the section constraint the twist matrices $U$ belong.
With the explicit dictionary between EFT and the original supergravities, 
given above for type IIB and in \cite{Hohm:2013vpa} for $D=11$ supergravity,
the simple factorization ansatz (\ref{SSembedding}), (\ref{SSfermions}) then translates into a
highly non-linear ansatz for the consistent embedding of the lower-dimensional theory.

\section{Summary and Outlook}
We have reviewed the E$_{6(6)}$ exceptional field theory and established the precise 
embedding of ten-dimensional type IIB supergravity upon picking the corresponding solution of 
the section constraint.  
Given that, as shown here, the resulting theory admits the full ten-dimensional 
diffeomorphism invariance, maximal supersymmetry and the global SL$(2,\mathbb{R})$ S-duality invariance,
its equivalence to type IIB supergravity is guaranteed on general grounds.
It is nevertheless useful to work out the explicit embedding.  We have done so in this review by 
first matching the gauge symmetries on both sides. On the type IIB supergravity side, 
this requires a number of field redefinitions, which are largely analogous to those needed in conventional 
Kaluza-Klein compactifications. On the exceptional field theory side, this requires a suitable 
parametrization of the E$_{6(6)}$ valued `27-bein'. We have then given the explicit dictionary from
the various components of the IIB fields to the EFT fields after solving the section constraint.
We also established the on-shell equivalence of both theories and in particular showed how the three- and four-forms 
of type IIB, originating from components of the self-dual four-form in ten dimensions, 
are reconstructed on-shell in exceptional field theory in which these fields are not present from the start. 

Having determined the precise embedding of type IIB into E$_{6(6)}$ exceptional field theory, 
we can use the results of \cite{Hohm:2014qga} on 
generalized Scherk-Schwarz compactifications in exceptional field theory 
to give the explicit embedding 
of various consistent Kaluza-Klein truncations of type IIB. 
The details will appear in \cite{Baguet:2015xxx}. In particular, this establishes the 
Kaluza-Klein consistency of AdS$_5\times {\rm S}^5$ in type IIB and, more importantly, 
gives the precise embedding formulas. 
This requires the precise interplay between various identities whose validity appears 
somewhat miraculous from the point of view of conventional geometry but which find a natural 
interpretation within the extended geometry of exceptional field theory.

\subsection*{Acknowledgements}
We would like to thank Bernard de Wit, Edvard Musaev, Hermann Nicolai, 
Mario Trigiante and Yi-Nan Wang  for helpful discussions.
The work of O.H.~is supported by the U.S. Department of Energy (DoE) under the cooperative research agreement DE-FG02-05ER41360 and a DFG Heisenberg fellowship.


\begin{thebibliography}{10}

\bibitem{Cremmer:1979up}
E.~Cremmer and B.~Julia, { The ${SO}(8)$ supergravity},  { Nucl. Phys.} { B159}
  (1979)
141.

\bibitem{Cremmer:1980gs}
E.~Cremmer, { Supergravities in 5 dimensions},  in { Superspace and
  supergravity : proceedings} (S.~Hawking and M.~Rocek., eds.), Cambridge Univ.
  Press, 1980.
\newblock Nuffield Gravity Workshop, Cambridge.

\bibitem{Hull:1994ys}
C.~Hull and P.~Townsend, { {Unity of superstring dualities}},  { Nucl.Phys.} {
  B438} (1995) 109--137,
[\href{http://xxx.lanl.gov/abs/hep-th/9410167}{{\tt hep-th/9410167}}].

\bibitem{deWit:1986mz}
B.~de~Wit and H.~Nicolai, { $d = 11$ supergravity with local {$SU(8)$}
  invariance},  { Nucl.Phys.} { B274} (1986)
363.

\bibitem{Koepsell:2000xg}
K.~Koepsell, H.~Nicolai, and H.~Samtleben, { An exceptional geometry for {$D =
  11$} supergravity?},  { Class.Quant.Grav.} { 17} (2000) 3689--3702,
[\href{http://xxx.lanl.gov/abs/hep-th/0006034}{{\tt hep-th/0006034}}].

\bibitem{West:2001as}
P.~C. West, { {${{E}}_{11}$ and {M} theory}},  { Class. Quant. Grav.} { 18}
  (2001) 4443--4460,
[\href{http://xxx.lanl.gov/abs/hep-th/0104081}{{\tt hep-th/0104081}}].

\bibitem{Damour:2002cu}
T.~Damour, M.~Henneaux, and H.~Nicolai, { {${E}_{10}$} and a `small tension
  expansion' of {M} theory},  { Phys. Rev. Lett.} { 89} (2002) 221601,
[\href{http://xxx.lanl.gov/abs/hep-th/0207267}{{\tt hep-th/0207267}}].

\bibitem{HenryLabordere:2002dk}
P.~Henry-Labord\`ere, B.~Julia, and L.~Paulot, { Borcherds symmetries in {M}
  theory},  { JHEP} { 0204} (2002) 049,
[\href{http://xxx.lanl.gov/abs/hep-th/0203070}{{\tt hep-th/0203070}}].

\bibitem{West:2003fc}
P.~C. West, { {$E_{11}$}, {$SL(32)$} and central charges},  { Phys.Lett.} {
  B575} (2003) 333--342,
[\href{http://xxx.lanl.gov/abs/hep-th/0307098}{{\tt hep-th/0307098}}].

\bibitem{West:2004iz}
P.~C. West, { Brane dynamics, central charges and {$E_{11}$}},  { JHEP} { 0503}
  (2005) 077,
[\href{http://xxx.lanl.gov/abs/hep-th/0412336}{{\tt hep-th/0412336}}].

\bibitem{Hillmann:2009ci}
C.~Hillmann, { Generalized ${E}_{7(7)}$ coset dynamics and {$D=11$}
  supergravity},  { JHEP} { 0903} (2009) 135,
[\href{http://xxx.lanl.gov/abs/0901.1581}{{\tt 0901.1581}}].

\bibitem{Hohm:2013pua}
O.~Hohm and H.~Samtleben, { Exceptional form of ${D}=11$ supergravity},  {
  Phys.Rev.Lett.} { 111} (2013) 231601,
[\href{http://xxx.lanl.gov/abs/1308.1673}{{\tt 1308.1673}}].

\bibitem{Hohm:2013vpa}
O.~Hohm and H.~Samtleben, { Exceptional field theory {I}: ${E}_{6(6)}$
  covariant form of {M}-theory and type {IIB}},  { Phys.Rev.} { D89} (2014)
  066016,
[\href{http://xxx.lanl.gov/abs/1312.0614}{{\tt 1312.0614}}].

\bibitem{Hohm:2013uia}
O.~Hohm and H.~Samtleben, { Exceptional field theory {II}: {E}$_{7(7)}$},  {
  Phys.Rev.} { D89} (2014) 066017,
[\href{http://xxx.lanl.gov/abs/1312.4542}{{\tt 1312.4542}}].

\bibitem{Hohm:2014fxa}
O.~Hohm and H.~Samtleben, { Exceptional field theory {III}: ${E}_{8(8)}$},  {
  Phys.Rev.} { D90} (2014) 066002,
[\href{http://xxx.lanl.gov/abs/1406.3348}{{\tt 1406.3348}}].

\bibitem{Siegel:1993th}
W.~Siegel, { {Superspace duality in low-energy superstrings}},  { Phys.Rev.} {
  D48} (1993) 2826--2837,
[\href{http://xxx.lanl.gov/abs/hep-th/9305073}{{\tt hep-th/9305073}}].

\bibitem{Hull:2009mi}
C.~Hull and B.~Zwiebach, { Double field theory},  { JHEP} { 0909} (2009) 099,
[\href{http://xxx.lanl.gov/abs/0904.4664}{{\tt 0904.4664}}].

\bibitem{Hull:2009zb}
C.~Hull and B.~Zwiebach, { The gauge algebra of double field theory and
  {C}ourant brackets},  { JHEP} { 0909} (2009) 090,
[\href{http://xxx.lanl.gov/abs/0908.1792}{{\tt 0908.1792}}].

\bibitem{Hohm:2010jy}
O.~Hohm, C.~Hull, and B.~Zwiebach, { {Background independent action for double
  field theory}},  { JHEP} { 1007} (2010) 016,
[\href{http://xxx.lanl.gov/abs/1003.5027}{{\tt 1003.5027}}].

\bibitem{Hohm:2010pp}
O.~Hohm, C.~Hull, and B.~Zwiebach, { {Generalized metric formulation of double
  field theory}},  { JHEP} { 1008} (2010) 008,
[\href{http://xxx.lanl.gov/abs/1006.4823}{{\tt 1006.4823}}].

\bibitem{Hohm:2010xe}
O.~Hohm and S.~K. Kwak, { Frame-like geometry of double field theory},  {
  J.Phys.} { A44} (2011) 085404,
[\href{http://xxx.lanl.gov/abs/1011.4101}{{\tt 1011.4101}}].

\bibitem{Berman:2010is}
D.~S. Berman and M.~J. Perry, { Generalized geometry and {M} theory},  { JHEP}
  { 1106} (2011) 074,
[\href{http://xxx.lanl.gov/abs/1008.1763}{{\tt 1008.1763}}].

\bibitem{Berman:2011pe}
D.~S. Berman, H.~Godazgar, and M.~J. Perry, { {$SO(5,5)$} duality in {M}-theory
  and generalized geometry},  { Phys.Lett.} { B700} (2011) 65--67,
[\href{http://xxx.lanl.gov/abs/1103.5733}{{\tt 1103.5733}}].

\bibitem{Berman:2011jh}
D.~S. Berman, H.~Godazgar, M.~J. Perry, and P.~West, { Duality invariant
  actions and generalised geometry},  { JHEP} { 1202} (2012) 108,
[\href{http://xxx.lanl.gov/abs/1111.0459}{{\tt 1111.0459}}].

\bibitem{Aldazabal:2013mya}
G.~Aldazabal, M.~Gra{\~n}a, D.~Marqu{\'e}s, and J.~Rosabal, { {Extended
  geometry and gauged maximal supergravity}},  { JHEP} { 1306} (2013) 046,
[\href{http://xxx.lanl.gov/abs/1302.5419}{{\tt 1302.5419}}].

\bibitem{Hull:2007zu}
C.~Hull, { Generalised geometry for {M}-theory},  { JHEP} { 0707} (2007) 079,
[\href{http://xxx.lanl.gov/abs/hep-th/0701203}{{\tt hep-th/0701203}}].

\bibitem{Pacheco:2008ps}
P.~P. Pacheco and D.~Waldram, { {M}-theory, exceptional generalised geometry
  and superpotentials},  { JHEP} { 0809} (2008) 123,
[\href{http://xxx.lanl.gov/abs/0804.1362}{{\tt 0804.1362}}].

\bibitem{Coimbra:2011ky}
A.~Coimbra, C.~Strickland-Constable, and D.~Waldram, { {$E_{d(d)} \times
  \mathbb{R}^+$ generalised geometry, connections and M theory}},  { JHEP} {
  1402} (2014) 054,
[\href{http://xxx.lanl.gov/abs/1112.3989}{{\tt 1112.3989}}].

\bibitem{Coimbra:2012af}
A.~Coimbra, C.~Strickland-Constable, and D.~Waldram, { Supergravity as
  generalised geometry {II}: {$E_{d(d)} \times \mathbb{R}^+$} and {M} theory},
  { JHEP} { 1403} (2014) 019,
[\href{http://xxx.lanl.gov/abs/1212.1586}{{\tt 1212.1586}}].

\bibitem{Hitchin:2004ut}
N.~Hitchin, { {Generalized Calabi-Yau manifolds}},  { Quart.J.Math.Oxford Ser.}
  { 54} (2003) 281--308,
[\href{http://xxx.lanl.gov/abs/math/0209099}{{\tt math/0209099}}].

\bibitem{Gualtieri:2003dx}
M.~Gualtieri, { Generalized complex geometry},  { Ann. of Math. (2)} { 174}
  (2011), no.~1 75--123, [\href{http://xxx.lanl.gov/abs/math/0401221}{{\tt
  math/0401221}}].

\bibitem{Hohm:2015xna}
O.~Hohm and Y.-N. Wang, { Tensor hierarchy and generalized {C}artan calculus in
  {SL}(3) $\times$ {SL}(2) exceptional field theory},  { JHEP} { 1504} (2015)
  050,
[\href{http://xxx.lanl.gov/abs/1501.01600}{{\tt 1501.01600}}].

\bibitem{Abzalov:2015ega}
A.~Abzalov, I.~Bakhmatov, and E.~T. Musaev, { Exceptional field theory:
  ${SO}(5,5)$},
\href{http://xxx.lanl.gov/abs/1504.01523}{{\tt 1504.01523}}.

\bibitem{Wang:2015hca}
Y.-N. Wang, { Generalized {C}artan {C}alculus in general dimension},
\href{http://xxx.lanl.gov/abs/1504.04780}{{\tt 1504.04780}}.

\bibitem{Godazgar:2014nqa}
H.~Godazgar, M.~Godazgar, O.~Hohm, H.~Nicolai, and H.~Samtleben, {
  Supersymmetric {E$_{7(7)}$} exceptional field theory},  { JHEP} { 1409}
  (2014) 044,
[\href{http://xxx.lanl.gov/abs/1406.3235}{{\tt 1406.3235}}].

\bibitem{Musaev:2014lna}
E.~Musaev and H.~Samtleben, { Fermions and supersymmetry in {E}$_{6(6)}$
  exceptional field theory},  { JHEP} { 1503} (2015) 027,
[\href{http://xxx.lanl.gov/abs/1412.7286}{{\tt 1412.7286}}].

\bibitem{Hohm:2011zr}
O.~Hohm, S.~K. Kwak, and B.~Zwiebach, { Unification of type {II} strings and
  {T}-duality},  { Phys.Rev.Lett.} { 107} (2011) 171603,
[\href{http://xxx.lanl.gov/abs/1106.5452}{{\tt 1106.5452}}].

\bibitem{Hohm:2011dv}
O.~Hohm, S.~K. Kwak, and B.~Zwiebach, { Double field theory of type {II}
  strings},  { JHEP} { 1109} (2011) 013,
[\href{http://xxx.lanl.gov/abs/1107.0008}{{\tt 1107.0008}}].

\bibitem{Hohm:2014qga}
O.~Hohm and H.~Samtleben, { Consistent {K}aluza-{K}lein truncations via
  exceptional field theory},  { JHEP} { 1501} (2015) 131,
[\href{http://xxx.lanl.gov/abs/1410.8145}{{\tt 1410.8145}}].

\bibitem{Hohm:2011cp}
O.~Hohm and S.~K. Kwak, { Massive type {II} in double field theory},  { JHEP} {
  1111} (2011) 086,
[\href{http://xxx.lanl.gov/abs/1108.4937}{{\tt 1108.4937}}].

\bibitem{Aldazabal:2011nj}
G.~Aldazabal, W.~Baron, D.~Marques, and C.~Nunez, { The effective action of
  double field theory},  { JHEP} { 1111} (2011) 052,
[\href{http://xxx.lanl.gov/abs/1109.0290}{{\tt 1109.0290}}].

\bibitem{Geissbuhler:2011mx}
D.~Geissb\"uhler, { Double field theory and {$N=4$} gauged supergravity},  {
  JHEP} { 1111} (2011) 116,
[\href{http://xxx.lanl.gov/abs/1109.4280}{{\tt 1109.4280}}].

\bibitem{Grana:2012rr}
M.~Gra{\~n}a and D.~Marqu{\'e}s, { Gauged double field theory},  { JHEP} {
  1204} (2012) 020,
[\href{http://xxx.lanl.gov/abs/1201.2924}{{\tt 1201.2924}}].

\bibitem{Dibitetto:2012rk}
G.~Dibitetto, J.~Fernandez-Melgarejo, D.~Marques, and D.~Roest, { Duality
  orbits of non-geometric fluxes},  { Fortsch.Phys.} { 60} (2012) 1123--1149,
[\href{http://xxx.lanl.gov/abs/1203.6562}{{\tt 1203.6562}}].

\bibitem{Berman:2012uy}
D.~S. Berman, E.~T. Musaev, and D.~C. Thompson, { Duality invariant {M}-theory:
  {G}auged supergravities and {S}cherk-{S}chwarz reductions},  { JHEP} { 1210}
  (2012) 174,
[\href{http://xxx.lanl.gov/abs/1208.0020}{{\tt 1208.0020}}].

\bibitem{Musaev:2013rq}
E.~T. Musaev, { Gauged supergravities in 5 and 6 dimensions from generalised
  {S}cherk-{S}chwarz reductions},  { JHEP} { 1305} (2013) 161,
[\href{http://xxx.lanl.gov/abs/1301.0467}{{\tt 1301.0467}}].

\bibitem{Lee:2014mla}
K.~Lee, C.~Strickland-Constable, and D.~Waldram, { {Spheres, generalised
  parallelisability and consistent truncations}},
\href{http://xxx.lanl.gov/abs/1401.3360}{{\tt 1401.3360}}.

\bibitem{Blair:2014zba}
C.~D.~A. Blair and E.~Malek, { Geometry and fluxes of {${\rm SL}(5)$}
  exceptional field theory},  { JHEP} { 1503} (2015) 144,
[\href{http://xxx.lanl.gov/abs/1412.0635}{{\tt 1412.0635}}].

\bibitem{deWit:1986iy}
B.~de~Wit and H.~Nicolai, { The consistency of the {$S^7$} truncation in
  {$D=11$} supergravity},  { Nucl.Phys.} { B281} (1987)
211.

\bibitem{Nastase:1999kf}
H.~Nastase, D.~Vaman, and P.~van Nieuwenhuizen, { Consistency of the {AdS}$_7
  \times {S}^4$ reduction and the origin of self-duality in odd dimensions},  {
  Nucl. Phys.} { B581} (2000) 179--239,
[\href{http://xxx.lanl.gov/abs/hep-th/9911238}{{\tt hep-th/9911238}}].

\bibitem{Hull:1988jw}
C.~M. Hull and N.~P. Warner, { Noncompact gaugings from higher dimensions},  {
  Class. Quant. Grav.} { 5} (1988)
1517.

\bibitem{Baguet:2015xxx}
A.~Baguet, O.~Hohm, and H.~Samtleben, { Consistent type {IIB} reductions to maximal 5D supergravity},  
  {Phys.Rev.} { D92} (2015) 065004,
[\href{http://xxx.lanl.gov/abs/1506.01385}{{\tt 1506.01385}}].

\bibitem{Berman:2012vc}
D.~S. Berman, M.~Cederwall, A.~Kleinschmidt, and D.~C. Thompson, { {The gauge
  structure of generalised diffeomorphisms}},  { JHEP} { 1301} (2013) 064,
[\href{http://xxx.lanl.gov/abs/1208.5884}{{\tt 1208.5884}}].

\bibitem{deWit:2005hv}
B.~de~Wit and H.~Samtleben, { Gauged maximal supergravities and hierarchies of
  nonabelian vector-tensor systems},  { Fortschr. Phys.} { 53} (2005) 442--449,
[\href{http://xxx.lanl.gov/abs/hep-th/0501243}{{\tt hep-th/0501243}}].

\bibitem{deWit:2008ta}
B.~de~Wit, H.~Nicolai, and H.~Samtleben, { Gauged supergravities, tensor
  hierarchies, and {M}-theory},  { JHEP} { 0802} (2008) 044,
[\href{http://xxx.lanl.gov/abs/arXiv:0801.1294}{{\tt arXiv:0801.1294}}].

\bibitem{deWit:2004nw}
B.~de~Wit, H.~Samtleben, and M.~Trigiante, { The maximal ${D} = 5$
  supergravities},  { Nucl. Phys.} { B716} (2005) 215--247,
[\href{http://xxx.lanl.gov/abs/hep-th/0412173}{{\tt hep-th/0412173}}].

\bibitem{Cederwall:2013naa}
M.~Cederwall, J.~Edlund, and A.~Karlsson, { Exceptional geometry and tensor
  fields},  { JHEP} { 1307} (2013) 028,
[\href{http://xxx.lanl.gov/abs/1302.6736}{{\tt 1302.6736}}].

\bibitem{Cremmer:1997ct}
E.~Cremmer, B.~Julia, H.~Lu, and C.~N. Pope, { {Dualisation of dualities. I}},
  { Nucl. Phys.} { B523} (1998) 73--144,
[\href{http://xxx.lanl.gov/abs/hep-th/9710119}{{\tt hep-th/9710119}}].

\bibitem{Schwarz:1983wa}
J.~H. Schwarz and P.~C. West, { {Symmetries and transformations of chiral
  ${N}=2$ ${D}=10$ supergravity}},  { Phys. Lett.} { B126} (1983)
301.

\bibitem{Schwarz:1983qr}
J.~H. Schwarz, { Covariant field equations of chiral {$N=2$, $D=10$}
  supergravity},  { Nucl.Phys.} { B226} (1983)
269.

\bibitem{Howe:1983sra}
P.~S. Howe and P.~C. West, { {The complete ${N}=2$, ${D}=10$ supergravity}},  {
  Nucl. Phys.} { B238} (1984)
181.

\bibitem{Ciceri:2014wya}
F.~Ciceri, B.~de~Wit, and O.~Varela, { {IIB} supergravity and the {E}$_{6(6)}$
  covariant vector-tensor hierarchy},  { JHEP} { 1504} (2015) 094,
[\href{http://xxx.lanl.gov/abs/1412.8297}{{\tt 1412.8297}}].

\bibitem{Scherk:1979zr}
J.~Scherk and J.~H. Schwarz, { How to get masses from extra dimensions},  {
  Nucl. Phys.} { B153} (1979)
61--88.

\bibitem{LeDiffon:2008sh}
A.~Le~Diffon and H.~Samtleben, { Supergravities without an action: {G}auging
  the trombone},  { Nucl. Phys.} { B811} (2009) 1--35,
[\href{http://xxx.lanl.gov/abs/0809.5180}{{\tt 0809.5180}}].

\bibitem{deWit:2002vt}
B.~de~Wit, H.~Samtleben, and M.~Trigiante, { On {L}agrangians and gaugings of
  maximal supergravities},  { Nucl. Phys.} { B655} (2003) 93--126,
[\href{http://xxx.lanl.gov/abs/hep-th/0212239}{{\tt hep-th/0212239}}].

\bibitem{Hohm:2012gk}
O.~Hohm and B.~Zwiebach, { Large gauge transformations in double field theory},
   { JHEP} { 1302} (2013) 075,
[\href{http://xxx.lanl.gov/abs/1207.4198}{{\tt 1207.4198}}].

\bibitem{Hohm:2013bwa}
O.~Hohm, D.~L{\"u}st, and B.~Zwiebach, { The spacetime of double field theory:
  {R}eview, remarks, and outlook},  { Fortsch.Phys.} { 61} (2013) 926--966,
[\href{http://xxx.lanl.gov/abs/1309.2977}{{\tt 1309.2977}}].

\bibitem{Berman:2014jba}
D.~S. Berman, M.~Cederwall, and M.~J. Perry, { {Global aspects of double
  geometry}},  { JHEP} { 1409} (2014) 066,
[\href{http://xxx.lanl.gov/abs/1401.1311}{{\tt 1401.1311}}].

\bibitem{Naseer:2015tia}
U.~Naseer, { {A note on large gauge transformations in double field theory}},
 JHEP {1506} (2015) 002,
[\href{http://xxx.lanl.gov/abs/1504.05913}{{\tt 1504.05913}}].

\bibitem{Grana:2008yw}
M.~Grana, R.~Minasian, M.~Petrini, and D.~Waldram, { {T}-duality, generalized
  geometry and non-geometric backgrounds},  { JHEP} { 0904} (2009) 075,
[\href{http://xxx.lanl.gov/abs/0807.4527}{{\tt 0807.4527}}].

\end{thebibliography}

\providecommand{\href}[2]{#2}\begingroup\raggedright\endgroup

\end{document}